\documentclass[prd,nofootinbib,preprintnumbers,twocolumn,
showpacs,floatfix,superscriptaddress]{revtex4-1}
\pdfoutput=1
\usepackage{amssymb}
\usepackage{braket}
\usepackage{amsthm}
\usepackage{mathtools}
\usepackage{physics}
\usepackage{hyperref}
\usepackage[usenames, dvipsnames]{color}

\usepackage{amssymb,mathtools}
\def\eq#1{{Eq.~(\ref{#1})}}
\def\eqs#1#2{{Eqs.~(\ref{#1})--(\ref{#2})}}
\def\fig#1{{Fig.~\ref{#1}}}

\def\Table#1{{Table~\ref{#1}}}

\def\sect#1{{Section~\ref{#1}}}

\def\app#1{{Appendix~\ref{#1}}}

\def\abs#1{\left| #1\right|}

\def\Re{\mbox{Re}\,}

\begin{document}


\title{One constraint to kill them all?}

\author{Luca Di Luzio}
\email{luca.di-luzio@durham.ac.uk}
\affiliation{\normalsize \it 
Institute for Particle Physics Phenomenology, Department of Physics, Durham University, DH1 3LE, Durham, United Kingdom}
\author{Matthew Kirk}
\email{m.j.kirk@durham.ac.uk}
\affiliation{\normalsize \it 
Institute for Particle Physics Phenomenology, Department of Physics, Durham University, DH1 3LE, Durham, United Kingdom}
\author{Alexander Lenz}
\email{alexander.lenz@durham.ac.uk}
\affiliation{\normalsize \it 
Institute for Particle Physics Phenomenology, Department of Physics, Durham University, DH1 3LE, Durham, United Kingdom}

\begin{abstract}
  {Many new physics models that explain the intriguing anomalies in the $b$-quark flavour sector
  are severely constrained by $B_s$-mixing, for which the Standard Model prediction and experiment agreed well until recently.
  The most recent FLAG average of lattice results for the non-perturbative matrix elements
  points, however, in the direction of a small discrepancy
  in this observable. Using up-to-date inputs
  from standard sources such as PDG, FLAG and one of the two leading CKM fitting groups
  to determine $\Delta M_s^{\rm SM}$,
  we find a severe reduction of the allowed parameter space of $Z'$ and leptoquark models explaining the $B$-anomalies.  
  Remarkably, in the former case the upper bound on the $Z'$ mass approaches dangerously close to 
  the energy scales already probed by the LHC.
  We finally identify some model building directions in order to alleviate the tension with $B_s$-mixing.}
\end{abstract}

\maketitle




\section{Introduction}
\label{sec:intro}
Direct searches for 
new physics (NP)
effects at the LHC have so far 
shown no discrepancies from the Standard Model (SM),
while we have an intriguing list of deviations between experiment and theory for flavour observables. In particular
$b \to s \ell^+ \ell^-$ transitions seem to be in tension with the SM expectations: branching ratios of hadronic 
$b \to s \mu^+ \mu^-$ decays \cite{Aaij:2014pli,Aaij:2015esa,Khachatryan:2015isa}
and the angular distributions for $B \to K^{(*)} \mu^+ \mu^-$ decay
\cite{Aaij:2015esa,Khachatryan:2015isa,Lees:2015ymt,Wei:2009zv,Aaltonen:2011ja,Aaij:2015oid,Abdesselam:2016llu,Wehle:2016yoi,Aaboud:2018krd,Sirunyan:2017dhj}
hint at a negative, beyond the SM (BSM) contribution to $C_9$
\cite{Descotes-Genon:2013wba,Beaujean:2013soa,Altmannshofer:2014rta,Descotes-Genon:2015uva,Hurth:2016fbr,Altmannshofer:2017fio,Ciuchini:2017mik,Geng:2017svp,Capdevila:2017bsm,Altmannshofer:2017yso,DAmico:2017mtc,Alok:2017sui}.
The significance of the effect is still under discussion because of the difficulty of determining the exact size of
the hadronic contributions
(see e.g.~\cite{Jager:2012uw,Jager:2014rwa,Descotes-Genon:2014uoa,Ciuchini:2015qxb,Chobanova:2017ghn,Capdevila:2017ert,Bobeth:2017vxj}). 
Estimates of the combined significance of all these deviations range between three and almost six standard deviations. 
A theoretically much cleaner observable is given by the lepton flavour universality (LFU) 
ratios $R_K$ and $R_{K^*}$ \cite{Hiller:2003js,Bordone:2016gaq},
where hadronic uncertainties drop out to a very large extent.
Here again a sizeable deviation from the SM expectation is found by LHCb
\cite{Aaij:2014ora,Aaij:2017vbb}.
Such an effect might arise for instance 
from new particles coupling to $b \bar{s}$ and $\mu^+ \mu^-$, while leaving the $e^+e^-$-coupling mainly unchanged
(see e.g.~\cite{Buras:2013qja,Gauld:2013qja,Buras:2013dea,Altmannshofer:2014cfa,Crivellin:2015mga,Crivellin:2015lwa,Celis:2015ara,Belanger:2015nma,Falkowski:2015zwa,Carmona:2015ena,Allanach:2015gkd,Chiang:2016qov,Boucenna:2016wpr,Megias:2016bde,Boucenna:2016qad,Altmannshofer:2016jzy,Crivellin:2016ejn,GarciaGarcia:2016nvr,Bhatia:2017tgo,Cline:2017lvv,Baek:2017sew,Cline:2017ihf,DiChiara:2017cjq,Kamenik:2017tnu,Ko:2017lzd,Ko:2017yrd,Alonso:2017bff,Ellis:2017nrp,Alonso:2017uky,Carmona:2017fsn}
for an arbitrary set of papers investigating $Z'$ models).
Any new $b \bar{s}$-coupling immediately leads to tree-level contributions to
$B_s$-mixing, which is severely constrained by experiment.
For quite some time the SM value for the mass difference $\Delta M_s$ of neutral 
$B_s$ mesons -- triggering the oscillation frequency -- was in perfect agreement with experiment,
see e.g. \cite{Artuso:2015swg} or \cite{Lenz:2011ti}.
Taking, however, the most recent lattice inputs, in particular the new average provided by the Flavour Lattice Averaging Group (FLAG) 
one gets a SM value considerably above the measurement. In this paper we investigate the drastic  
consequences of this new theory prediction.
In Section \ref{BmixingSM} we review the SM prediction of $B_s$-mixing, whose consequences for BSM models trying to explain the 
$B$-anomalies are studied in Section \ref{BmixingBSM}. We conclude in Section \ref{conclusions}. 
In the Appendices we give further details of the SM prediction as well
as a more critical discussion of the theoretical uncertainties.

\section{$B_s$-mixing in the SM}
\label{BmixingSM}

The mass difference of the mass eigenstates of the neutral $B_s$ mesons is given by
\begin{equation}
\Delta M_s \equiv M_H^s - M_L^s
=
2 \left|M_{12}^s\right| \, .
\label{DMdef}
\end{equation}
\begin{figure}[ht]
\includegraphics[width=0.48 \textwidth]{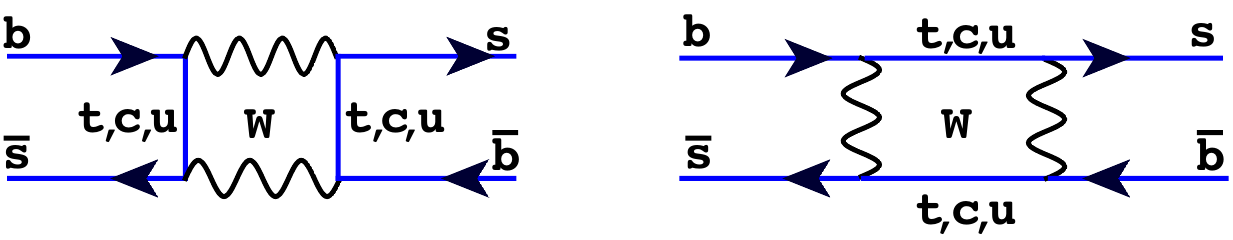}
\caption{\label{box} SM diagrams for the transition between $B_s$ and $\bar{B}_s$ mesons.
 The contribution of internal off-shell particles is denoted by $M_{12}^s$. 
}
\end{figure}
The calculation of the box diagrams in Fig.~\ref{box} gives the SM value for $M_{12}^s$,
see e.g.~\cite{Artuso:2015swg} for a brief review, and one gets
\begin{equation}
M_{12}^s = \frac{G_F^2}{12 \pi^2}
\lambda_t^2 M_W^2 S_0(x_t)
B f_{B_s}^2  M_{B_s} \hat{\eta }_B\, ,
\label{M12}
\end{equation}
with the Fermi constant $G_F$, the masses of the $W$ boson,
$M_W$, and of the $B_s$ meson, $M_{B_s}$. 
Using CKM unitarity one finds only one contributing CKM structure 
$ \lambda_t = V_{ts}^* V_{tb}$.
The CKM elements are the only place in Eq.~(\ref{M12}) where an imaginary part can arise.
The result of the 1-loop diagrams given in Fig.~\ref{box}
is denoted by the Inami-Lim function \cite{Inami:1980fz}
$ S_0(x_t= (\bar{m}_t(\bar{m}_t))^2/M_W^2) \approx 2.36853$, where $\bar{m}_t(\bar{m}_t)$ is
the $\overline{\rm MS}$-mass \cite{Bardeen:1978yd} of the top quark.
Perturbative 2-loop QCD corrections are compressed in the factor
$\hat{\eta }_B \approx 0.83798$, they have been calculated by
\cite{Buras:1990fn}.
In the SM calculation of $M_{12}^s$ one four quark $\Delta B=2$ operator
arises
\begin{equation}
Q = \bar s^\alpha \gamma_\mu (1- \gamma_5) b^\alpha
        \times
        \bar s^\beta  \gamma^\mu (1- \gamma_5) b^\beta \, .
      \label{Q}
\end{equation}
The hadronic matrix element of this operator is parametrised in terms of a decay constant $f_{B_s}$ and a bag parameter
$B$:
\begin{equation}
\langle Q \rangle \equiv \langle B_s^0|Q |\bar{B}_s^0 \rangle
       =
      \frac{8}{3}  M_{B_s}^2  f_{B_s}^2 B (\mu) \, ,
\label{ME1}
\end{equation}
We also indicated the renormalisation scale dependence
of the bag parameter; in our analysis we take $\mu = \bar{m}_b(\bar{m}_b)$. 
\\
Sometimes a different notation for the QCD corrections and the bag parameter
is used in the literature
(e.g.~by FLAG: \cite{Aoki:2016frl}),
$(\eta_B, \hat{B})$ instead of $(\hat{\eta}_B, B)$ with
\begin{align}
\hat{\eta}_B B & \equiv \eta_B \hat{B} = \eta_B   \alpha_s(\mu)^{-\frac{6}{23}} \left[ 1 + \frac{\alpha_s(\mu)}{4 \pi} \frac{5165}{3174} \right] B
\; ,
\\
\hat{B} & = 1.51926 \, B \; .
\end{align}
The parameter $\hat{B}$ has the advantage
of being renormalisation scale and scheme independent.
\\
A commonly used SM prediction of $\Delta M_s$ was given by \cite{Lenz:2011ti,Artuso:2015swg}
\begin{align}
\label{DeltaMold2011}
\Delta M_s^{\rm SM,\, 2011} &= \left(17.3 \pm 2.6 \right)\, \mbox{ps}^{-1} \, , \\
\label{DeltaMold2015}
\Delta M_s^{\rm SM,\, 2015} &= \left(18.3 \pm 2.7 \right)
\, \mbox{ps}^{-1} \; .
\end{align}
Both predictions agreed very well with the experimental measurement \cite{Amhis:2016xyh}
\begin{equation}
\Delta M_s^{\rm Exp} = \left(17.757 \pm 0.021 \right)\; \mbox{ps}^{-1} \; .
\label{DeltaMExp}
\end{equation}
In 2016 Fermilab/MILC presented a new calculation \cite{Bazavov:2016nty}, which gave considerably larger values for
the non-perturbative parameter, resulting in values around 20 ps$^{-1}$ for the mass difference
\cite{Bazavov:2016nty,Blanke:2016bhf,Jubb:2016mvq,Buras:2016dxz,Altmannshofer:2017uvs} and being thus larger than experiment. An independent confirmation
of these large values would of course be desirable; a first step in that direction has been done by the HQET sum rule
calculation of \cite{Kirk:2017juj} which is in agreement with Fermilab/MILC for the bag parameters.

Using the most recent numerical inputs listed in \app{app:input}
we predict the mass difference of the neutral $B_s$ mesons to be\footnote{A more conservative determination of the SM value of 
the mass difference using only tree-level inputs for the CKM parameters can be found in \eq{CKMtreeonly}.}
\begin{equation}
\boxed{
\Delta M_s^{\rm SM,\, 2017} = \left(20.01 \pm 1.25 \right) \; \mbox{ps}^{-1} \; .
}
\label{DeltaM2017}
\end{equation}
Here the dominant uncertainty still comes  from the lattice predictions for the non-perturbative
parameters  $B$ and $ f_{B_s}$, giving a relative error of $5.8 \%$.
The uncertainty in the CKM elements
contributes $2.1\%$ to the error budget. The CKM parameters were determined assuming
unitarity of the $3 \times 3$ CKM matrix. 
The uncertainties due to ${m}_t$, ${m}_b$ and $\alpha_s$
can be safely neglected at the current stage.
A detailed discussion of the input parameters and the
error budget is given in \app{app:input} and \app{app:error}, respectively.
The new central value for the mass difference in Eq.~(\ref{DeltaM2017}) is 1.8 $\sigma$ above 
the experimental one given in Eq.~(\ref{DeltaMExp}). This difference has profound implications for NP models 
that predict sizeable positive contributions to $B_s$-mixing. The new value for the SM prediction depends strongly on the
non-perturbative input as well as the values of the CKM elements. We use the averages that are provided by
the lattice community (FLAG) and by one of the two leading CKM fitting groups (CKMfitter) -- see 
\app{app:Lattice} and \app{app:CKM} for a further discussion of these inputs.

\section{$B_s$-mixing beyond the SM}
\label{BmixingBSM}
To determine the allowed space for NP effects in $B_s$-mixing we compare the experimental
measurement of the mass difference with the prediction in the SM plus NP: 
\begin{equation}
\Delta M_s^{\rm Exp} = 2 \left| M_{12}^{\rm SM} + M_{12}^{\rm NP} \right|
= \Delta M_s^{\rm SM} \left| 1 +  \frac{M_{12}^{\rm NP}}{M_{12}^{\rm SM}}\right|  \, .
\label{MsExpvsNP}
\end{equation}
For this equation we will use in the SM part the CKM elements, which have been determined  assuming the validity
of the SM. In the presence of BSM effects the CKM elements used in the prediction of $M_{12}^{\rm SM}$ could
in general differ from the ones we use -- see e.g.~the case of a fourth chiral fermion generation \cite{Bobrowski:2009ng}.
In the following, 
we will assume that NP effects do not involve sizeable shifts in the CKM elements.

A simple estimate shows that the improvement of the SM prediction from \eqs{DeltaMold2011}{DeltaMold2015} to 
\eq{DeltaM2017} can have a drastic impact on the size of the allowed BSM effects on $B_s$-mixing. 
For a generic NP model we can parametrise 
\begin{equation}
\label{LambdaNP}
\frac{\Delta M_s^{\rm Exp}}{\Delta M_s^{\rm SM}} = \abs{1 + \frac{\kappa}{\Lambda^2_{\rm NP}}} \, ,
\end{equation}
where $\Lambda_{\rm NP}$ denotes the mass scale of the NP mediator and $\kappa$ is a dimensionful quantity 
which encodes NP couplings and the SM contribution. If $\kappa > 0$, which is often the case in 
many BSM scenarios for $B$-anomalies considered in the literature, and since 
$\Delta M_s^{\rm SM} > \Delta M_s^{\rm Exp}$, 
the 2$\sigma$ bound on $\Lambda_{\rm NP}$ scales like
\begin{equation}
\label{scalingNP}
\frac{\Lambda^{\rm 2017}_{\rm NP}}{\Lambda^{\rm 2015}_{\rm NP}} 
= \sqrt{
\frac{\frac{\Delta M_s^{\rm Exp}}{\left(\Delta M_s^{\rm SM} - 2 \delta\Delta M_s^{\rm SM}\right)^{2015}} - 1}
{\frac{\Delta M_s^{\rm Exp}}{\left(\Delta M_s^{\rm SM} - 2 \delta\Delta M_s^{\rm SM}\right)^{2017}} - 1}}
\simeq 5.2 \, ,
\end{equation}
where $\delta \Delta M_s^{\rm SM}$ denotes the 1$\sigma$ error of the SM prediction. 
Hence, in models where $\kappa > 0$, the limit on the mass of the NP mediators is strengthened by a factor 5. 
On the other hand, if the tension between the SM prediction and $\Delta M_s^{\rm Exp}$ increases 
in the future, a NP contribution with $\kappa < 0$ would be required in order to accommodate the 
discrepancy. 

A typical example where $\kappa > 0$ is that of a purely LH vector-current operator, 
which arises from the exchange 
of a single mediator featuring real couplings, 
cf.~\sect{BmixvsBanom}.\footnote{Similar scenarios leading to $\kappa > 0$ 
were considered in 2016 by Blanke and Buras \cite{Blanke:2016bhf} 
in the context of CMFV models.} 
In such a case, the short-distance contribution to $B_s$-mixing is described by the effective Lagrangian 
\begin{equation}
\label{LNPDB2}
\mathcal{L}^{\rm NP}_{\Delta B = 2} = - \frac{4 G_F}{\sqrt{2}} \left( V_{tb} V^*_{ts} \right)^2 
\left[ C^{LL}_{bs} \left( \bar s_L \gamma_\mu b_L \right)^2 + \text{h.c.} \right] \, ,
\end{equation}
where $C^{LL}_{bs}$ is a Wilson coefficient to be matched with ultraviolet (UV) models, 
and which enters Eq.~(\ref{MsExpvsNP}) as  
\begin{equation}
\frac{\Delta M_s^{\rm Exp}}{\Delta M_s^{\rm SM}}
= \abs{1 + \frac{C^{LL}_{bs}}{R^{\rm loop}_{\rm SM}}}
\, , 
\end{equation}
where 
\begin{equation}
R^{\rm loop}_{\rm SM} 
= \frac{\sqrt{2} G_F M_W^2 \hat\eta_B S_0(x_t)}{16 \pi^2} = 1.3397 \times 10^{-3} \, .
\end{equation}

In the following,
we will show how the updated bound from $\Delta M_s$ impacts the parameter space 
of simplified models (with $\kappa > 0$) put forth for the explanation of the recent discrepancies in 
semi-leptonic B-physics data (\sect{BmixvsBanom}) 
and then discuss some model-building directions in order to achieve $\kappa < 0$ (\sect{NPmodelskl0}).

\subsection{Impact of $B_s$-mixing on NP models for B-anomalies}
\label{BmixvsBanom}

A useful application of the refined SM prediction in Eq.~(\ref{DeltaM2017}) 
is in the context of the recent hints of LFU violation in semi-leptonic $B$-meson decays, 
both in neutral and charged currents.  
Focussing first on neutral current anomalies, the main observables are 
the LFU violating ratios $R_{K^{(*)}} \equiv \mathcal{B}(B \to K^{(*)} \mu^+ \mu^-) / \mathcal{B}(B \to K^{(*)} e^+ e^-)$ 
\cite{Aaij:2014ora,Aaij:2017vbb}, together with the angular distributions of $B \to K^{(*)} \mu^+ \mu^-$ \cite{Aaij:2015esa,Khachatryan:2015isa,Lees:2015ymt,Wei:2009zv,Aaltonen:2011ja,Aaij:2015oid,Abdesselam:2016llu,Wehle:2016yoi,Aaboud:2018krd,Sirunyan:2017dhj}
and the branching ratios of hadronic $b \to s \mu^+ \mu^-$ decays \cite{Aaij:2014pli,Aaij:2015esa,Khachatryan:2015isa}. 
As hinted by various recent global fits 
\cite{Capdevila:2017bsm,Altmannshofer:2017yso,DAmico:2017mtc,Alok:2017sui,Geng:2017svp,Ciuchini:2017mik}, 
and in order to simplify a bit the discussion, 
we assume NP contributions only in purely LH vector currents involving muons. 
The generalisation to different type of operators is straightforward. 
The effective Lagrangian for semi-leptonic $b \to s \mu^+ \mu^-$ transitions contains the terms
\begin{equation}
\label{Leffbsmumu}
\mathcal{L}^{\rm NP}_{b \to s \mu \mu} \supset \frac{4 G_F}{\sqrt{2}} V_{tb} V^*_{ts} \left( \delta C^\mu_9 O^\mu_9 + \delta C^\mu_{10} O^\mu_{10} \right) + \text{h.c.} \, ,
\end{equation}
with 
\begin{align}
O^\mu_9 &= \frac{\alpha}{4 \pi} (\bar s_L \gamma_\mu b_L) (\bar \mu \gamma^\mu \mu) \, , \\
O^\mu_{10} &= \frac{\alpha}{4 \pi} (\bar s_L \gamma_\mu b_L) (\bar \mu \gamma^\mu \gamma_5 \mu) \, . 
\end{align}
Assuming purely LH currents
and real Wilson coefficients the 
best-fit of $R_{K}$ and $R_{K^*}$ yields 
(from e.g.~\cite{Altmannshofer:2017yso}):
$\Re (\delta C^\mu_9) = - \Re (\delta C^\mu_{10}) \in [-0.81, -0.48]$ ($[-1.00, -0.32]$) at 1$\sigma $ (2$\sigma$). 
Adding also the data on $B \to K^{(*)} \mu^+\mu^-$ angular distributions and 
other $b \to s \mu^+ \mu^-$ observables\footnote{These include for instance $\mathcal{B}(B_s \to \mu^+ \mu^-)$ 
which is particularly constraining in the case of pseudo-scalar mediated quark transitions (see e.g.~\cite{Golowich:2011cx}).} 
improves the statistical significance of the fit, 
but does not necessarily imply larger deviations of $\Re (\delta C^\mu_9)$ from zero (see e.g.~\cite{Capdevila:2017bsm}). 
In the following we will stick only to the $R_{K}$ and $R_{K^*}$ observables and denote this benchmark as ``$R_{K^{(*)}}$''.  

\subsubsection{Z'}
A paradigmatic NP model for explaining the $B$-anomalies in neutral currents 
is that of a $Z'$ dominantly coupled via LH currents.  
Here, we focus only on the part of the Lagrangian relevant for $b \to s \mu^+ \mu^-$ transitions and $B_s$-mixing, namely 
\begin{equation}
\label{LZp}
\mathcal{L}_{Z'} = 
\frac{1}{2} M^2_{Z'} (Z'_\mu)^2 +
\left( \lambda^Q_{ij} \, \bar d_L^i \gamma^\mu d_L^j + \lambda^L_{\alpha\beta} \, \bar \ell_L^\alpha \gamma^\mu \ell_L^\beta \right) Z'_\mu 
\, , 
\end{equation}
where $d^i$ and $\ell^\alpha$ denote down-quark and charged-lepton mass eigenstates, and 
$\lambda^{Q,L}$ are 
hermitian matrices in flavour space. 
Of course, any full-fledged (i.e.~$SU(2)_L \times U(1)_Y$ 
gauge invariant and anomaly free) $Z'$ model attempting an explanation of $R_{K^{(*)}}$ 
via LH currents can be mapped into Eq.~(\ref{LZp}). 
After integrating out the $Z'$ at tree level, we obtain the 
effective Lagrangian
\begin{align}
\mathcal{L}_{Z'}^{\rm eff} &= -\frac{1}{2 M^2_{Z'}} \left( \lambda^Q_{ij} \, \bar d_L^i \gamma_\mu d_L^j 
+ \lambda^L_{\alpha\beta} \, \bar \ell_L^\alpha \gamma_\mu \ell_L^\beta \right)^2  \\
&\supset -\frac{1}{2 M^2_{Z'}} 
\left[ 
(\lambda^Q_{23})^2 \left( \bar s_L \gamma_\mu b_L \right)^2 \right. \nonumber \\
& \left. + 2 \lambda^Q_{23} \lambda^L_{22} (\bar s_L \gamma_\mu b_L) (\bar \mu_L \gamma^\mu \mu_L)
+ \text{h.c.}
\right]
\, . \nonumber
\end{align}
Matching with Eq.~(\ref{Leffbsmumu}) and (\ref{LNPDB2}) we get 
\begin{equation}
\delta C^\mu_9 = -\delta C^\mu_{10} = - \frac{\pi}{\sqrt{2} G_F M^2_{Z'} \alpha} \left( \frac{\lambda^Q_{23} \lambda^L_{22}}{V_{tb} V^*_{ts}} \right) \, ,
\end{equation}
and 
\begin{equation} 
\label{CbsZp}
C^{LL}_{bs} = \frac{\eta^{LL}(M_{Z'})}{4 \sqrt{2} G_F M^2_{Z'}} \left( \frac{\lambda^Q_{23}}{V_{tb} V^*_{ts}} \right)^2
\, , 
\end{equation}
where $\eta^{LL}(M_{Z'})$ 
encodes the running down to the bottom mass scale using NLO anomalous dimensions \cite{Ciuchini:1997bw,Buras:2000if}. 
E.g.~for $M_{Z'} \in [1,10]$ TeV we find $\eta^{LL}(M_{Z'}) \in [0.79,0.75]$.   

Here we consider the case of a real coupling $\lambda^Q_{23}$, so that 
$C^{LL}_{bs} > 0$ and $\delta C^\mu_9 = -\delta C^\mu_{10}$ 
is also real. This assumption is consistent with the fact that nearly all the 
groups performing global fits \cite{Descotes-Genon:2013wba,Beaujean:2013soa,Altmannshofer:2014rta,Descotes-Genon:2015uva,Hurth:2016fbr,Altmannshofer:2017fio,Ciuchini:2017mik,Geng:2017svp,Capdevila:2017bsm,Altmannshofer:2017yso,DAmico:2017mtc,Alok:2017sui} (see however \cite{Alok:2017jgr} for an exception) 
assumed so far real Wilson coefficients in \eq{Leffbsmumu} and also follows the standard 
approach adopted in the literature for the $Z'$ models aiming at an explanation of the $b \to s \mu^+ \mu^-$ anomalies
(for an incomplete list, see \cite{Buras:2013qja,Gauld:2013qja,Buras:2013dea,Altmannshofer:2014cfa,Crivellin:2015mga,Crivellin:2015lwa,Celis:2015ara,Belanger:2015nma,Falkowski:2015zwa,Carmona:2015ena,Allanach:2015gkd,Chiang:2016qov,Boucenna:2016wpr,Megias:2016bde,Boucenna:2016qad,Altmannshofer:2016jzy,Crivellin:2016ejn,GarciaGarcia:2016nvr,Bhatia:2017tgo,Cline:2017lvv,Baek:2017sew,Cline:2017ihf,DiChiara:2017cjq,Kamenik:2017tnu,Ko:2017lzd,Ko:2017yrd,Alonso:2017bff,Ellis:2017nrp,Alonso:2017uky,Carmona:2017fsn}).
In fact, complex $Z'$ couplings can arise via fermion mixing, 
but are subject to additional constraints from CP-violating observables (cf.~\sect{NPmodelskl0}). 

\begin{figure}[htb!]
\center
\includegraphics[width=.45\textwidth]{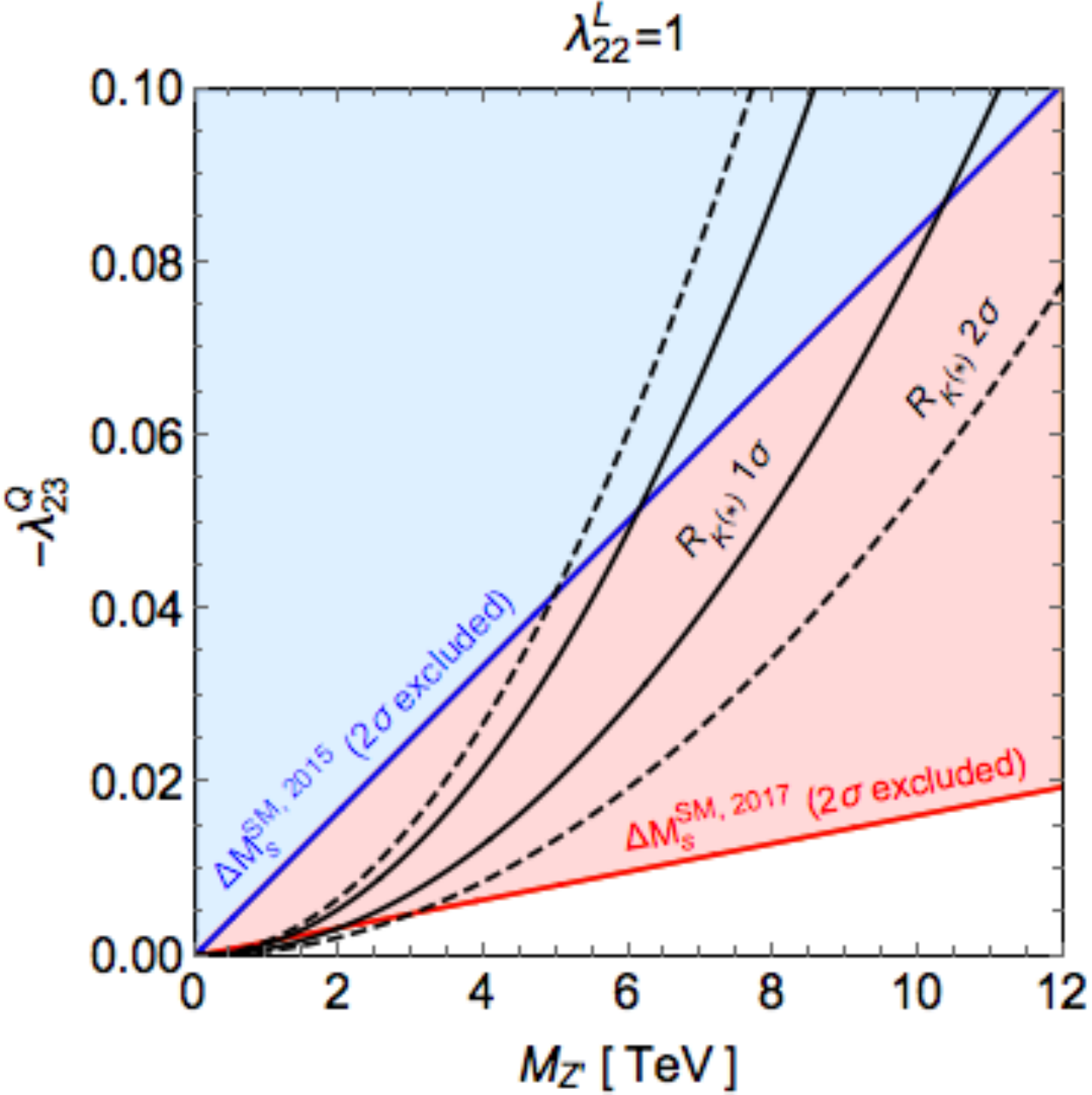} 
\caption{\label{fig:BsmixvsRK}
Bounds from $B_s$-mixing on the parameter space of the 
simplified $Z'$ model of \eq{LZp}, for real $\lambda^Q_{23}$ and $\lambda^L_{22} =1$. 
The blue and red shaded areas correspond respectively to the 2$\sigma$ exclusions from $\Delta M_s^{\rm SM,\,2015}$ and 
$\Delta M_s^{\rm SM,\,2017}$, while the solid (dashed) black curves encompass the 1$\sigma$ (2$\sigma$) 
best-fit region from $R_{K^{(*)}}$. 
}
\end{figure}

The impact of the improved SM calculation of $B_s$-mixing on the parameter space of the $Z'$ explanation of 
$R_{K^{(*)}}$ is displayed in Fig.~\ref{fig:BsmixvsRK}, for the reference value $\lambda^L_{22} =1$.\footnote{For 
$m_{Z'} \lesssim 1$ TeV the coupling $\lambda^L_{22}$ is bounded by the $Z\to 4\mu$ measurement at LHC and 
by neutrino trident production \cite{Altmannshofer:2014pba}. 
See for instance Fig.~1 in \cite{Falkowski:2018dsl} for a recent analysis.}
Note that the old SM determination, $\Delta M_s^{\rm SM,\,2015}$, allowed for $M_Z'$ as heavy as 
$\approx 10$ TeV in order to explain $R_{K^{(*)}}$ at 1$\sigma$. 
In contrast, $\Delta M_s^{\rm SM,\,2017}$ implies now $M_Z' \lesssim 2$ TeV.
Remarkably, even for $\lambda^L_{22} =\sqrt{4 \pi}$, which saturates the perturbative unitarity bound \cite{DiLuzio:2017chi,DiLuzio:2016sur}, 
we find that the updated limit from $B_s$-mixing requires 
$M_Z' \lesssim 8$ TeV for the 1$\sigma$ explanation of $R_{K^{(*)}}$. 
Whether a few TeV $Z'$ is ruled out or not by direct searches at LHC depends however on the details of the $Z'$ model.  
For instance, the stringent constraints from di-lepton searches \cite{Aaboud:2017buh} are tamed in models where the $Z'$ 
couples mainly third generation fermions (as e.g.~in \cite{Alonso:2017uky}).  
This notwithstanding, the updated limit from $B_s$-mixing cuts dramatically into the parameter space 
of the $Z'$ explanation of the $b \to s \mu^+ \mu^-$ anomalies, 
with important implications for LHC direct searches and future colliders \cite{Allanach:2017bta}.

\subsubsection{Leptoquarks}
\begin{figure}[htb!]
\center
\includegraphics[width=.45\textwidth]{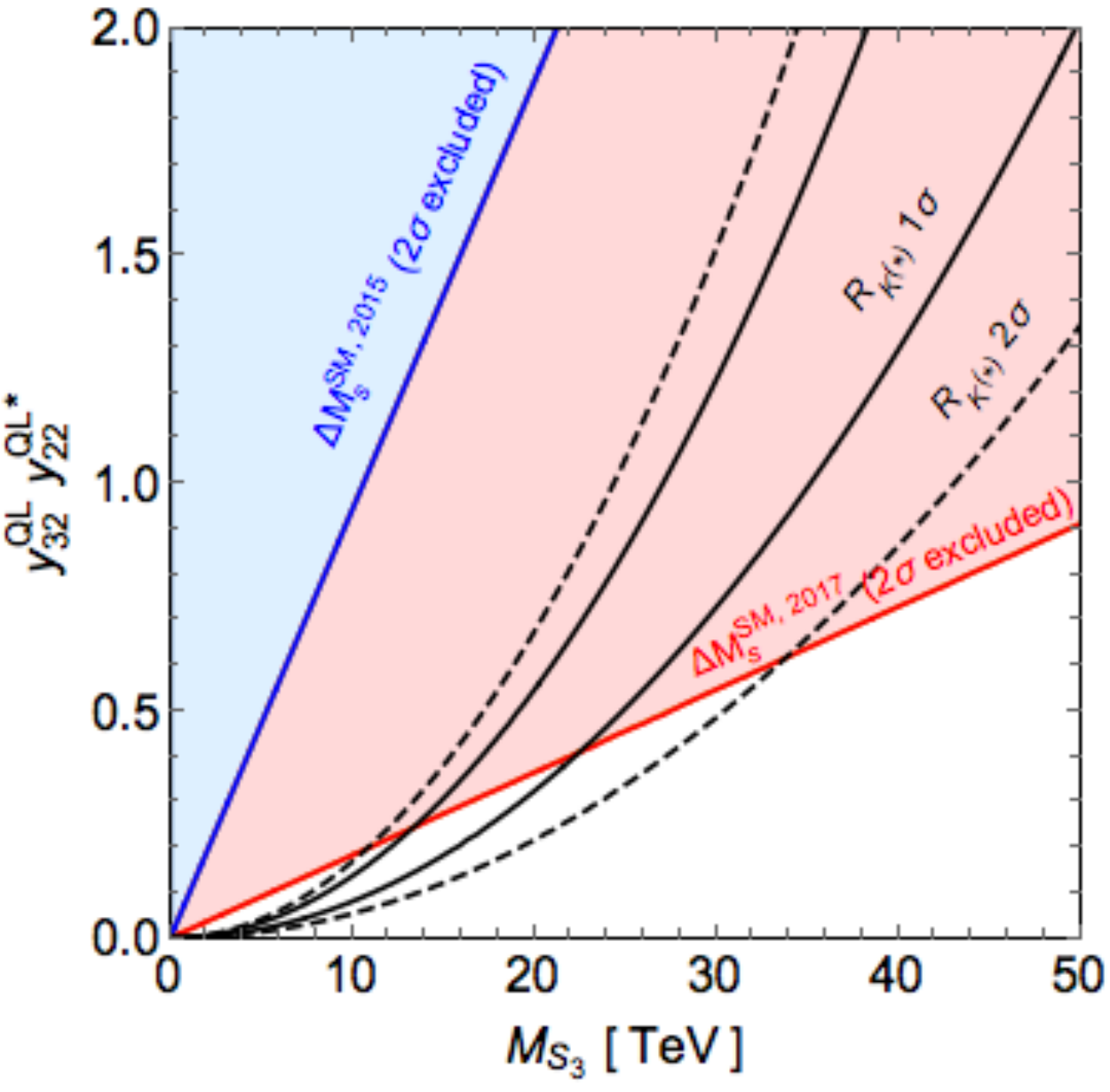} 
\caption{\label{fig:BsmixvsRK_LQ}
Bounds from $B_s$-mixing on the parameter space of the 
scalar leptoquark model of \eq{LLQS3}, for 
real $y_{32}^{QL} y_{22}^{QL*}$ couplings. 
Meaning of shaded areas and curves as in \fig{fig:BsmixvsRK}. 
}
\end{figure}

Another popular class of simplified models which has been proposed in order 
to address the $b \to s \mu^+ \mu^-$ anomalies consists 
in leptoquark mediators 
(see e.g.~\cite{Hiller:2014yaa,Gripaios:2014tna,Varzielas:2015iva,Becirevic:2015asa,Alonso:2015sja,Bauer:2015knc,Fajfer:2015ycq,Barbieri:2015yvd,Becirevic:2016oho,Becirevic:2016yqi,Crivellin:2017zlb,Hiller:2017bzc,Becirevic:2017jtw,Dorsner:2017ufx,Assad:2017iib,DiLuzio:2017vat,Calibbi:2017qbu,Bordone:2017bld}). 
Although $B_s$-mixing is generated at one loop \cite{Davidson:1993qk,Dorsner:2016wpm},\footnote{The scalar leptoquark model proposed 
in Ref.~\cite{Becirevic:2017jtw} is a notable exception.}  
and hence the constraints are expected to be milder compared to the $Z'$ case, 
the connection with the 
anomalies is more direct due to the structure of the leptoquark couplings. 
For instance, let us consider the scalar leptoquark $S_3 \sim (\bar 3,3,1/3)$,\footnote{Similar considerations 
apply to the vector leptoquarks $U^\mu_1 \sim (3,1,2/3)$ and $U^\mu_3 \sim (3,3,2/3)$, which also provide 
a good fit for $R_{K^{(*)}}$. The case of massive vectors is however subtler, since the calculability 
of loop observables depends upon the UV completion (for a recent discussion, see e.g.~\cite{Biggio:2016wyy}). 
} 
with the Lagrangian
\begin{equation}
\label{LLQS3}
\mathcal{L}_{S_3} = 
- M^2_{S_3} |S_3^a|^2 +
y_{i\alpha}^{QL} \overline{Q^c}^i (\epsilon\sigma^a) L^\alpha \, S^a_3  + \text{h.c.} \, ,
\end{equation}
where 
$\sigma^a$ (for $a=1,2,3$) are Pauli matrices, $\epsilon = i \sigma^2$, 
and we employed the quark 
$Q^i = (V^{*}_{ji} u^j_L \ d^i_L)^T$ and lepton $L^\alpha = (\nu^\alpha_L \
\ell^\alpha_L)^T$ doublet representations ($V$ being the CKM matrix).
The contribution to the Wilson coefficients in \eq{Leffbsmumu} 
arises at tree level and reads
\begin{equation}
\delta C^\mu_9 = -\delta C^\mu_{10} = \frac{\pi}{\sqrt{2} G_F M^2_{S_3} \alpha} \left( \frac{y_{32}^{QL} y_{22}^{QL*}}{V_{tb} V^*_{ts}} \right) \, ,
\end{equation}
while that to $B_s$--mixing in \eq{LNPDB2} is induced at one loop \cite{Bobeth:2017ecx}
\begin{equation} 
\label{CbsLQ}
C^{LL}_{bs} = \frac{\eta^{LL}(M_{S_3})}{4 \sqrt{2} G_F M^2_{S_3}} \frac{5}{64\pi^2} 
\left( \frac{y^{QL} _{3\alpha} y^{QL*} _{2\alpha}}{V_{tb} V^*_{ts}} \right)^2 \, , 
\end{equation}
where the sum over the leptonic index $\alpha = 1,2,3$ is understood.  
In order to compare the two observables we consider in \fig{fig:BsmixvsRK_LQ} 
the case in which only the couplings $y_{32}^{QL} y_{22}^{QL*}$ (namely those directly connected to $R_{K^{(*)}}$) 
contribute to $B_s$-mixing and further assume real couplings, so that we can use the results 
of global fits which apply to real $\delta C^\mu_9 = -\delta C^\mu_{10}$. 

The bound on $M_{S_3}$ from $B_s$-mixing is strengthened by a factor 5 thanks to the new 
determination of $\Delta M_s$, which yields $M_{S_3} \lesssim 22$ TeV, in order to explain $R_{K^{(*)}}$ 
at 1$\sigma$ (cf.~\fig{fig:BsmixvsRK_LQ}). 
On the other hand, in flavour models predicting a hierarchical structure for the 
leptoquark couplings one rather expects $y_{i3}^{QL} \gg y_{i2}^{QL}$, 
so that the dominant contribution to $\Delta M_s$ is given by $y^{QL} _{33} y^{QL*} _{23}$.    
For example, $y_{i3}^{QL} / y_{i2}^{QL} \sim \sqrt{m_\tau / m_\mu} \approx 4$ in the partial compositeness framework 
of Ref.~\cite{Gripaios:2014tna}, so that the upper bound on $M_{S_3}$ is strengthened by a factor 
$y^{QL} _{33} y^{QL*} _{23} / y^{QL} _{32} y^{QL*} _{22} \sim 16$.  
The latter can then easily approach the limits from LHC direct searches which imply $M_{S_3} \gtrsim 900$ GeV, 
e.g.~for a QCD pair-produced $S_3$ dominantly coupled to third generation fermions \cite{Sirunyan:2017yrk}.  

\subsubsection{Combined $R_{K^{(*)}}$ and $R_{D^{(*)}}$ explanations}

Another set of intriguing anomalies in $B$-physics data is that 
related to the LFU violating ratios $R_{D^{(*)}} \equiv \mathcal{B}(B \to D^{(*)} \tau \bar\nu) / \mathcal{B}(B \to D^{(*)} \ell \bar\nu)$ 
(here, $\ell = e,\mu$), 
which turn out to be larger than the SM \cite{Lees:2013uzd,Aaij:2015yra,Hirose:2016wfn}. 
Notably, in this case NP must compete with a tree-level SM charged current, 
thus requiring a sizeably larger effect compared to neutral current anomalies. 
The conditions under which a combined explanation of  
$R_{K^{(*)}}$ and $R_{D^{(*)}}$ can be obtained, compatibly with a plethora 
of other indirect constraints (as e.g.~those pointed out in \cite{Feruglio:2016gvd,Feruglio:2017rjo}), 
have been recently reassessed at the EFT level in Ref.~\cite{Buttazzo:2017ixm}.  
Regarding $B_s$-mixing, dimensional analysis (see e.g.~Eq.~(6) in \cite{Buttazzo:2017ixm}) 
shows that models without some additional dynamical suppression (compared to semi-leptonic operators) 
are severely constrained already with the old $\Delta M_s$ value. 
For instance, solutions based on a vector triplet $V' \sim (1,3,0)$ \cite{Greljo:2015mma}, 
where $B_s$-mixing arises at tree level, are in serious tension with data unless one invokes e.g.~a percent level cancellation 
from extra contributions \cite{Buttazzo:2017ixm}. The updated value of $\Delta M_s$ in \eq{DeltaM2017} 
makes the tuning required to achieve that even worse. 
On the other hand, leptoquark solutions (e.g.~the vector $U^\mu_1 \sim (3,1,2/3)$) 
comply better with the bound due to the fact that $B_s$-mixing arises at one loop, 
but the contribution to $\Delta M_s$ should be actually addressed 
in specific UV models whenever calculable \cite{DiLuzio:2017vat}.

\subsection{Model building directions for $\Delta M_s^{\rm NP}<0$}
\label{NPmodelskl0}

Given the fact that $\Delta M_s^{\rm SM} > \Delta M_s^{\rm exp}$ at about 2$\sigma$, 
it is interesting to speculate about possible ways to obtain a negative NP contribution 
to $\Delta M_s$, thus relaxing the tension between 
the SM and the experimental measurement. 

Sticking to the simplified models of \sect{BmixvsBanom} ($Z'$ and leptoquarks 
coupled only to LH currents), an obvious solution in order to achieve $C^{LL}_{bs}<0$ 
is to allow for complex couplings 
(cf.~\eq{CbsZp} and \eq{CbsLQ}). 
For instance, in $Z'$ models this could happen as a consequence of fermion 
mixing if the $Z'$ does not couple universally in the gauge-current basis. 
A similar mechanism could be at play for 
vector leptoquarks arising from a spontaneously broken gauge theory, 
while scalar-leptoquark couplings to SM fermions 
are in general complex even before going in the mass basis. 

\begin{figure}[ht!]
\center
\includegraphics[width=.45\textwidth]{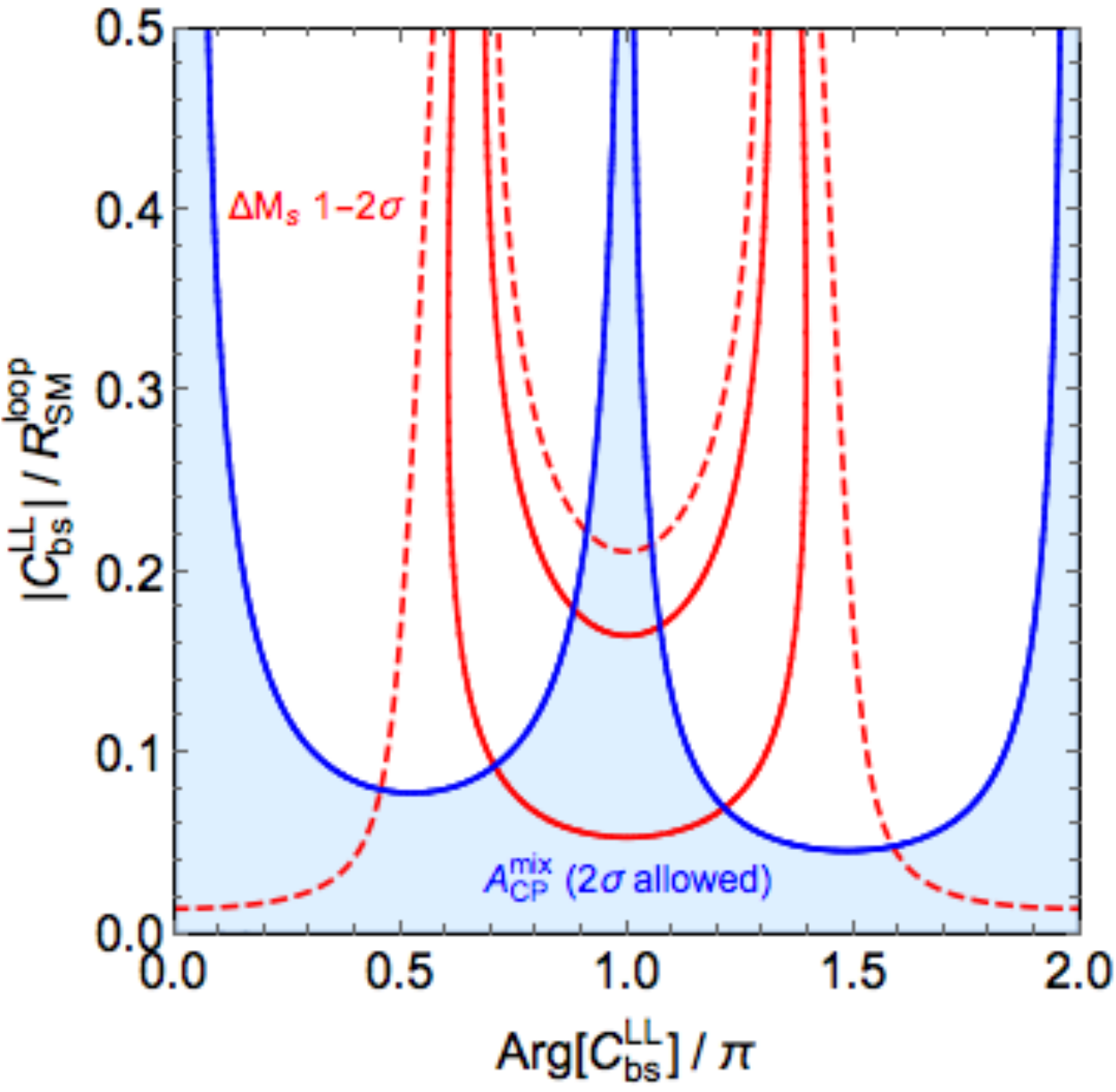}
\caption{\label{fig:Cbs_complex}
Combined 
constraints on the complex Wilson coefficient $C^{LL}_{bs}$.
The blue shaded area is the 2$\sigma$ allowed region from $A^{\rm mix}_{\rm CP}$, 
while the solid (dashed) red curves enclose the 1$\sigma$ (2$\sigma$) 
regions from $\Delta M_s^{\rm SM,\,2017}$.
}
\end{figure}

Extra phases in the couplings are constrained by CP-violating observables, that we discuss in turn.   
In order to quantify the allowed parameter space for a generic, complex coefficient $C^{LL}_{bs}$ in \eq{LNPDB2},  
we parametrise NP effects in $B_s$-mixing via
\begin{equation}
\frac{M^{\rm SM + NP}_{12}}{M^{\rm SM}_{12}} \equiv \abs{\Delta} e^{i \phi_\Delta} 
\, ,
\end{equation}
where 
\begin{equation} 
\abs{\Delta} = \abs{1 + \frac{C^{LL}_{bs}}{R^{\rm loop}_{\rm SM}}} \, , \qquad 
\phi_{\Delta} = \text{Arg} \left( 1 + \frac{C^{LL}_{bs}}{R^{\rm loop}_{\rm SM}} \right) \, .
\end{equation}
The former is constrained by $\Delta M^{\rm Exp}_s / \Delta M^{\rm SM}_s = \abs{\Delta}$, while the latter by the mixing-induced CP 
asymmetry \cite{Lenz:2006hd,Artuso:2015swg}\footnote{The semi-leptonic CP asymmetries for 
flavour-specific decays, $a^s_{\rm sl}$, 
do not pose serious constraints since the experimental errors are still too large \cite{Artuso:2015swg}.}
\begin{equation}
A^{\rm mix}_{\rm CP} (B_s \to J/\psi \phi) = \sin{\left( \phi_\Delta - 2 \beta_s \right)} \, ,
\end{equation}
where $A^{\rm mix}_{\rm CP} = -0.021 \pm 0.031$ \cite{Amhis:2016xyh}, $\beta_s = 0.01852 \pm 0.00032$ \cite{Charles:2004jd}, 
and we neglected penguin contributions \cite{Artuso:2015swg}. 
The combined $2\sigma$ constraints 
on the Wilson coefficient 
$C^{LL}_{bs}$ are displayed in Fig.~(\ref{fig:Cbs_complex}).  

For $\text{Arg}(C^{LL}_{bs}) = 0$ we recover the $2\sigma$ bound 
$\abs{C^{LL}_{bs}} / R^{\rm loop}_{\rm SM} \lesssim 0.014$,  
which basically corresponds to the case discussed in \sect{BmixvsBanom}
where we assumed a nearly real $C^{LL}_{bs}$ (up to a small imaginary part due to $V_{ts}$).  
On the other hand, a non-zero phase of $C^{LL}_{bs}$ allows to relax the bound from $\Delta M_s$, 
or even accommodate $\Delta M_s$ at 1$\sigma$ (region between the two solid red curves in \fig{fig:Cbs_complex}), 
compatibly with the $2\sigma$ allowed region from $A^{\rm mix}_{\rm CP}$ (blue shaded area in \fig{fig:Cbs_complex}).  
For $\text{Arg}(C^{LL}_{bs}) \approx \pi$ values of $\abs{C^{LL}_{bs}} / R^{\rm loop}_{\rm SM}$ as high as $0.21$ are allowed at 2$\sigma$, relaxing 
the bound on the modulus of the Wilson coefficient by a factor 15 
with respect to the $\text{Arg}(C^{LL}_{bs}) = 0$ case. 
Note, however, that the limit $\text{Arg}(C^{LL}_{bs}) = \pi$ corresponds to a nearly imaginary
$\delta C^{\mu}_9 = -\delta C^{\mu}_{10}$ which would presumably spoil the fit of $R_{K^{(*)}}$, since the 
interference with the SM contribution would be strongly suppressed. 
Nevertheless, it would be interesting to perform a global fit of $R_{K^{(*)}}$, 
together with $\Delta M_s$ and $A^{\rm mix}_{\rm CP}$ while allowing 
for non-zero values of the phase, in order to see whether a better agreement with the data can be obtained. 
Non-zero weak phases can potentially reveal themselves also via their contribution to triple product CP asymmetries in 
$B \to K^{(*)} \mu^+ \mu^-$ angular distributions \cite{Alok:2017jgr}. 
This is however beyond the scope of the present paper and we leave it for a future work. 

An alternative way to achieve a negative contribution for $\Delta M_s^{\rm NP}$ is to go beyond 
the simplified models of \sect{BmixvsBanom} and contemplate generalised chirality structures. 
Let us consider for definiteness the case of a $Z'$ coupled both to LH and RH down-quark currents
\begin{equation}
\mathcal{L}_{Z'} \supset \frac{1}{2} M^2_{Z'} (Z'_\mu)^2 + \left( \lambda^Q_{ij} \, \bar d_L^i \gamma^\mu d_L^j + 
\lambda^d_{ij} \, \bar d_R^i \gamma^\mu d_R^j \right) Z'_\mu \, .
\end{equation}
Upon integrating out the $Z'$ one obtains 
\begin{align}
\mathcal{L}_{Z'}^{\rm eff} 
& \supset -\frac{1}{2 M^2_{Z'}} 
 \left[ 
(\lambda^Q_{23})^2 \left( \bar s_L \gamma_\mu b_L \right)^2 
+ (\lambda^d_{23})^2 \left( \bar s_R \gamma_\mu b_R \right)^2 \right. \nonumber \\
& \left. + 2 \lambda^Q_{23} \lambda^d_{23} (\bar s_L \gamma_\mu b_L) (\bar s_R \gamma_\mu b_R)
+ \text{h.c.}
\right]
\, . 
\end{align}
The LR vector operator can clearly have any sign, even for real couplings. 
Moreover, since it gets strongly enhanced by renormalisation-group effects compared to LL and RR vector operators \cite{Buras:2001ra}, 
it can easily dominate the contribution to $\Delta M_s^{\rm NP}$. Note, however, that $\lambda^d_{23}$ 
contributes to $R_{K^{(*)}}$ via RH quark currents whose presence is disfavoured by global fits, 
since they break the approximate relation $R_{K} \approx R_{K^{*}}$ that is observed experimentally (see e.g.~\cite{DAmico:2017mtc}). 
Hence, also in this case, a careful study would be required in order to assess the simultaneous explanation 
of $R_{K^{(*)}}$ and $\Delta M_s$.

\section{Conclusions}
\label{conclusions}

In this paper, we have updated the SM prediction for the $B_s$-mixing observable \(\Delta M_s\) (\eq{DeltaM2017}) using the
most recent values for the input parameters, in particular new results from the lattice averaging group FLAG.
Our update shifts the central value of the SM theory prediction upwards and away from experiment by 13\%, 
while reducing the theory uncertainty compared to the previous SM determination by a factor of two. 
This implies a 1.8 $\sigma$ discrepancy from the SM.

We further discussed an important application of the \(\Delta M_s\) update for 
NP models aimed at explaining the recent anomalies in semi-leptonic \(B_s\) decays.
The latter typically predict a positive shift in the NP contribution to 
\(\Delta M_s\), thus making the discrepancy with respect to the experimental value even worse.   
As a generic result we have shown that, whenever the NP contribution to \(\Delta M_s\) is positive, 
the limit on the mass of the NP mediators that must be invoked to explain any of the anomalies is strengthened
by a factor of five (for a given size of couplings) compared to using the 2015 SM calculation for $\Delta M_s$.

In particular, we considered two representative examples of NP models 
featuring purely LH current and real couplings
-- that of a \(Z^\prime\) with the minimal couplings needed to explain 
\(R_{K^{(*)}}\)
anomalies, and a scalar ($SU(2)_L$ triplet) leptoquark model.
For the \(Z^\prime\) case we get an upper bound on the \(Z^\prime\) mass of 2 TeV (for unit $Z'$ coupling to muons, cf.~\fig{fig:BsmixvsRK}), 
an energy scale that is already probed by direct searches at LHC. 
On the other hand, the bounds on leptoquark models from $B_s$-mixing are generically milder, being the latter loop suppressed. 
For instance, taking only the contribution of the couplings needed to fit $R_{K^{(*)}}$ for the evaluation 
of $\Delta M_s$ we find that the upper bound on the scalar leptoquark 
mass is brought down to about \(20\)~TeV (cf.~\fig{fig:BsmixvsRK_LQ}). This limit gets however strengthened in flavour models 
predicting a hierarchical structure of the leptoquark couplings to SM fermions and can easily approach 
the region probed by the LHC. 
Trying in addition to solve the deviations in $R_{D^{(*)}}$ implies very 
severe bounds from $B_s$-mixing as well, since the overall scale of NP must be lowered 
compared to the case of only neutral current anomalies. 

Given the current status of a higher theory value for \(\Delta M_s\) compared to experiment, we also have looked at possible
ways in which NP can provide a \emph{negative} contribution that lessens the tension.
A non-zero phase in the NP couplings is one such way, and we have shown how extra constraints from the CP violating
observable \(A^\text{mix}_\text{CP}\) in \(B_s \to J/\psi \phi\) decays cuts out parameter space where otherwise a
significant NP contribution could be present. However, a large phase can potentially worsen the fit for $R_{K^{(*)}}$ -- here
a global combined fit of \(\Delta M_s\), \(A^\text{mix}_\text{CP}\) and $R_{K^{(*)}}$ seems to be an important next step.
Another possibility is to consider NP models with a generalised chirality structure. In particular, $\Delta B = 2$ 
LR vector operators, which are renormalisation-group enhanced, can accommodate any sign for \(\Delta M^{\rm NP}_s\),  
even for real couplings. Large contributions from RH currents are however disfavoured by the $R_{K^{(*)}}$ fit, 
hence also here a more careful analysis is needed.

Finally, a confirmation of our results, by further lattice groups confirming the large FNAL/MILC results for the four quark matrix
elements, as well as a definite solution of the $V_{cb}$ puzzle, would give further confidence in the extraordinary
strength of the bounds presented in this paper.

\section*{Acknowledgments}
We thank S\'{e}bastien Descotes-Genon for providing the unpublished {\it tree-level only} CKM values obtained by CKMfitter, 
Tomomi Ishikawa and  Andreas J{\"u}ttner for advise on lattice inputs and Marco Nardecchia for helpful feedback on the BSM section. 
This work was supported by the STFC through the IPPP grant.

\appendix

\section{Numerical input for theory predictions}
\label{app:input}
We use the following input for our numerical evaluations. The values in
Table \ref{parameter1} are taken from the PDG \cite{Patrignani:2016xqp}, from non-relativistic sum rules (NRSR)
\cite{Beneke:2014pta,Beneke:2016oox}, from the CKMfitter group (web-update of \cite{Charles:2004jd} -- 
similar values can be taken from the UTfit group \cite{Bona:2006ah}) and
the non-perturbative parameters from FLAG (web-update of \cite{Aoki:2016frl}).
For $\alpha_s$ we use RunDec \cite{Herren:2017osy}
  with 5-loop accuracy  \cite{Baikov:2016tgj,Herzog:2017ohr,Luthe:2017ttc,Luthe:2017ttg,Chetyrkin:2017bjc},
    running from
$M_Z$ down to the bottom mass scale. At the low scale we use 2-loop accuracy to determine $\Lambda^{(5)}$.
 
\begin{center}
\begin{table}[htb!]
\begin{center}
\begin{tabular}[c]{|l|l|l|}
\hline
Parameter & Value & Reference
\\
\hline
\hline
$M_W$   & $80.385(15)  \; \mbox{GeV}$ & $\mbox{PDG 2017} $
\\
\hline
$G_F $  & $1.1663787(6)   10^{-5}   \; \mbox{GeV}^{-2} $&$\mbox{PDG 2017}$
\\
\hline
$\hbar$ & $6.582119514(40) 10^{-25} \; \mbox{GeV s}$ & $\mbox{PDG 2017}$
\\
\hline
$M_{B_s} $& $  5.36689(19) \ \mbox{GeV} $& $\mbox{PDG 2017}$
\\
\hline
${m}_t $  & $ 173.1(0.6)  \; \mbox{GeV} $& $\mbox{PDG 2017}$
\\
\hline
$\bar{m}_t  (\bar{m}_t) $  & $ 165.6 5(57)   \; \mbox{GeV} $& $\mbox{own evaluation}$
\\
\hline
$\bar{m}_b  (\bar{m}_b) $  & $  4.203(25) \; \mbox{GeV} $& $\mbox{NRSR}$
\\
\hline
$\alpha_s(M_Z) $ & $0.1181(11)  $& $ \mbox{PDG 2017}$
\\
\hline
$\alpha_s(\overline{m}_b) $ & $0.2246(21)  $& $ \mbox{own evaluation}$
\\
\hline
$\Lambda^{(5)} $ & $0.2259(68) \, \mbox{GeV}  $& $ \mbox{own evaluation}$
\\
\hline
$V_{us}$ &  $0.22508^{+0.00030}_{-0.00028}$ & $\mbox{CKMfitter}$
\\
\hline
$V_{cb}$ &  $0.04181^{+0.00028}_{-0.00060}$ & $\mbox{CKMfitter}$
\\
\hline
$|V_{ub}/V_{cb}|$ &  $0.0889(14)$ & $\mbox{CKMfitter}$
\\
\hline
$\gamma_{\rm CKM}$ &  $1.141^{+0.017}_{-0.020} 	   $ & $\mbox{CKMfitter}$
\\
\hline
$f_{B_s} \sqrt{\hat{B}} $ &  $274(8) \,\mbox{MeV} $ & $\mbox{FLAG}$
\\
\hline
\end{tabular}
\end{center}
\caption{List of input parameters needed for an update of the theory
prediction of different mixing observables.}
\label{parameter1}
\end{table}
\end{center}

\section{Error budget of the theory predictions}
\label{app:error}
In this appendix we compare the error budget of our new SM prediction for  $\Delta M_s^{\rm SM}$
with the ones given in 2015 by \cite{Artuso:2015swg},
in 2011 by \cite{Lenz:2011ti} and 2006 by \cite{Lenz:2006hd}. The numbers are given
in
Table \ref{error1}.
\begin{center}
\begin{table*}
\begin{center}
 \begin{tabular}{|c||c|c|c|c|}
\hline
$\Delta M_s^{\rm SM} $   &   $\mbox{This work}  $& $\mbox{ABL 2015} \ \text{\cite{Artuso:2015swg}} $&  $\mbox{LN 2011} \ \text{\cite{Lenz:2011ti}} $  &  $ {\mbox{LN 2006}} \ \text{\cite{Lenz:2006hd}} $
\\
\hline
\hline
 $\mbox{Central Value} $   &   $20.01 \, \mbox{ps}^{-1} $&   $18.3 \, \mbox{ps}^{-1} $ &  $ 17.3 \, \mbox{ps}^{-1 } $ &  $ 19.3 \, \mbox{ps}^{-1} $
\\
\hline
 $\delta (f_{B_s} \sqrt{B}) $  &  $ 5.8\%  $ & $ 13.9\%  $          & $  13.5 \%   $              & $  34.1 \% $
\\
\hline
 $\delta (V_{cb})    $     & $  2.1 \%   $         &$  4.9 \%   $         &  $   3.4 \%   $             &  $ 4.9 \% $
\\
\hline
 $\delta (m_t)     $       & $ 0.7 \%   $          &$ 0.7 \%   $          &  $  1.1 \%    $             & $  1.8 \% $
\\ 
\hline 
 $\delta (\alpha_s)  $     & $ 0.1 \%    $        & $ 0.1 \%    $        &  $  0.4 \%      $           &  $ 2.0 \% $
\\
\hline
 $\delta (\gamma_{\rm CKM})   $      &  $0.1 \%  $           &  $0.1 \%  $           &  $  0.3 \%      $           &  $ 1.0 \% $
\\
\hline
 $\delta (|V_{ub}/V_{cb}|) $ &  $<0.1 \%   $        & $0.1 \%   $        &  $  0.2 \%    $             &   $0.5 \% $
\\
\hline
 $\delta (\overline{ m}_b)  $      &  $<0.1 \%    $        & $<0.1 \%    $        &  $  0.1 \%   $              &  $ --- $
\\
\hline
\hline
 $\sum \delta  $           &  $ 6.2 \%   $        & $ 14.8 \%   $        &  $14.0 \%    $             &  $34.6 \% $
\\
\hline
\end{tabular}
\end{center}
 \caption{List of the individual contributions to the theoretical error of the mass difference 
   $\Delta M_s$ within the SM and comparison with the values obtained in \cite{Artuso:2015swg},
   \cite{Lenz:2011ti} and \cite{Lenz:2006hd}. In the last row, the errors are summed in quadrature.}
\label{error1}
\end{table*}
\end{center}
We observe a considerable improvement in accuracy and a sizeable shift
compared to the 2015 prediction,
mostly stemming from the new lattice results for $f_{B_s} \sqrt{B}$, which still is responsible
for the largest error contribution of about $ 6 \%$.
The next important uncertainty is the accuracy of the CKM element $V_{cb}$, which
contributes about $2 \%$ to the error budget. If one gives up the assumption of the
unitarity of the $3 \times 3$ CKM matrix, the uncertainty can go up.
The uncertainties due to the remaining parameters play a less important role.
All in all we are left with an overall uncertainty of about $6 \%$, which
has to be compared to the experimental uncertainty of about 1 per mille.

\section{Non-perturbative inputs}
\label{app:Lattice}

As a word of caution we present here a wider range of non-perturbative determinations of the
matrix elements of the four-quark operators including also the
corresponding predictions for the mass differences, see Table \ref{ME}:
\begin{center}
  \begin{table*}
    \begin{center}
\begin{tabular}{|c||c|c|}
\hline
$\mbox{Source}$     & $f_{B_s} \sqrt{\hat{B}} $           & $\Delta M_s^{\rm SM} $   
\\
\hline
\hline
HPQCD14 \cite{Dowdall:2014qka}    &  $ (247 \pm 12) \; {\rm MeV} $  & $(16.2 \pm 1.7) \, \mbox{ps}^{-1} $
\\\hline
ETMC13 \cite{Carrasco:2013zta}    &  $ (262 \pm 10) \; {\rm MeV} $  & $(18.3 \pm 1.5) \, \mbox{ps}^{-1} $
\\
\hline
HPQCD09 \cite{Gamiz:2009ku} = FLAG13 \cite{Aoki:2013ldr}  &  $ (266 \pm 18) \; {\rm MeV} $  & $(18.9 \pm 2.6)\, \mbox{ps}^{-1} $
\\
\hline
\textbf{FLAG17} \cite{Aoki:2016frl} & $\mathbf{(274 \pm 8) \; \text{\textbf{MeV}}}$ & $\mathbf{(20.01 \pm 1.25)\,ps^{-1}}$
\\
\hline
Fermilab16 \cite{Bazavov:2016nty}  &  $ (274.6 \pm 8.8) \; {\rm MeV} $  & $(20.1 \pm 1.5) \, \mbox{ps}^{-1} $
\\
\hline
HQET-SR \cite{Kirk:2017juj,Gelhausen:2013wia}    &  $ (278^{+28}_{-24}) \; {\rm MeV} $  & $(20.6^{+4.4}_{-3.4} )\, \mbox{ps}^{-1} $
\\
\hline
HPQCD06  \cite{Dalgic:2006gp} &  $ (281 \pm 20) \; {\rm MeV} $  & $(21.0 \pm 3.0) \, \mbox{ps}^{-1} $
\\
\hline
RBC/UKQCD14  \cite{Aoki:2014nga} &  $ (290 \pm 20)\; {\rm MeV} $  & $(22.4 \pm 3.4) \, \mbox{ps}^{-1} $
\\
\hline
Fermilab11 \cite{Bouchard:2011xj}  &  $ (291 \pm 18) \; {\rm MeV} $  & $(22.6 \pm 2.8) \, \mbox{ps}^{-1} $
\\
\hline
\end{tabular}
\end{center}
    \caption{List of predictions for the non-perturbative parameter $f_{B_s} \sqrt{\hat{B}}$ and the corresponding SM prediction
      for $\Delta M_s$. The current FLAG average is dominated by the FERMILAB/MILC value from 2016.}
\label{ME}
\end{table*}
\end{center}

HPQCD presented in 2014 preliminary results for $N_f = 2+1$ in \cite{Dowdall:2014qka} and for our numerical estimate in
Table~(\ref{ME}) we had to read off the numbers from Fig.~3
in their proceedings \cite{Dowdall:2014qka}. When finalised, this new calculation will supersede the 2006 \cite{Dalgic:2006gp}
and 2009 \cite{Gamiz:2009ku} values.
The ETMC $N_f = 2$ number stems from 2013
\cite{Carrasco:2013zta}, it is obtained with only two active flavours in the lattice simulation.
The Fermilab/MILC $N_f = 2+1$  number stems from 2016 \cite{Bazavov:2016nty} and it supersedes
the 2011 value \cite{Bouchard:2011xj}. This precise value is currently dominating the
FLAG average. The numerical effect of these new inputs on mixing observables was e.g.~studied in \cite{Jubb:2016mvq}.
The previous FLAG average from 2013 \cite{Aoki:2013ldr} was considerably lower.
There is  also a large $N_f = 2+1$ value from RBC-UKQCD presented at LATTICE 2015 (update of \cite{Aoki:2014nga}).
However, this number is obtained in the static limit and currently missing $1/m_b$
corrections are expected to be very sizeable.\footnote{Private communication with Tomomi Ishikawa.}
The HQET sum rules estimate for the Bag parameter \cite{Kirk:2017juj} can also be combined with the decay constant from lattice.
\\
Here clearly a convergence of these determinations, in particular an independent confirmation of the Fermilab/MILC
result which is currently dominating the FLAG average, would be very desirable.

\section{CKM-dependence}
\label{app:CKM}

\begin{figure}
\includegraphics[width=0.49 \textwidth]{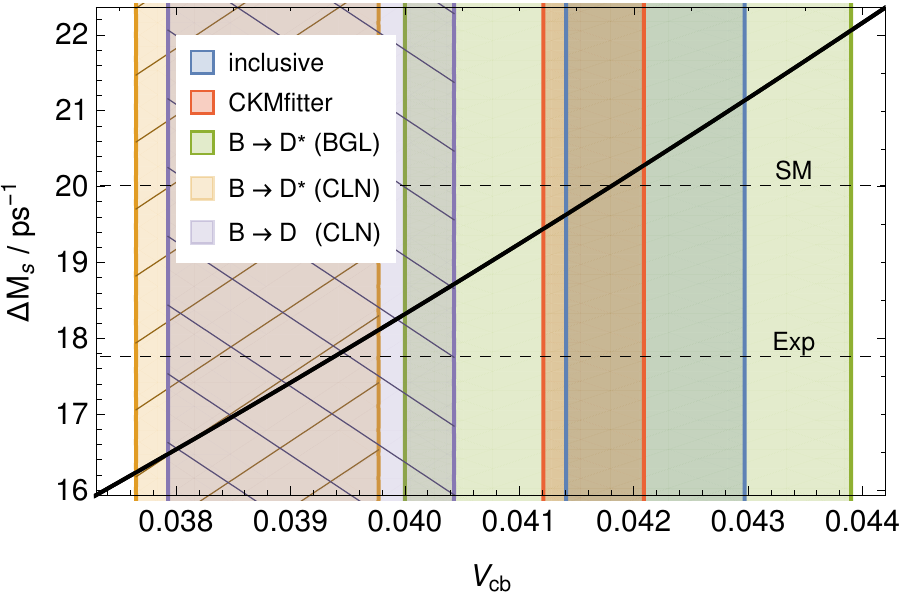}
\caption{\label{Vcb} Dependence of the SM prediction for $\Delta M_s$ from
  the value of the CKM-element $V_{cb}$. See text for details.
}
\end{figure}

The second most important input parameter for the prediction of $\Delta M_s$ is the CKM parameter $V_{cb}$.
There is a longstanding discrepancy between the inclusive determination and values obtained from
studying exclusive $B$ decays, see \cite{Patrignani:2016xqp}. Recent studies found, however, that the low exclusive value might
actually be a problem originating in the use of a certain form factor parametrisation in the experimental analysis.\footnote{The form factor
  models are denoted by CLN \cite{Caprini:1997mu} and BGL \cite{Boyd:1994tt}. Traditionally experiments were using CLN. It turned out, however,
  that CLN might underestimate some uncertainties.}
Using the BGL parametrisation one finds values that lie
considerably closer to the inclusive one, see \cite{Grinstein:2017nlq,Bigi:2017jbd,Bernlochner:2017xyx,Jaiswal:2017rve}.
Currently, there are various determinations of $V_{cb}$ available:
\begin{align}
  &\!V_{cb}^\text{Inclusive} = 0.04219 \pm 0.00078 \ \ \text{\cite{Amhis:2016xyh}} \, ,
  \\
  &\!V_{cb}^{B \to D}  = 0.03918 \pm 0.00094 \pm 0.00031 \ \ \text{\cite{Amhis:2016xyh}} \, ,
  \\
  &\!V_{cb}^{B \to D^*,\, \text{CLN}}  = 0.03871 \pm 0.00047 \pm 0.00059 \ \ \text{\cite{Amhis:2016xyh}} \, ,
  \\
  &\!V_{cb}^{B \to D^*,\, \text{BGL}}  = 0.0419^{+0.0020}_{-0.0019} \ \ \text{\cite{Grinstein:2017nlq}} \, .
\end{align}
In Fig.~\ref{Vcb} we plot the dependence of the SM prediction of 
$\Delta M_s$ on $V_{cb}$, and show the regions predicted by the above inclusive and exclusive determinations. 
We use the CKMfitter result for $V_{cb}$ (see \Table{parameter1}) for our new SM prediction of $\Delta M_s$ (see 
\eq{DeltaM2017} and the (upper) horizontal dashed line denoted with ``SM''), the corresponding error band is shown in orange. 
The predictions obtained by using the inclusive value of $V_{cb}$ only is given by the blue region. For completeness we show also the regions obtained by using the exclusive extractions of $V_{cb}$. The disfavoured CLN values result in much lower values for the mass difference (hatched areas), while the BGL value agrees well with the inclusive region, albeit with a higher uncertainty. The experimental value of $\Delta M_s$ is shown by the (lower) horizontal dashed line denoted with ``Exp''. 

The preference for the inclusive determination agrees also with the value obtained from the CKM fit 
(which we use in our SM estimate), as well as with the fit
value that is found if the direct measurements of $V_{cb}$ are not included in the fit 
\begin{align}
  V_{cb}^\text{CKM-fitter (no direct)
  } & =  0.04235^{+0.00074}_{-0.00069} \ \ \text{\cite{Charles:2004jd}} \, .
\end{align}
We also note that the CKMfitter determinations take into account loop-mediated processes, where potentially NP can arise.
Taking only tree-level inputs, they find:\footnote{Private communication with S\'{e}bastien Descotes-Genon --
  see also \cite{Koppenburg:2017mad}.}
\begin{align}
V_{us} &= 0.22520^{+0.00012}_{-0.00038} \, , \\
V_{cb} &= 0.04175^{+0.00033}_{-0.00172} \, , \\
|V_{ub} / V_{cb} | &= 0.092^{+0.004}_{-0.005} \, , \\
\gamma_{\rm CKM} &= 1.223^{+0.017}_{-0.030} \, ,
\end{align}
and using these inputs we find
\begin{equation}
\label{CKMtreeonly}
\Delta M_s^\text{SM, 2017 (tree)} = (19.9 \pm 1.5) \; \text{ps}^{-1} \, , 
\end{equation}
which shows an overall consistency with the prediction in \eq{DeltaM2017}.

\bibliographystyle{apsrev4-1.bst}
\bibliography{bibliography}

\begin{thebibliography}{146}%
\makeatletter
\providecommand \@ifxundefined [1]{%
 \@ifx{#1\undefined}
}%
\providecommand \@ifnum [1]{%
 \ifnum #1\expandafter \@firstoftwo
 \else \expandafter \@secondoftwo
 \fi
}%
\providecommand \@ifx [1]{%
 \ifx #1\expandafter \@firstoftwo
 \else \expandafter \@secondoftwo
 \fi
}%
\providecommand \natexlab [1]{#1}%
\providecommand \enquote  [1]{``#1''}%
\providecommand \bibnamefont  [1]{#1}%
\providecommand \bibfnamefont [1]{#1}%
\providecommand \citenamefont [1]{#1}%
\providecommand \href@noop [0]{\@secondoftwo}%
\providecommand \href [0]{\begingroup \@sanitize@url \@href}%
\providecommand \@href[1]{\@@startlink{#1}\@@href}%
\providecommand \@@href[1]{\endgroup#1\@@endlink}%
\providecommand \@sanitize@url [0]{\catcode `\\12\catcode `\$12\catcode
  `\&12\catcode `\#12\catcode `\^12\catcode `\_12\catcode `\%12\relax}%
\providecommand \@@startlink[1]{}%
\providecommand \@@endlink[0]{}%
\providecommand \url  [0]{\begingroup\@sanitize@url \@url }%
\providecommand \@url [1]{\endgroup\@href {#1}{\urlprefix }}%
\providecommand \urlprefix  [0]{URL }%
\providecommand \Eprint [0]{\href }%
\providecommand \doibase [0]{http://dx.doi.org/}%
\providecommand \selectlanguage [0]{\@gobble}%
\providecommand \bibinfo  [0]{\@secondoftwo}%
\providecommand \bibfield  [0]{\@secondoftwo}%
\providecommand \translation [1]{[#1]}%
\providecommand \BibitemOpen [0]{}%
\providecommand \bibitemStop [0]{}%
\providecommand \bibitemNoStop [0]{.\EOS\space}%
\providecommand \EOS [0]{\spacefactor3000\relax}%
\providecommand \BibitemShut  [1]{\csname bibitem#1\endcsname}%
\let\auto@bib@innerbib\@empty
\bibitem [{\citenamefont {Aaij}\ \emph
  {et~al.}(2014{\natexlab{a}})\citenamefont {Aaij} \emph
  {et~al.}}]{Aaij:2014pli}%
  \BibitemOpen
  \bibfield  {author} {\bibinfo {author} {\bibfnamefont {R.}~\bibnamefont
  {Aaij}} \emph {et~al.} (\bibinfo {collaboration} {LHCb}),\ }\href {\doibase
  10.1007/JHEP06(2014)133} {\bibfield  {journal} {\bibinfo  {journal} {JHEP}\
  }\textbf {\bibinfo {volume} {06}},\ \bibinfo {pages} {133} (\bibinfo {year}
  {2014}{\natexlab{a}})},\ \Eprint {http://arxiv.org/abs/1403.8044}
  {arXiv:1403.8044 [hep-ex]} \BibitemShut {NoStop}%
\bibitem [{\citenamefont {Aaij}\ \emph
  {et~al.}(2015{\natexlab{a}})\citenamefont {Aaij} \emph
  {et~al.}}]{Aaij:2015esa}%
  \BibitemOpen
  \bibfield  {author} {\bibinfo {author} {\bibfnamefont {R.}~\bibnamefont
  {Aaij}} \emph {et~al.} (\bibinfo {collaboration} {LHCb}),\ }\href {\doibase
  10.1007/JHEP09(2015)179} {\bibfield  {journal} {\bibinfo  {journal} {JHEP}\
  }\textbf {\bibinfo {volume} {09}},\ \bibinfo {pages} {179} (\bibinfo {year}
  {2015}{\natexlab{a}})},\ \Eprint {http://arxiv.org/abs/1506.08777}
  {arXiv:1506.08777 [hep-ex]} \BibitemShut {NoStop}%
\bibitem [{\citenamefont {Khachatryan}\ \emph {et~al.}(2016)\citenamefont
  {Khachatryan} \emph {et~al.}}]{Khachatryan:2015isa}%
  \BibitemOpen
  \bibfield  {author} {\bibinfo {author} {\bibfnamefont {V.}~\bibnamefont
  {Khachatryan}} \emph {et~al.} (\bibinfo {collaboration} {CMS}),\ }\href
  {\doibase 10.1016/j.physletb.2015.12.020} {\bibfield  {journal} {\bibinfo
  {journal} {Phys. Lett.}\ }\textbf {\bibinfo {volume} {B753}},\ \bibinfo
  {pages} {424} (\bibinfo {year} {2016})},\ \Eprint
  {http://arxiv.org/abs/1507.08126} {arXiv:1507.08126 [hep-ex]} \BibitemShut
  {NoStop}%
\bibitem [{\citenamefont {Lees}\ \emph {et~al.}(2016)\citenamefont {Lees} \emph
  {et~al.}}]{Lees:2015ymt}%
  \BibitemOpen
  \bibfield  {author} {\bibinfo {author} {\bibfnamefont {J.~P.}\ \bibnamefont
  {Lees}} \emph {et~al.} (\bibinfo {collaboration} {BaBar}),\ }\href {\doibase
  10.1103/PhysRevD.93.052015} {\bibfield  {journal} {\bibinfo  {journal} {Phys.
  Rev.}\ }\textbf {\bibinfo {volume} {D93}},\ \bibinfo {pages} {052015}
  (\bibinfo {year} {2016})},\ \Eprint {http://arxiv.org/abs/1508.07960}
  {arXiv:1508.07960 [hep-ex]} \BibitemShut {NoStop}%
\bibitem [{\citenamefont {Wei}\ \emph {et~al.}(2009)\citenamefont {Wei} \emph
  {et~al.}}]{Wei:2009zv}%
  \BibitemOpen
  \bibfield  {author} {\bibinfo {author} {\bibfnamefont {J.~T.}\ \bibnamefont
  {Wei}} \emph {et~al.} (\bibinfo {collaboration} {Belle}),\ }\href {\doibase
  10.1103/PhysRevLett.103.171801} {\bibfield  {journal} {\bibinfo  {journal}
  {Phys. Rev. Lett.}\ }\textbf {\bibinfo {volume} {103}},\ \bibinfo {pages}
  {171801} (\bibinfo {year} {2009})},\ \Eprint {http://arxiv.org/abs/0904.0770}
  {arXiv:0904.0770 [hep-ex]} \BibitemShut {NoStop}%
\bibitem [{\citenamefont {Aaltonen}\ \emph {et~al.}(2012)\citenamefont
  {Aaltonen} \emph {et~al.}}]{Aaltonen:2011ja}%
  \BibitemOpen
  \bibfield  {author} {\bibinfo {author} {\bibfnamefont {T.}~\bibnamefont
  {Aaltonen}} \emph {et~al.} (\bibinfo {collaboration} {CDF}),\ }\href
  {\doibase 10.1103/PhysRevLett.108.081807} {\bibfield  {journal} {\bibinfo
  {journal} {Phys. Rev. Lett.}\ }\textbf {\bibinfo {volume} {108}},\ \bibinfo
  {pages} {081807} (\bibinfo {year} {2012})},\ \Eprint
  {http://arxiv.org/abs/1108.0695} {arXiv:1108.0695 [hep-ex]} \BibitemShut
  {NoStop}%
\bibitem [{\citenamefont {Aaij}\ \emph {et~al.}(2016)\citenamefont {Aaij} \emph
  {et~al.}}]{Aaij:2015oid}%
  \BibitemOpen
  \bibfield  {author} {\bibinfo {author} {\bibfnamefont {R.}~\bibnamefont
  {Aaij}} \emph {et~al.} (\bibinfo {collaboration} {LHCb}),\ }\href {\doibase
  10.1007/JHEP02(2016)104} {\bibfield  {journal} {\bibinfo  {journal} {JHEP}\
  }\textbf {\bibinfo {volume} {02}},\ \bibinfo {pages} {104} (\bibinfo {year}
  {2016})},\ \Eprint {http://arxiv.org/abs/1512.04442} {arXiv:1512.04442
  [hep-ex]} \BibitemShut {NoStop}%
\bibitem [{\citenamefont {Abdesselam}\ \emph {et~al.}(2016)\citenamefont
  {Abdesselam} \emph {et~al.}}]{Abdesselam:2016llu}%
  \BibitemOpen
  \bibfield  {author} {\bibinfo {author} {\bibfnamefont {A.}~\bibnamefont
  {Abdesselam}} \emph {et~al.} (\bibinfo {collaboration} {Belle}),\ }in\ \href
  {https://inspirehep.net/record/1446979/files/arXiv:1604.04042.pdf} {\emph
  {\bibinfo {booktitle} {{Proceedings, LHCSki 2016 - A First Discussion of 13
  TeV Results: Obergurgl, Austria, April 10-15, 2016}}}}\ (\bibinfo {year}
  {2016})\ \Eprint {http://arxiv.org/abs/1604.04042} {arXiv:1604.04042
  [hep-ex]} \BibitemShut {NoStop}%
\bibitem [{\citenamefont {Wehle}\ \emph {et~al.}(2017)\citenamefont {Wehle}
  \emph {et~al.}}]{Wehle:2016yoi}%
  \BibitemOpen
  \bibfield  {author} {\bibinfo {author} {\bibfnamefont {S.}~\bibnamefont
  {Wehle}} \emph {et~al.} (\bibinfo {collaboration} {Belle}),\ }\href {\doibase
  10.1103/PhysRevLett.118.111801} {\bibfield  {journal} {\bibinfo  {journal}
  {Phys. Rev. Lett.}\ }\textbf {\bibinfo {volume} {118}},\ \bibinfo {pages}
  {111801} (\bibinfo {year} {2017})},\ \Eprint
  {http://arxiv.org/abs/1612.05014} {arXiv:1612.05014 [hep-ex]} \BibitemShut
  {NoStop}%
\bibitem [{\citenamefont {Aaboud}\ \emph {et~al.}(2018)\citenamefont {Aaboud}
  \emph {et~al.}}]{Aaboud:2018krd}%
  \BibitemOpen
  \bibfield  {author} {\bibinfo {author} {\bibfnamefont {M.}~\bibnamefont
  {Aaboud}} \emph {et~al.} (\bibinfo {collaboration} {ATLAS}),\ }\href@noop {}
  {\  (\bibinfo {year} {2018})},\ \Eprint {http://arxiv.org/abs/1805.04000}
  {arXiv:1805.04000 [hep-ex]} \BibitemShut {NoStop}%
\bibitem [{\citenamefont {Sirunyan}\ \emph
  {et~al.}(2017{\natexlab{a}})\citenamefont {Sirunyan} \emph
  {et~al.}}]{Sirunyan:2017dhj}%
  \BibitemOpen
  \bibfield  {author} {\bibinfo {author} {\bibfnamefont {A.~M.}\ \bibnamefont
  {Sirunyan}} \emph {et~al.} (\bibinfo {collaboration} {CMS}),\ }\href@noop {}
  {\  (\bibinfo {year} {2017}{\natexlab{a}})},\ \Eprint
  {http://arxiv.org/abs/1710.02846} {arXiv:1710.02846 [hep-ex]} \BibitemShut
  {NoStop}%
\bibitem [{\citenamefont {Descotes-Genon}\ \emph {et~al.}(2013)\citenamefont
  {Descotes-Genon}, \citenamefont {Matias},\ and\ \citenamefont
  {Virto}}]{Descotes-Genon:2013wba}%
  \BibitemOpen
  \bibfield  {author} {\bibinfo {author} {\bibfnamefont {S.}~\bibnamefont
  {Descotes-Genon}}, \bibinfo {author} {\bibfnamefont {J.}~\bibnamefont
  {Matias}}, \ and\ \bibinfo {author} {\bibfnamefont {J.}~\bibnamefont
  {Virto}},\ }\href {\doibase 10.1103/PhysRevD.88.074002} {\bibfield  {journal}
  {\bibinfo  {journal} {Phys. Rev.}\ }\textbf {\bibinfo {volume} {D88}},\
  \bibinfo {pages} {074002} (\bibinfo {year} {2013})},\ \Eprint
  {http://arxiv.org/abs/1307.5683} {arXiv:1307.5683 [hep-ph]} \BibitemShut
  {NoStop}%
\bibitem [{\citenamefont {Beaujean}\ \emph {et~al.}(2014)\citenamefont
  {Beaujean}, \citenamefont {Bobeth},\ and\ \citenamefont {van
  Dyk}}]{Beaujean:2013soa}%
  \BibitemOpen
  \bibfield  {author} {\bibinfo {author} {\bibfnamefont {F.}~\bibnamefont
  {Beaujean}}, \bibinfo {author} {\bibfnamefont {C.}~\bibnamefont {Bobeth}}, \
  and\ \bibinfo {author} {\bibfnamefont {D.}~\bibnamefont {van Dyk}},\ }\href
  {\doibase 10.1140/epjc/s10052-014-2897-0, 10.1140/epjc/s10052-014-3179-6}
  {\bibfield  {journal} {\bibinfo  {journal} {Eur. Phys. J.}\ }\textbf
  {\bibinfo {volume} {C74}},\ \bibinfo {pages} {2897} (\bibinfo {year}
  {2014})},\ \bibinfo {note} {[Erratum: Eur. Phys. J.C74,3179(2014)]},\ \Eprint
  {http://arxiv.org/abs/1310.2478} {arXiv:1310.2478 [hep-ph]} \BibitemShut
  {NoStop}%
\bibitem [{\citenamefont {Altmannshofer}\ and\ \citenamefont
  {Straub}(2015)}]{Altmannshofer:2014rta}%
  \BibitemOpen
  \bibfield  {author} {\bibinfo {author} {\bibfnamefont {W.}~\bibnamefont
  {Altmannshofer}}\ and\ \bibinfo {author} {\bibfnamefont {D.~M.}\ \bibnamefont
  {Straub}},\ }\href {\doibase 10.1140/epjc/s10052-015-3602-7} {\bibfield
  {journal} {\bibinfo  {journal} {Eur. Phys. J.}\ }\textbf {\bibinfo {volume}
  {C75}},\ \bibinfo {pages} {382} (\bibinfo {year} {2015})},\ \Eprint
  {http://arxiv.org/abs/1411.3161} {arXiv:1411.3161 [hep-ph]} \BibitemShut
  {NoStop}%
\bibitem [{\citenamefont {Descotes-Genon}\ \emph {et~al.}(2016)\citenamefont
  {Descotes-Genon}, \citenamefont {Hofer}, \citenamefont {Matias},\ and\
  \citenamefont {Virto}}]{Descotes-Genon:2015uva}%
  \BibitemOpen
  \bibfield  {author} {\bibinfo {author} {\bibfnamefont {S.}~\bibnamefont
  {Descotes-Genon}}, \bibinfo {author} {\bibfnamefont {L.}~\bibnamefont
  {Hofer}}, \bibinfo {author} {\bibfnamefont {J.}~\bibnamefont {Matias}}, \
  and\ \bibinfo {author} {\bibfnamefont {J.}~\bibnamefont {Virto}},\ }\href
  {\doibase 10.1007/JHEP06(2016)092} {\bibfield  {journal} {\bibinfo  {journal}
  {JHEP}\ }\textbf {\bibinfo {volume} {06}},\ \bibinfo {pages} {092} (\bibinfo
  {year} {2016})},\ \Eprint {http://arxiv.org/abs/1510.04239} {arXiv:1510.04239
  [hep-ph]} \BibitemShut {NoStop}%
\bibitem [{\citenamefont {Hurth}\ \emph {et~al.}(2016)\citenamefont {Hurth},
  \citenamefont {Mahmoudi},\ and\ \citenamefont {Neshatpour}}]{Hurth:2016fbr}%
  \BibitemOpen
  \bibfield  {author} {\bibinfo {author} {\bibfnamefont {T.}~\bibnamefont
  {Hurth}}, \bibinfo {author} {\bibfnamefont {F.}~\bibnamefont {Mahmoudi}}, \
  and\ \bibinfo {author} {\bibfnamefont {S.}~\bibnamefont {Neshatpour}},\
  }\href {\doibase 10.1016/j.nuclphysb.2016.05.022} {\bibfield  {journal}
  {\bibinfo  {journal} {Nucl. Phys.}\ }\textbf {\bibinfo {volume} {B909}},\
  \bibinfo {pages} {737} (\bibinfo {year} {2016})},\ \Eprint
  {http://arxiv.org/abs/1603.00865} {arXiv:1603.00865 [hep-ph]} \BibitemShut
  {NoStop}%
\bibitem [{\citenamefont {Altmannshofer}\ \emph
  {et~al.}(2017{\natexlab{a}})\citenamefont {Altmannshofer}, \citenamefont
  {Niehoff}, \citenamefont {Stangl},\ and\ \citenamefont
  {Straub}}]{Altmannshofer:2017fio}%
  \BibitemOpen
  \bibfield  {author} {\bibinfo {author} {\bibfnamefont {W.}~\bibnamefont
  {Altmannshofer}}, \bibinfo {author} {\bibfnamefont {C.}~\bibnamefont
  {Niehoff}}, \bibinfo {author} {\bibfnamefont {P.}~\bibnamefont {Stangl}}, \
  and\ \bibinfo {author} {\bibfnamefont {D.~M.}\ \bibnamefont {Straub}},\
  }\href {\doibase 10.1140/epjc/s10052-017-4952-0} {\bibfield  {journal}
  {\bibinfo  {journal} {Eur. Phys. J.}\ }\textbf {\bibinfo {volume} {C77}},\
  \bibinfo {pages} {377} (\bibinfo {year} {2017}{\natexlab{a}})},\ \Eprint
  {http://arxiv.org/abs/1703.09189} {arXiv:1703.09189 [hep-ph]} \BibitemShut
  {NoStop}%
\bibitem [{\citenamefont {Ciuchini}\ \emph {et~al.}(2017)\citenamefont
  {Ciuchini}, \citenamefont {Coutinho}, \citenamefont {Fedele}, \citenamefont
  {Franco}, \citenamefont {Paul}, \citenamefont {Silvestrini},\ and\
  \citenamefont {Valli}}]{Ciuchini:2017mik}%
  \BibitemOpen
  \bibfield  {author} {\bibinfo {author} {\bibfnamefont {M.}~\bibnamefont
  {Ciuchini}}, \bibinfo {author} {\bibfnamefont {A.~M.}\ \bibnamefont
  {Coutinho}}, \bibinfo {author} {\bibfnamefont {M.}~\bibnamefont {Fedele}},
  \bibinfo {author} {\bibfnamefont {E.}~\bibnamefont {Franco}}, \bibinfo
  {author} {\bibfnamefont {A.}~\bibnamefont {Paul}}, \bibinfo {author}
  {\bibfnamefont {L.}~\bibnamefont {Silvestrini}}, \ and\ \bibinfo {author}
  {\bibfnamefont {M.}~\bibnamefont {Valli}},\ }\href {\doibase
  10.1140/epjc/s10052-017-5270-2} {\bibfield  {journal} {\bibinfo  {journal}
  {Eur. Phys. J.}\ }\textbf {\bibinfo {volume} {C77}},\ \bibinfo {pages} {688}
  (\bibinfo {year} {2017})},\ \Eprint {http://arxiv.org/abs/1704.05447}
  {arXiv:1704.05447 [hep-ph]} \BibitemShut {NoStop}%
\bibitem [{\citenamefont {Geng}\ \emph {et~al.}(2017)\citenamefont {Geng},
  \citenamefont {Grinstein}, \citenamefont {Jäger}, \citenamefont
  {Martin~Camalich}, \citenamefont {Ren},\ and\ \citenamefont
  {Shi}}]{Geng:2017svp}%
  \BibitemOpen
  \bibfield  {author} {\bibinfo {author} {\bibfnamefont {L.-S.}\ \bibnamefont
  {Geng}}, \bibinfo {author} {\bibfnamefont {B.}~\bibnamefont {Grinstein}},
  \bibinfo {author} {\bibfnamefont {S.}~\bibnamefont {Jäger}}, \bibinfo
  {author} {\bibfnamefont {J.}~\bibnamefont {Martin~Camalich}}, \bibinfo
  {author} {\bibfnamefont {X.-L.}\ \bibnamefont {Ren}}, \ and\ \bibinfo
  {author} {\bibfnamefont {R.-X.}\ \bibnamefont {Shi}},\ }\href {\doibase
  10.1103/PhysRevD.96.093006} {\bibfield  {journal} {\bibinfo  {journal} {Phys.
  Rev.}\ }\textbf {\bibinfo {volume} {D96}},\ \bibinfo {pages} {093006}
  (\bibinfo {year} {2017})},\ \Eprint {http://arxiv.org/abs/1704.05446}
  {arXiv:1704.05446 [hep-ph]} \BibitemShut {NoStop}%
\bibitem [{\citenamefont {Capdevila}\ \emph
  {et~al.}(2017{\natexlab{a}})\citenamefont {Capdevila}, \citenamefont
  {Crivellin}, \citenamefont {Descotes-Genon}, \citenamefont {Matias},\ and\
  \citenamefont {Virto}}]{Capdevila:2017bsm}%
  \BibitemOpen
  \bibfield  {author} {\bibinfo {author} {\bibfnamefont {B.}~\bibnamefont
  {Capdevila}}, \bibinfo {author} {\bibfnamefont {A.}~\bibnamefont
  {Crivellin}}, \bibinfo {author} {\bibfnamefont {S.}~\bibnamefont
  {Descotes-Genon}}, \bibinfo {author} {\bibfnamefont {J.}~\bibnamefont
  {Matias}}, \ and\ \bibinfo {author} {\bibfnamefont {J.}~\bibnamefont
  {Virto}},\ }\href@noop {} {\  (\bibinfo {year} {2017}{\natexlab{a}})},\
  \Eprint {http://arxiv.org/abs/1704.05340} {arXiv:1704.05340 [hep-ph]}
  \BibitemShut {NoStop}%
\bibitem [{\citenamefont {Altmannshofer}\ \emph
  {et~al.}(2017{\natexlab{b}})\citenamefont {Altmannshofer}, \citenamefont
  {Stangl},\ and\ \citenamefont {Straub}}]{Altmannshofer:2017yso}%
  \BibitemOpen
  \bibfield  {author} {\bibinfo {author} {\bibfnamefont {W.}~\bibnamefont
  {Altmannshofer}}, \bibinfo {author} {\bibfnamefont {P.}~\bibnamefont
  {Stangl}}, \ and\ \bibinfo {author} {\bibfnamefont {D.~M.}\ \bibnamefont
  {Straub}},\ }\href {\doibase 10.1103/PhysRevD.96.055008} {\bibfield
  {journal} {\bibinfo  {journal} {Phys. Rev.}\ }\textbf {\bibinfo {volume}
  {D96}},\ \bibinfo {pages} {055008} (\bibinfo {year} {2017}{\natexlab{b}})},\
  \Eprint {http://arxiv.org/abs/1704.05435} {arXiv:1704.05435 [hep-ph]}
  \BibitemShut {NoStop}%
\bibitem [{\citenamefont {D'Amico}\ \emph {et~al.}(2017)\citenamefont
  {D'Amico}, \citenamefont {Nardecchia}, \citenamefont {Panci}, \citenamefont
  {Sannino}, \citenamefont {Strumia}, \citenamefont {Torre},\ and\
  \citenamefont {Urbano}}]{DAmico:2017mtc}%
  \BibitemOpen
  \bibfield  {author} {\bibinfo {author} {\bibfnamefont {G.}~\bibnamefont
  {D'Amico}}, \bibinfo {author} {\bibfnamefont {M.}~\bibnamefont {Nardecchia}},
  \bibinfo {author} {\bibfnamefont {P.}~\bibnamefont {Panci}}, \bibinfo
  {author} {\bibfnamefont {F.}~\bibnamefont {Sannino}}, \bibinfo {author}
  {\bibfnamefont {A.}~\bibnamefont {Strumia}}, \bibinfo {author} {\bibfnamefont
  {R.}~\bibnamefont {Torre}}, \ and\ \bibinfo {author} {\bibfnamefont
  {A.}~\bibnamefont {Urbano}},\ }\href {\doibase 10.1007/JHEP09(2017)010}
  {\bibfield  {journal} {\bibinfo  {journal} {JHEP}\ }\textbf {\bibinfo
  {volume} {09}},\ \bibinfo {pages} {010} (\bibinfo {year} {2017})},\ \Eprint
  {http://arxiv.org/abs/1704.05438} {arXiv:1704.05438 [hep-ph]} \BibitemShut
  {NoStop}%
\bibitem [{\citenamefont {Alok}\ \emph
  {et~al.}(2017{\natexlab{a}})\citenamefont {Alok}, \citenamefont
  {Bhattacharya}, \citenamefont {Datta}, \citenamefont {Kumar}, \citenamefont
  {Kumar},\ and\ \citenamefont {London}}]{Alok:2017sui}%
  \BibitemOpen
  \bibfield  {author} {\bibinfo {author} {\bibfnamefont {A.~K.}\ \bibnamefont
  {Alok}}, \bibinfo {author} {\bibfnamefont {B.}~\bibnamefont {Bhattacharya}},
  \bibinfo {author} {\bibfnamefont {A.}~\bibnamefont {Datta}}, \bibinfo
  {author} {\bibfnamefont {D.}~\bibnamefont {Kumar}}, \bibinfo {author}
  {\bibfnamefont {J.}~\bibnamefont {Kumar}}, \ and\ \bibinfo {author}
  {\bibfnamefont {D.}~\bibnamefont {London}},\ }\href {\doibase
  10.1103/PhysRevD.96.095009} {\bibfield  {journal} {\bibinfo  {journal} {Phys.
  Rev.}\ }\textbf {\bibinfo {volume} {D96}},\ \bibinfo {pages} {095009}
  (\bibinfo {year} {2017}{\natexlab{a}})},\ \Eprint
  {http://arxiv.org/abs/1704.07397} {arXiv:1704.07397 [hep-ph]} \BibitemShut
  {NoStop}%
\bibitem [{\citenamefont {Jäger}\ and\ \citenamefont
  {Martin~Camalich}(2013)}]{Jager:2012uw}%
  \BibitemOpen
  \bibfield  {author} {\bibinfo {author} {\bibfnamefont {S.}~\bibnamefont
  {Jäger}}\ and\ \bibinfo {author} {\bibfnamefont {J.}~\bibnamefont
  {Martin~Camalich}},\ }\href {\doibase 10.1007/JHEP05(2013)043} {\bibfield
  {journal} {\bibinfo  {journal} {JHEP}\ }\textbf {\bibinfo {volume} {05}},\
  \bibinfo {pages} {043} (\bibinfo {year} {2013})},\ \Eprint
  {http://arxiv.org/abs/1212.2263} {arXiv:1212.2263 [hep-ph]} \BibitemShut
  {NoStop}%
\bibitem [{\citenamefont {Jäger}\ and\ \citenamefont
  {Martin~Camalich}(2016)}]{Jager:2014rwa}%
  \BibitemOpen
  \bibfield  {author} {\bibinfo {author} {\bibfnamefont {S.}~\bibnamefont
  {Jäger}}\ and\ \bibinfo {author} {\bibfnamefont {J.}~\bibnamefont
  {Martin~Camalich}},\ }\href {\doibase 10.1103/PhysRevD.93.014028} {\bibfield
  {journal} {\bibinfo  {journal} {Phys. Rev.}\ }\textbf {\bibinfo {volume}
  {D93}},\ \bibinfo {pages} {014028} (\bibinfo {year} {2016})},\ \Eprint
  {http://arxiv.org/abs/1412.3183} {arXiv:1412.3183 [hep-ph]} \BibitemShut
  {NoStop}%
\bibitem [{\citenamefont {Descotes-Genon}\ \emph {et~al.}(2014)\citenamefont
  {Descotes-Genon}, \citenamefont {Hofer}, \citenamefont {Matias},\ and\
  \citenamefont {Virto}}]{Descotes-Genon:2014uoa}%
  \BibitemOpen
  \bibfield  {author} {\bibinfo {author} {\bibfnamefont {S.}~\bibnamefont
  {Descotes-Genon}}, \bibinfo {author} {\bibfnamefont {L.}~\bibnamefont
  {Hofer}}, \bibinfo {author} {\bibfnamefont {J.}~\bibnamefont {Matias}}, \
  and\ \bibinfo {author} {\bibfnamefont {J.}~\bibnamefont {Virto}},\ }\href
  {\doibase 10.1007/JHEP12(2014)125} {\bibfield  {journal} {\bibinfo  {journal}
  {JHEP}\ }\textbf {\bibinfo {volume} {12}},\ \bibinfo {pages} {125} (\bibinfo
  {year} {2014})},\ \Eprint {http://arxiv.org/abs/1407.8526} {arXiv:1407.8526
  [hep-ph]} \BibitemShut {NoStop}%
\bibitem [{\citenamefont {Ciuchini}\ \emph {et~al.}(2016)\citenamefont
  {Ciuchini}, \citenamefont {Fedele}, \citenamefont {Franco}, \citenamefont
  {Mishima}, \citenamefont {Paul}, \citenamefont {Silvestrini},\ and\
  \citenamefont {Valli}}]{Ciuchini:2015qxb}%
  \BibitemOpen
  \bibfield  {author} {\bibinfo {author} {\bibfnamefont {M.}~\bibnamefont
  {Ciuchini}}, \bibinfo {author} {\bibfnamefont {M.}~\bibnamefont {Fedele}},
  \bibinfo {author} {\bibfnamefont {E.}~\bibnamefont {Franco}}, \bibinfo
  {author} {\bibfnamefont {S.}~\bibnamefont {Mishima}}, \bibinfo {author}
  {\bibfnamefont {A.}~\bibnamefont {Paul}}, \bibinfo {author} {\bibfnamefont
  {L.}~\bibnamefont {Silvestrini}}, \ and\ \bibinfo {author} {\bibfnamefont
  {M.}~\bibnamefont {Valli}},\ }\href {\doibase 10.1007/JHEP06(2016)116}
  {\bibfield  {journal} {\bibinfo  {journal} {JHEP}\ }\textbf {\bibinfo
  {volume} {06}},\ \bibinfo {pages} {116} (\bibinfo {year} {2016})},\ \Eprint
  {http://arxiv.org/abs/1512.07157} {arXiv:1512.07157 [hep-ph]} \BibitemShut
  {NoStop}%
\bibitem [{\citenamefont {Chobanova}\ \emph {et~al.}(2017)\citenamefont
  {Chobanova}, \citenamefont {Hurth}, \citenamefont {Mahmoudi}, \citenamefont
  {Martinez~Santos},\ and\ \citenamefont {Neshatpour}}]{Chobanova:2017ghn}%
  \BibitemOpen
  \bibfield  {author} {\bibinfo {author} {\bibfnamefont {V.~G.}\ \bibnamefont
  {Chobanova}}, \bibinfo {author} {\bibfnamefont {T.}~\bibnamefont {Hurth}},
  \bibinfo {author} {\bibfnamefont {F.}~\bibnamefont {Mahmoudi}}, \bibinfo
  {author} {\bibfnamefont {D.}~\bibnamefont {Martinez~Santos}}, \ and\ \bibinfo
  {author} {\bibfnamefont {S.}~\bibnamefont {Neshatpour}},\ }\href {\doibase
  10.1007/JHEP07(2017)025} {\bibfield  {journal} {\bibinfo  {journal} {JHEP}\
  }\textbf {\bibinfo {volume} {07}},\ \bibinfo {pages} {025} (\bibinfo {year}
  {2017})},\ \Eprint {http://arxiv.org/abs/1702.02234} {arXiv:1702.02234
  [hep-ph]} \BibitemShut {NoStop}%
\bibitem [{\citenamefont {Capdevila}\ \emph
  {et~al.}(2017{\natexlab{b}})\citenamefont {Capdevila}, \citenamefont
  {Descotes-Genon}, \citenamefont {Hofer},\ and\ \citenamefont
  {Matias}}]{Capdevila:2017ert}%
  \BibitemOpen
  \bibfield  {author} {\bibinfo {author} {\bibfnamefont {B.}~\bibnamefont
  {Capdevila}}, \bibinfo {author} {\bibfnamefont {S.}~\bibnamefont
  {Descotes-Genon}}, \bibinfo {author} {\bibfnamefont {L.}~\bibnamefont
  {Hofer}}, \ and\ \bibinfo {author} {\bibfnamefont {J.}~\bibnamefont
  {Matias}},\ }\href {\doibase 10.1007/JHEP04(2017)016} {\bibfield  {journal}
  {\bibinfo  {journal} {JHEP}\ }\textbf {\bibinfo {volume} {04}},\ \bibinfo
  {pages} {016} (\bibinfo {year} {2017}{\natexlab{b}})},\ \Eprint
  {http://arxiv.org/abs/1701.08672} {arXiv:1701.08672 [hep-ph]} \BibitemShut
  {NoStop}%
\bibitem [{\citenamefont {Bobeth}\ \emph {et~al.}(2017)\citenamefont {Bobeth},
  \citenamefont {Chrzaszcz}, \citenamefont {van Dyk},\ and\ \citenamefont
  {Virto}}]{Bobeth:2017vxj}%
  \BibitemOpen
  \bibfield  {author} {\bibinfo {author} {\bibfnamefont {C.}~\bibnamefont
  {Bobeth}}, \bibinfo {author} {\bibfnamefont {M.}~\bibnamefont {Chrzaszcz}},
  \bibinfo {author} {\bibfnamefont {D.}~\bibnamefont {van Dyk}}, \ and\
  \bibinfo {author} {\bibfnamefont {J.}~\bibnamefont {Virto}},\ }\href@noop {}
  {\  (\bibinfo {year} {2017})},\ \Eprint {http://arxiv.org/abs/1707.07305}
  {arXiv:1707.07305 [hep-ph]} \BibitemShut {NoStop}%
\bibitem [{\citenamefont {Hiller}\ and\ \citenamefont
  {Kruger}(2004)}]{Hiller:2003js}%
  \BibitemOpen
  \bibfield  {author} {\bibinfo {author} {\bibfnamefont {G.}~\bibnamefont
  {Hiller}}\ and\ \bibinfo {author} {\bibfnamefont {F.}~\bibnamefont
  {Kruger}},\ }\href {\doibase 10.1103/PhysRevD.69.074020} {\bibfield
  {journal} {\bibinfo  {journal} {Phys. Rev.}\ }\textbf {\bibinfo {volume}
  {D69}},\ \bibinfo {pages} {074020} (\bibinfo {year} {2004})},\ \Eprint
  {http://arxiv.org/abs/hep-ph/0310219} {arXiv:hep-ph/0310219 [hep-ph]}
  \BibitemShut {NoStop}%
\bibitem [{\citenamefont {Bordone}\ \emph {et~al.}(2016)\citenamefont
  {Bordone}, \citenamefont {Isidori},\ and\ \citenamefont
  {Pattori}}]{Bordone:2016gaq}%
  \BibitemOpen
  \bibfield  {author} {\bibinfo {author} {\bibfnamefont {M.}~\bibnamefont
  {Bordone}}, \bibinfo {author} {\bibfnamefont {G.}~\bibnamefont {Isidori}}, \
  and\ \bibinfo {author} {\bibfnamefont {A.}~\bibnamefont {Pattori}},\ }\href
  {\doibase 10.1140/epjc/s10052-016-4274-7} {\bibfield  {journal} {\bibinfo
  {journal} {Eur. Phys. J.}\ }\textbf {\bibinfo {volume} {C76}},\ \bibinfo
  {pages} {440} (\bibinfo {year} {2016})},\ \Eprint
  {http://arxiv.org/abs/1605.07633} {arXiv:1605.07633 [hep-ph]} \BibitemShut
  {NoStop}%
\bibitem [{\citenamefont {Aaij}\ \emph
  {et~al.}(2014{\natexlab{b}})\citenamefont {Aaij} \emph
  {et~al.}}]{Aaij:2014ora}%
  \BibitemOpen
  \bibfield  {author} {\bibinfo {author} {\bibfnamefont {R.}~\bibnamefont
  {Aaij}} \emph {et~al.} (\bibinfo {collaboration} {LHCb}),\ }\href {\doibase
  10.1103/PhysRevLett.113.151601} {\bibfield  {journal} {\bibinfo  {journal}
  {Phys. Rev. Lett.}\ }\textbf {\bibinfo {volume} {113}},\ \bibinfo {pages}
  {151601} (\bibinfo {year} {2014}{\natexlab{b}})},\ \Eprint
  {http://arxiv.org/abs/1406.6482} {arXiv:1406.6482 [hep-ex]} \BibitemShut
  {NoStop}%
\bibitem [{\citenamefont {Aaij}\ \emph {et~al.}(2017)\citenamefont {Aaij} \emph
  {et~al.}}]{Aaij:2017vbb}%
  \BibitemOpen
  \bibfield  {author} {\bibinfo {author} {\bibfnamefont {R.}~\bibnamefont
  {Aaij}} \emph {et~al.} (\bibinfo {collaboration} {LHCb}),\ }\href {\doibase
  10.1007/JHEP08(2017)055} {\bibfield  {journal} {\bibinfo  {journal} {JHEP}\
  }\textbf {\bibinfo {volume} {08}},\ \bibinfo {pages} {055} (\bibinfo {year}
  {2017})},\ \Eprint {http://arxiv.org/abs/1705.05802} {arXiv:1705.05802
  [hep-ex]} \BibitemShut {NoStop}%
\bibitem [{\citenamefont {Buras}\ and\ \citenamefont
  {Girrbach}(2013)}]{Buras:2013qja}%
  \BibitemOpen
  \bibfield  {author} {\bibinfo {author} {\bibfnamefont {A.~J.}\ \bibnamefont
  {Buras}}\ and\ \bibinfo {author} {\bibfnamefont {J.}~\bibnamefont
  {Girrbach}},\ }\href {\doibase 10.1007/JHEP12(2013)009} {\bibfield  {journal}
  {\bibinfo  {journal} {JHEP}\ }\textbf {\bibinfo {volume} {12}},\ \bibinfo
  {pages} {009} (\bibinfo {year} {2013})},\ \Eprint
  {http://arxiv.org/abs/1309.2466} {arXiv:1309.2466 [hep-ph]} \BibitemShut
  {NoStop}%
\bibitem [{\citenamefont {Gauld}\ \emph {et~al.}(2014)\citenamefont {Gauld},
  \citenamefont {Goertz},\ and\ \citenamefont {Haisch}}]{Gauld:2013qja}%
  \BibitemOpen
  \bibfield  {author} {\bibinfo {author} {\bibfnamefont {R.}~\bibnamefont
  {Gauld}}, \bibinfo {author} {\bibfnamefont {F.}~\bibnamefont {Goertz}}, \
  and\ \bibinfo {author} {\bibfnamefont {U.}~\bibnamefont {Haisch}},\ }\href
  {\doibase 10.1007/JHEP01(2014)069} {\bibfield  {journal} {\bibinfo  {journal}
  {JHEP}\ }\textbf {\bibinfo {volume} {01}},\ \bibinfo {pages} {069} (\bibinfo
  {year} {2014})},\ \Eprint {http://arxiv.org/abs/1310.1082} {arXiv:1310.1082
  [hep-ph]} \BibitemShut {NoStop}%
\bibitem [{\citenamefont {Buras}\ \emph {et~al.}(2014)\citenamefont {Buras},
  \citenamefont {De~Fazio},\ and\ \citenamefont {Girrbach}}]{Buras:2013dea}%
  \BibitemOpen
  \bibfield  {author} {\bibinfo {author} {\bibfnamefont {A.~J.}\ \bibnamefont
  {Buras}}, \bibinfo {author} {\bibfnamefont {F.}~\bibnamefont {De~Fazio}}, \
  and\ \bibinfo {author} {\bibfnamefont {J.}~\bibnamefont {Girrbach}},\ }\href
  {\doibase 10.1007/JHEP02(2014)112} {\bibfield  {journal} {\bibinfo  {journal}
  {JHEP}\ }\textbf {\bibinfo {volume} {02}},\ \bibinfo {pages} {112} (\bibinfo
  {year} {2014})},\ \Eprint {http://arxiv.org/abs/1311.6729} {arXiv:1311.6729
  [hep-ph]} \BibitemShut {NoStop}%
\bibitem [{\citenamefont {Altmannshofer}\ \emph
  {et~al.}(2014{\natexlab{a}})\citenamefont {Altmannshofer}, \citenamefont
  {Gori}, \citenamefont {Pospelov},\ and\ \citenamefont
  {Yavin}}]{Altmannshofer:2014cfa}%
  \BibitemOpen
  \bibfield  {author} {\bibinfo {author} {\bibfnamefont {W.}~\bibnamefont
  {Altmannshofer}}, \bibinfo {author} {\bibfnamefont {S.}~\bibnamefont {Gori}},
  \bibinfo {author} {\bibfnamefont {M.}~\bibnamefont {Pospelov}}, \ and\
  \bibinfo {author} {\bibfnamefont {I.}~\bibnamefont {Yavin}},\ }\href
  {\doibase 10.1103/PhysRevD.89.095033} {\bibfield  {journal} {\bibinfo
  {journal} {Phys. Rev.}\ }\textbf {\bibinfo {volume} {D89}},\ \bibinfo {pages}
  {095033} (\bibinfo {year} {2014}{\natexlab{a}})},\ \Eprint
  {http://arxiv.org/abs/1403.1269} {arXiv:1403.1269 [hep-ph]} \BibitemShut
  {NoStop}%
\bibitem [{\citenamefont {Crivellin}\ \emph
  {et~al.}(2015{\natexlab{a}})\citenamefont {Crivellin}, \citenamefont
  {D'Ambrosio},\ and\ \citenamefont {Heeck}}]{Crivellin:2015mga}%
  \BibitemOpen
  \bibfield  {author} {\bibinfo {author} {\bibfnamefont {A.}~\bibnamefont
  {Crivellin}}, \bibinfo {author} {\bibfnamefont {G.}~\bibnamefont
  {D'Ambrosio}}, \ and\ \bibinfo {author} {\bibfnamefont {J.}~\bibnamefont
  {Heeck}},\ }\href {\doibase 10.1103/PhysRevLett.114.151801} {\bibfield
  {journal} {\bibinfo  {journal} {Phys. Rev. Lett.}\ }\textbf {\bibinfo
  {volume} {114}},\ \bibinfo {pages} {151801} (\bibinfo {year}
  {2015}{\natexlab{a}})},\ \Eprint {http://arxiv.org/abs/1501.00993}
  {arXiv:1501.00993 [hep-ph]} \BibitemShut {NoStop}%
\bibitem [{\citenamefont {Crivellin}\ \emph
  {et~al.}(2015{\natexlab{b}})\citenamefont {Crivellin}, \citenamefont
  {D'Ambrosio},\ and\ \citenamefont {Heeck}}]{Crivellin:2015lwa}%
  \BibitemOpen
  \bibfield  {author} {\bibinfo {author} {\bibfnamefont {A.}~\bibnamefont
  {Crivellin}}, \bibinfo {author} {\bibfnamefont {G.}~\bibnamefont
  {D'Ambrosio}}, \ and\ \bibinfo {author} {\bibfnamefont {J.}~\bibnamefont
  {Heeck}},\ }\href {\doibase 10.1103/PhysRevD.91.075006} {\bibfield  {journal}
  {\bibinfo  {journal} {Phys. Rev.}\ }\textbf {\bibinfo {volume} {D91}},\
  \bibinfo {pages} {075006} (\bibinfo {year} {2015}{\natexlab{b}})},\ \Eprint
  {http://arxiv.org/abs/1503.03477} {arXiv:1503.03477 [hep-ph]} \BibitemShut
  {NoStop}%
\bibitem [{\citenamefont {Celis}\ \emph {et~al.}(2015)\citenamefont {Celis},
  \citenamefont {Fuentes-Martin}, \citenamefont {Jung},\ and\ \citenamefont
  {Serodio}}]{Celis:2015ara}%
  \BibitemOpen
  \bibfield  {author} {\bibinfo {author} {\bibfnamefont {A.}~\bibnamefont
  {Celis}}, \bibinfo {author} {\bibfnamefont {J.}~\bibnamefont
  {Fuentes-Martin}}, \bibinfo {author} {\bibfnamefont {M.}~\bibnamefont
  {Jung}}, \ and\ \bibinfo {author} {\bibfnamefont {H.}~\bibnamefont
  {Serodio}},\ }\href {\doibase 10.1103/PhysRevD.92.015007} {\bibfield
  {journal} {\bibinfo  {journal} {Phys. Rev.}\ }\textbf {\bibinfo {volume}
  {D92}},\ \bibinfo {pages} {015007} (\bibinfo {year} {2015})},\ \Eprint
  {http://arxiv.org/abs/1505.03079} {arXiv:1505.03079 [hep-ph]} \BibitemShut
  {NoStop}%
\bibitem [{\citenamefont {Belanger}\ \emph {et~al.}(2015)\citenamefont
  {Belanger}, \citenamefont {Delaunay},\ and\ \citenamefont
  {Westhoff}}]{Belanger:2015nma}%
  \BibitemOpen
  \bibfield  {author} {\bibinfo {author} {\bibfnamefont {G.}~\bibnamefont
  {Belanger}}, \bibinfo {author} {\bibfnamefont {C.}~\bibnamefont {Delaunay}},
  \ and\ \bibinfo {author} {\bibfnamefont {S.}~\bibnamefont {Westhoff}},\
  }\href {\doibase 10.1103/PhysRevD.92.055021} {\bibfield  {journal} {\bibinfo
  {journal} {Phys. Rev.}\ }\textbf {\bibinfo {volume} {D92}},\ \bibinfo {pages}
  {055021} (\bibinfo {year} {2015})},\ \Eprint
  {http://arxiv.org/abs/1507.06660} {arXiv:1507.06660 [hep-ph]} \BibitemShut
  {NoStop}%
\bibitem [{\citenamefont {Falkowski}\ \emph {et~al.}(2015)\citenamefont
  {Falkowski}, \citenamefont {Nardecchia},\ and\ \citenamefont
  {Ziegler}}]{Falkowski:2015zwa}%
  \BibitemOpen
  \bibfield  {author} {\bibinfo {author} {\bibfnamefont {A.}~\bibnamefont
  {Falkowski}}, \bibinfo {author} {\bibfnamefont {M.}~\bibnamefont
  {Nardecchia}}, \ and\ \bibinfo {author} {\bibfnamefont {R.}~\bibnamefont
  {Ziegler}},\ }\href {\doibase 10.1007/JHEP11(2015)173} {\bibfield  {journal}
  {\bibinfo  {journal} {JHEP}\ }\textbf {\bibinfo {volume} {11}},\ \bibinfo
  {pages} {173} (\bibinfo {year} {2015})},\ \Eprint
  {http://arxiv.org/abs/1509.01249} {arXiv:1509.01249 [hep-ph]} \BibitemShut
  {NoStop}%
\bibitem [{\citenamefont {Carmona}\ and\ \citenamefont
  {Goertz}(2016)}]{Carmona:2015ena}%
  \BibitemOpen
  \bibfield  {author} {\bibinfo {author} {\bibfnamefont {A.}~\bibnamefont
  {Carmona}}\ and\ \bibinfo {author} {\bibfnamefont {F.}~\bibnamefont
  {Goertz}},\ }\href {\doibase 10.1103/PhysRevLett.116.251801} {\bibfield
  {journal} {\bibinfo  {journal} {Phys. Rev. Lett.}\ }\textbf {\bibinfo
  {volume} {116}},\ \bibinfo {pages} {251801} (\bibinfo {year} {2016})},\
  \Eprint {http://arxiv.org/abs/1510.07658} {arXiv:1510.07658 [hep-ph]}
  \BibitemShut {NoStop}%
\bibitem [{\citenamefont {Allanach}\ \emph {et~al.}(2016)\citenamefont
  {Allanach}, \citenamefont {Queiroz}, \citenamefont {Strumia},\ and\
  \citenamefont {Sun}}]{Allanach:2015gkd}%
  \BibitemOpen
  \bibfield  {author} {\bibinfo {author} {\bibfnamefont {B.}~\bibnamefont
  {Allanach}}, \bibinfo {author} {\bibfnamefont {F.~S.}\ \bibnamefont
  {Queiroz}}, \bibinfo {author} {\bibfnamefont {A.}~\bibnamefont {Strumia}}, \
  and\ \bibinfo {author} {\bibfnamefont {S.}~\bibnamefont {Sun}},\ }\href
  {\doibase 10.1103/PhysRevD.93.055045, 10.1103/PhysRevD.95.119902} {\bibfield
  {journal} {\bibinfo  {journal} {Phys. Rev.}\ }\textbf {\bibinfo {volume}
  {D93}},\ \bibinfo {pages} {055045} (\bibinfo {year} {2016})},\ \bibinfo
  {note} {[Erratum: Phys. Rev.D95,no.11,119902(2017)]},\ \Eprint
  {http://arxiv.org/abs/1511.07447} {arXiv:1511.07447 [hep-ph]} \BibitemShut
  {NoStop}%
\bibitem [{\citenamefont {Chiang}\ \emph {et~al.}(2016)\citenamefont {Chiang},
  \citenamefont {He},\ and\ \citenamefont {Valencia}}]{Chiang:2016qov}%
  \BibitemOpen
  \bibfield  {author} {\bibinfo {author} {\bibfnamefont {C.-W.}\ \bibnamefont
  {Chiang}}, \bibinfo {author} {\bibfnamefont {X.-G.}\ \bibnamefont {He}}, \
  and\ \bibinfo {author} {\bibfnamefont {G.}~\bibnamefont {Valencia}},\ }\href
  {\doibase 10.1103/PhysRevD.93.074003} {\bibfield  {journal} {\bibinfo
  {journal} {Phys. Rev.}\ }\textbf {\bibinfo {volume} {D93}},\ \bibinfo {pages}
  {074003} (\bibinfo {year} {2016})},\ \Eprint
  {http://arxiv.org/abs/1601.07328} {arXiv:1601.07328 [hep-ph]} \BibitemShut
  {NoStop}%
\bibitem [{\citenamefont {Boucenna}\ \emph
  {et~al.}(2016{\natexlab{a}})\citenamefont {Boucenna}, \citenamefont {Celis},
  \citenamefont {Fuentes-Martin}, \citenamefont {Vicente},\ and\ \citenamefont
  {Virto}}]{Boucenna:2016wpr}%
  \BibitemOpen
  \bibfield  {author} {\bibinfo {author} {\bibfnamefont {S.~M.}\ \bibnamefont
  {Boucenna}}, \bibinfo {author} {\bibfnamefont {A.}~\bibnamefont {Celis}},
  \bibinfo {author} {\bibfnamefont {J.}~\bibnamefont {Fuentes-Martin}},
  \bibinfo {author} {\bibfnamefont {A.}~\bibnamefont {Vicente}}, \ and\
  \bibinfo {author} {\bibfnamefont {J.}~\bibnamefont {Virto}},\ }\href
  {\doibase 10.1016/j.physletb.2016.06.067} {\bibfield  {journal} {\bibinfo
  {journal} {Phys. Lett.}\ }\textbf {\bibinfo {volume} {B760}},\ \bibinfo
  {pages} {214} (\bibinfo {year} {2016}{\natexlab{a}})},\ \Eprint
  {http://arxiv.org/abs/1604.03088} {arXiv:1604.03088 [hep-ph]} \BibitemShut
  {NoStop}%
\bibitem [{\citenamefont {Megias}\ \emph {et~al.}(2016)\citenamefont {Megias},
  \citenamefont {Panico}, \citenamefont {Pujolas},\ and\ \citenamefont
  {Quiros}}]{Megias:2016bde}%
  \BibitemOpen
  \bibfield  {author} {\bibinfo {author} {\bibfnamefont {E.}~\bibnamefont
  {Megias}}, \bibinfo {author} {\bibfnamefont {G.}~\bibnamefont {Panico}},
  \bibinfo {author} {\bibfnamefont {O.}~\bibnamefont {Pujolas}}, \ and\
  \bibinfo {author} {\bibfnamefont {M.}~\bibnamefont {Quiros}},\ }\href
  {\doibase 10.1007/JHEP09(2016)118} {\bibfield  {journal} {\bibinfo  {journal}
  {JHEP}\ }\textbf {\bibinfo {volume} {09}},\ \bibinfo {pages} {118} (\bibinfo
  {year} {2016})},\ \Eprint {http://arxiv.org/abs/1608.02362} {arXiv:1608.02362
  [hep-ph]} \BibitemShut {NoStop}%
\bibitem [{\citenamefont {Boucenna}\ \emph
  {et~al.}(2016{\natexlab{b}})\citenamefont {Boucenna}, \citenamefont {Celis},
  \citenamefont {Fuentes-Martin}, \citenamefont {Vicente},\ and\ \citenamefont
  {Virto}}]{Boucenna:2016qad}%
  \BibitemOpen
  \bibfield  {author} {\bibinfo {author} {\bibfnamefont {S.~M.}\ \bibnamefont
  {Boucenna}}, \bibinfo {author} {\bibfnamefont {A.}~\bibnamefont {Celis}},
  \bibinfo {author} {\bibfnamefont {J.}~\bibnamefont {Fuentes-Martin}},
  \bibinfo {author} {\bibfnamefont {A.}~\bibnamefont {Vicente}}, \ and\
  \bibinfo {author} {\bibfnamefont {J.}~\bibnamefont {Virto}},\ }\href
  {\doibase 10.1007/JHEP12(2016)059} {\bibfield  {journal} {\bibinfo  {journal}
  {JHEP}\ }\textbf {\bibinfo {volume} {12}},\ \bibinfo {pages} {059} (\bibinfo
  {year} {2016}{\natexlab{b}})},\ \Eprint {http://arxiv.org/abs/1608.01349}
  {arXiv:1608.01349 [hep-ph]} \BibitemShut {NoStop}%
\bibitem [{\citenamefont {Altmannshofer}\ \emph {et~al.}(2016)\citenamefont
  {Altmannshofer}, \citenamefont {Gori}, \citenamefont {Profumo},\ and\
  \citenamefont {Queiroz}}]{Altmannshofer:2016jzy}%
  \BibitemOpen
  \bibfield  {author} {\bibinfo {author} {\bibfnamefont {W.}~\bibnamefont
  {Altmannshofer}}, \bibinfo {author} {\bibfnamefont {S.}~\bibnamefont {Gori}},
  \bibinfo {author} {\bibfnamefont {S.}~\bibnamefont {Profumo}}, \ and\
  \bibinfo {author} {\bibfnamefont {F.~S.}\ \bibnamefont {Queiroz}},\ }\href
  {\doibase 10.1007/JHEP12(2016)106} {\bibfield  {journal} {\bibinfo  {journal}
  {JHEP}\ }\textbf {\bibinfo {volume} {12}},\ \bibinfo {pages} {106} (\bibinfo
  {year} {2016})},\ \Eprint {http://arxiv.org/abs/1609.04026} {arXiv:1609.04026
  [hep-ph]} \BibitemShut {NoStop}%
\bibitem [{\citenamefont {Crivellin}\ \emph
  {et~al.}(2017{\natexlab{a}})\citenamefont {Crivellin}, \citenamefont
  {Fuentes-Martin}, \citenamefont {Greljo},\ and\ \citenamefont
  {Isidori}}]{Crivellin:2016ejn}%
  \BibitemOpen
  \bibfield  {author} {\bibinfo {author} {\bibfnamefont {A.}~\bibnamefont
  {Crivellin}}, \bibinfo {author} {\bibfnamefont {J.}~\bibnamefont
  {Fuentes-Martin}}, \bibinfo {author} {\bibfnamefont {A.}~\bibnamefont
  {Greljo}}, \ and\ \bibinfo {author} {\bibfnamefont {G.}~\bibnamefont
  {Isidori}},\ }\href {\doibase 10.1016/j.physletb.2016.12.057} {\bibfield
  {journal} {\bibinfo  {journal} {Phys. Lett.}\ }\textbf {\bibinfo {volume}
  {B766}},\ \bibinfo {pages} {77} (\bibinfo {year} {2017}{\natexlab{a}})},\
  \Eprint {http://arxiv.org/abs/1611.02703} {arXiv:1611.02703 [hep-ph]}
  \BibitemShut {NoStop}%
\bibitem [{\citenamefont {Garcia~Garcia}(2017)}]{GarciaGarcia:2016nvr}%
  \BibitemOpen
  \bibfield  {author} {\bibinfo {author} {\bibfnamefont {I.}~\bibnamefont
  {Garcia~Garcia}},\ }\href {\doibase 10.1007/JHEP03(2017)040} {\bibfield
  {journal} {\bibinfo  {journal} {JHEP}\ }\textbf {\bibinfo {volume} {03}},\
  \bibinfo {pages} {040} (\bibinfo {year} {2017})},\ \Eprint
  {http://arxiv.org/abs/1611.03507} {arXiv:1611.03507 [hep-ph]} \BibitemShut
  {NoStop}%
\bibitem [{\citenamefont {Bhatia}\ \emph {et~al.}(2017)\citenamefont {Bhatia},
  \citenamefont {Chakraborty},\ and\ \citenamefont {Dighe}}]{Bhatia:2017tgo}%
  \BibitemOpen
  \bibfield  {author} {\bibinfo {author} {\bibfnamefont {D.}~\bibnamefont
  {Bhatia}}, \bibinfo {author} {\bibfnamefont {S.}~\bibnamefont {Chakraborty}},
  \ and\ \bibinfo {author} {\bibfnamefont {A.}~\bibnamefont {Dighe}},\ }\href
  {\doibase 10.1007/JHEP03(2017)117} {\bibfield  {journal} {\bibinfo  {journal}
  {JHEP}\ }\textbf {\bibinfo {volume} {03}},\ \bibinfo {pages} {117} (\bibinfo
  {year} {2017})},\ \Eprint {http://arxiv.org/abs/1701.05825} {arXiv:1701.05825
  [hep-ph]} \BibitemShut {NoStop}%
\bibitem [{\citenamefont {Cline}\ \emph {et~al.}(2017)\citenamefont {Cline},
  \citenamefont {Cornell}, \citenamefont {London},\ and\ \citenamefont
  {Watanabe}}]{Cline:2017lvv}%
  \BibitemOpen
  \bibfield  {author} {\bibinfo {author} {\bibfnamefont {J.~M.}\ \bibnamefont
  {Cline}}, \bibinfo {author} {\bibfnamefont {J.~M.}\ \bibnamefont {Cornell}},
  \bibinfo {author} {\bibfnamefont {D.}~\bibnamefont {London}}, \ and\ \bibinfo
  {author} {\bibfnamefont {R.}~\bibnamefont {Watanabe}},\ }\href {\doibase
  10.1103/PhysRevD.95.095015} {\bibfield  {journal} {\bibinfo  {journal} {Phys.
  Rev.}\ }\textbf {\bibinfo {volume} {D95}},\ \bibinfo {pages} {095015}
  (\bibinfo {year} {2017})},\ \Eprint {http://arxiv.org/abs/1702.00395}
  {arXiv:1702.00395 [hep-ph]} \BibitemShut {NoStop}%
\bibitem [{\citenamefont {Baek}(2017)}]{Baek:2017sew}%
  \BibitemOpen
  \bibfield  {author} {\bibinfo {author} {\bibfnamefont {S.}~\bibnamefont
  {Baek}},\ }\href@noop {} {\  (\bibinfo {year} {2017})},\ \Eprint
  {http://arxiv.org/abs/1707.04573} {arXiv:1707.04573 [hep-ph]} \BibitemShut
  {NoStop}%
\bibitem [{\citenamefont {Cline}\ and\ \citenamefont
  {Martin~Camalich}(2017)}]{Cline:2017ihf}%
  \BibitemOpen
  \bibfield  {author} {\bibinfo {author} {\bibfnamefont {J.~M.}\ \bibnamefont
  {Cline}}\ and\ \bibinfo {author} {\bibfnamefont {J.}~\bibnamefont
  {Martin~Camalich}},\ }\href {\doibase 10.1103/PhysRevD.96.055036} {\bibfield
  {journal} {\bibinfo  {journal} {Phys. Rev.}\ }\textbf {\bibinfo {volume}
  {D96}},\ \bibinfo {pages} {055036} (\bibinfo {year} {2017})},\ \Eprint
  {http://arxiv.org/abs/1706.08510} {arXiv:1706.08510 [hep-ph]} \BibitemShut
  {NoStop}%
\bibitem [{\citenamefont {Di~Chiara}\ \emph {et~al.}(2017)\citenamefont
  {Di~Chiara}, \citenamefont {Fowlie}, \citenamefont {Fraser}, \citenamefont
  {Marzo}, \citenamefont {Marzola}, \citenamefont {Raidal},\ and\ \citenamefont
  {Spethmann}}]{DiChiara:2017cjq}%
  \BibitemOpen
  \bibfield  {author} {\bibinfo {author} {\bibfnamefont {S.}~\bibnamefont
  {Di~Chiara}}, \bibinfo {author} {\bibfnamefont {A.}~\bibnamefont {Fowlie}},
  \bibinfo {author} {\bibfnamefont {S.}~\bibnamefont {Fraser}}, \bibinfo
  {author} {\bibfnamefont {C.}~\bibnamefont {Marzo}}, \bibinfo {author}
  {\bibfnamefont {L.}~\bibnamefont {Marzola}}, \bibinfo {author} {\bibfnamefont
  {M.}~\bibnamefont {Raidal}}, \ and\ \bibinfo {author} {\bibfnamefont
  {C.}~\bibnamefont {Spethmann}},\ }\href {\doibase
  10.1016/j.nuclphysb.2017.08.003} {\bibfield  {journal} {\bibinfo  {journal}
  {Nucl. Phys.}\ }\textbf {\bibinfo {volume} {B923}},\ \bibinfo {pages} {245}
  (\bibinfo {year} {2017})},\ \Eprint {http://arxiv.org/abs/1704.06200}
  {arXiv:1704.06200 [hep-ph]} \BibitemShut {NoStop}%
\bibitem [{\citenamefont {Kamenik}\ \emph {et~al.}(2017)\citenamefont
  {Kamenik}, \citenamefont {Soreq},\ and\ \citenamefont
  {Zupan}}]{Kamenik:2017tnu}%
  \BibitemOpen
  \bibfield  {author} {\bibinfo {author} {\bibfnamefont {J.~F.}\ \bibnamefont
  {Kamenik}}, \bibinfo {author} {\bibfnamefont {Y.}~\bibnamefont {Soreq}}, \
  and\ \bibinfo {author} {\bibfnamefont {J.}~\bibnamefont {Zupan}},\
  }\href@noop {} {\  (\bibinfo {year} {2017})},\ \Eprint
  {http://arxiv.org/abs/1704.06005} {arXiv:1704.06005 [hep-ph]} \BibitemShut
  {NoStop}%
\bibitem [{\citenamefont {Ko}\ \emph {et~al.}(2017{\natexlab{a}})\citenamefont
  {Ko}, \citenamefont {Omura}, \citenamefont {Shigekami},\ and\ \citenamefont
  {Yu}}]{Ko:2017lzd}%
  \BibitemOpen
  \bibfield  {author} {\bibinfo {author} {\bibfnamefont {P.}~\bibnamefont
  {Ko}}, \bibinfo {author} {\bibfnamefont {Y.}~\bibnamefont {Omura}}, \bibinfo
  {author} {\bibfnamefont {Y.}~\bibnamefont {Shigekami}}, \ and\ \bibinfo
  {author} {\bibfnamefont {C.}~\bibnamefont {Yu}},\ }\href {\doibase
  10.1103/PhysRevD.95.115040} {\bibfield  {journal} {\bibinfo  {journal} {Phys.
  Rev.}\ }\textbf {\bibinfo {volume} {D95}},\ \bibinfo {pages} {115040}
  (\bibinfo {year} {2017}{\natexlab{a}})},\ \Eprint
  {http://arxiv.org/abs/1702.08666} {arXiv:1702.08666 [hep-ph]} \BibitemShut
  {NoStop}%
\bibitem [{\citenamefont {Ko}\ \emph {et~al.}(2017{\natexlab{b}})\citenamefont
  {Ko}, \citenamefont {Nomura},\ and\ \citenamefont {Okada}}]{Ko:2017yrd}%
  \BibitemOpen
  \bibfield  {author} {\bibinfo {author} {\bibfnamefont {P.}~\bibnamefont
  {Ko}}, \bibinfo {author} {\bibfnamefont {T.}~\bibnamefont {Nomura}}, \ and\
  \bibinfo {author} {\bibfnamefont {H.}~\bibnamefont {Okada}},\ }\href
  {\doibase 10.1103/PhysRevD.95.111701} {\bibfield  {journal} {\bibinfo
  {journal} {Phys. Rev.}\ }\textbf {\bibinfo {volume} {D95}},\ \bibinfo {pages}
  {111701} (\bibinfo {year} {2017}{\natexlab{b}})},\ \Eprint
  {http://arxiv.org/abs/1702.02699} {arXiv:1702.02699 [hep-ph]} \BibitemShut
  {NoStop}%
\bibitem [{\citenamefont {Alonso}\ \emph
  {et~al.}(2017{\natexlab{a}})\citenamefont {Alonso}, \citenamefont {Cox},
  \citenamefont {Han},\ and\ \citenamefont {Yanagida}}]{Alonso:2017bff}%
  \BibitemOpen
  \bibfield  {author} {\bibinfo {author} {\bibfnamefont {R.}~\bibnamefont
  {Alonso}}, \bibinfo {author} {\bibfnamefont {P.}~\bibnamefont {Cox}},
  \bibinfo {author} {\bibfnamefont {C.}~\bibnamefont {Han}}, \ and\ \bibinfo
  {author} {\bibfnamefont {T.~T.}\ \bibnamefont {Yanagida}},\ }\href {\doibase
  10.1103/PhysRevD.96.071701} {\bibfield  {journal} {\bibinfo  {journal} {Phys.
  Rev.}\ }\textbf {\bibinfo {volume} {D96}},\ \bibinfo {pages} {071701}
  (\bibinfo {year} {2017}{\natexlab{a}})},\ \Eprint
  {http://arxiv.org/abs/1704.08158} {arXiv:1704.08158 [hep-ph]} \BibitemShut
  {NoStop}%
\bibitem [{\citenamefont {Ellis}\ \emph {et~al.}(2017)\citenamefont {Ellis},
  \citenamefont {Fairbairn},\ and\ \citenamefont {Tunney}}]{Ellis:2017nrp}%
  \BibitemOpen
  \bibfield  {author} {\bibinfo {author} {\bibfnamefont {J.}~\bibnamefont
  {Ellis}}, \bibinfo {author} {\bibfnamefont {M.}~\bibnamefont {Fairbairn}}, \
  and\ \bibinfo {author} {\bibfnamefont {P.}~\bibnamefont {Tunney}},\
  }\href@noop {} {\  (\bibinfo {year} {2017})},\ \Eprint
  {http://arxiv.org/abs/1705.03447} {arXiv:1705.03447 [hep-ph]} \BibitemShut
  {NoStop}%
\bibitem [{\citenamefont {Alonso}\ \emph
  {et~al.}(2017{\natexlab{b}})\citenamefont {Alonso}, \citenamefont {Cox},
  \citenamefont {Han},\ and\ \citenamefont {Yanagida}}]{Alonso:2017uky}%
  \BibitemOpen
  \bibfield  {author} {\bibinfo {author} {\bibfnamefont {R.}~\bibnamefont
  {Alonso}}, \bibinfo {author} {\bibfnamefont {P.}~\bibnamefont {Cox}},
  \bibinfo {author} {\bibfnamefont {C.}~\bibnamefont {Han}}, \ and\ \bibinfo
  {author} {\bibfnamefont {T.~T.}\ \bibnamefont {Yanagida}},\ }\href {\doibase
  10.1016/j.physletb.2017.10.027} {\bibfield  {journal} {\bibinfo  {journal}
  {Phys. Lett.}\ }\textbf {\bibinfo {volume} {B774}},\ \bibinfo {pages} {643}
  (\bibinfo {year} {2017}{\natexlab{b}})},\ \Eprint
  {http://arxiv.org/abs/1705.03858} {arXiv:1705.03858 [hep-ph]} \BibitemShut
  {NoStop}%
\bibitem [{\citenamefont {Carmona}\ and\ \citenamefont
  {Goertz}(2017)}]{Carmona:2017fsn}%
  \BibitemOpen
  \bibfield  {author} {\bibinfo {author} {\bibfnamefont {A.}~\bibnamefont
  {Carmona}}\ and\ \bibinfo {author} {\bibfnamefont {F.}~\bibnamefont
  {Goertz}},\ }\href@noop {} {\  (\bibinfo {year} {2017})},\ \Eprint
  {http://arxiv.org/abs/1712.02536} {arXiv:1712.02536 [hep-ph]} \BibitemShut
  {NoStop}%
\bibitem [{\citenamefont {Artuso}\ \emph {et~al.}(2016)\citenamefont {Artuso},
  \citenamefont {Borissov},\ and\ \citenamefont {Lenz}}]{Artuso:2015swg}%
  \BibitemOpen
  \bibfield  {author} {\bibinfo {author} {\bibfnamefont {M.}~\bibnamefont
  {Artuso}}, \bibinfo {author} {\bibfnamefont {G.}~\bibnamefont {Borissov}}, \
  and\ \bibinfo {author} {\bibfnamefont {A.}~\bibnamefont {Lenz}},\ }\href
  {\doibase 10.1103/RevModPhys.88.045002} {\bibfield  {journal} {\bibinfo
  {journal} {Rev. Mod. Phys.}\ }\textbf {\bibinfo {volume} {88}},\ \bibinfo
  {pages} {045002} (\bibinfo {year} {2016})},\ \Eprint
  {http://arxiv.org/abs/1511.09466} {arXiv:1511.09466 [hep-ph]} \BibitemShut
  {NoStop}%
\bibitem [{\citenamefont {Lenz}\ and\ \citenamefont
  {Nierste}(2011)}]{Lenz:2011ti}%
  \BibitemOpen
  \bibfield  {author} {\bibinfo {author} {\bibfnamefont {A.}~\bibnamefont
  {Lenz}}\ and\ \bibinfo {author} {\bibfnamefont {U.}~\bibnamefont {Nierste}},\
  }in\ \href {http://inspirehep.net/record/890169/files/arXiv:1102.4274.pdf}
  {\emph {\bibinfo {booktitle} {{CKM unitarity triangle. Proceedings, 6th
  International Workshop, CKM 2010, Warwick, UK, September 6-10, 2010}}}}\
  (\bibinfo {year} {2011})\ \Eprint {http://arxiv.org/abs/1102.4274}
  {arXiv:1102.4274 [hep-ph]} \BibitemShut {NoStop}%
\bibitem [{\citenamefont {Inami}\ and\ \citenamefont
  {Lim}(1981)}]{Inami:1980fz}%
  \BibitemOpen
  \bibfield  {author} {\bibinfo {author} {\bibfnamefont {T.}~\bibnamefont
  {Inami}}\ and\ \bibinfo {author} {\bibfnamefont {C.~S.}\ \bibnamefont
  {Lim}},\ }\href {\doibase 10.1143/PTP.65.297} {\bibfield  {journal} {\bibinfo
   {journal} {Prog. Theor. Phys.}\ }\textbf {\bibinfo {volume} {65}},\ \bibinfo
  {pages} {297} (\bibinfo {year} {1981})},\ \bibinfo {note} {[Erratum: Prog.
  Theor. Phys.65,1772(1981)]}\BibitemShut {NoStop}%
\bibitem [{\citenamefont {Bardeen}\ \emph {et~al.}(1978)\citenamefont
  {Bardeen}, \citenamefont {Buras}, \citenamefont {Duke},\ and\ \citenamefont
  {Muta}}]{Bardeen:1978yd}%
  \BibitemOpen
  \bibfield  {author} {\bibinfo {author} {\bibfnamefont {W.~A.}\ \bibnamefont
  {Bardeen}}, \bibinfo {author} {\bibfnamefont {A.~J.}\ \bibnamefont {Buras}},
  \bibinfo {author} {\bibfnamefont {D.~W.}\ \bibnamefont {Duke}}, \ and\
  \bibinfo {author} {\bibfnamefont {T.}~\bibnamefont {Muta}},\ }\href {\doibase
  10.1103/PhysRevD.18.3998} {\bibfield  {journal} {\bibinfo  {journal} {Phys.
  Rev.}\ }\textbf {\bibinfo {volume} {D18}},\ \bibinfo {pages} {3998} (\bibinfo
  {year} {1978})}\BibitemShut {NoStop}%
\bibitem [{\citenamefont {Buras}\ \emph {et~al.}(1990)\citenamefont {Buras},
  \citenamefont {Jamin},\ and\ \citenamefont {Weisz}}]{Buras:1990fn}%
  \BibitemOpen
  \bibfield  {author} {\bibinfo {author} {\bibfnamefont {A.~J.}\ \bibnamefont
  {Buras}}, \bibinfo {author} {\bibfnamefont {M.}~\bibnamefont {Jamin}}, \ and\
  \bibinfo {author} {\bibfnamefont {P.~H.}\ \bibnamefont {Weisz}},\ }\href
  {\doibase 10.1016/0550-3213(90)90373-L} {\bibfield  {journal} {\bibinfo
  {journal} {Nucl. Phys.}\ }\textbf {\bibinfo {volume} {B347}},\ \bibinfo
  {pages} {491} (\bibinfo {year} {1990})}\BibitemShut {NoStop}%
\bibitem [{\citenamefont {Aoki}\ \emph {et~al.}(2017)\citenamefont {Aoki} \emph
  {et~al.}}]{Aoki:2016frl}%
  \BibitemOpen
  \bibfield  {author} {\bibinfo {author} {\bibfnamefont {S.}~\bibnamefont
  {Aoki}} \emph {et~al.},\ }\href {\doibase 10.1140/epjc/s10052-016-4509-7}
  {\bibfield  {journal} {\bibinfo  {journal} {Eur. Phys. J.}\ }\textbf
  {\bibinfo {volume} {C77}},\ \bibinfo {pages} {112} (\bibinfo {year}
  {2017})},\ \Eprint {http://arxiv.org/abs/1607.00299} {arXiv:1607.00299
  [hep-lat]} \BibitemShut {NoStop}%
\bibitem [{\citenamefont {Amhis}\ \emph {et~al.}(2016)\citenamefont {Amhis}
  \emph {et~al.}}]{Amhis:2016xyh}%
  \BibitemOpen
  \bibfield  {author} {\bibinfo {author} {\bibfnamefont {Y.}~\bibnamefont
  {Amhis}} \emph {et~al.},\ }\href@noop {} {\  (\bibinfo {year} {2016})},\
  \Eprint {http://arxiv.org/abs/1612.07233} {arXiv:1612.07233 [hep-ex]}
  \BibitemShut {NoStop}%
\bibitem [{\citenamefont {Bazavov}\ \emph {et~al.}(2016)\citenamefont {Bazavov}
  \emph {et~al.}}]{Bazavov:2016nty}%
  \BibitemOpen
  \bibfield  {author} {\bibinfo {author} {\bibfnamefont {A.}~\bibnamefont
  {Bazavov}} \emph {et~al.} (\bibinfo {collaboration} {Fermilab Lattice,
  MILC}),\ }\href {\doibase 10.1103/PhysRevD.93.113016} {\bibfield  {journal}
  {\bibinfo  {journal} {Phys. Rev.}\ }\textbf {\bibinfo {volume} {D93}},\
  \bibinfo {pages} {113016} (\bibinfo {year} {2016})},\ \Eprint
  {http://arxiv.org/abs/1602.03560} {arXiv:1602.03560 [hep-lat]} \BibitemShut
  {NoStop}%
\bibitem [{\citenamefont {Blanke}\ and\ \citenamefont
  {Buras}(2016)}]{Blanke:2016bhf}%
  \BibitemOpen
  \bibfield  {author} {\bibinfo {author} {\bibfnamefont {M.}~\bibnamefont
  {Blanke}}\ and\ \bibinfo {author} {\bibfnamefont {A.~J.}\ \bibnamefont
  {Buras}},\ }\href {\doibase 10.1140/epjc/s10052-016-4044-6} {\bibfield
  {journal} {\bibinfo  {journal} {Eur. Phys. J.}\ }\textbf {\bibinfo {volume}
  {C76}},\ \bibinfo {pages} {197} (\bibinfo {year} {2016})},\ \Eprint
  {http://arxiv.org/abs/1602.04020} {arXiv:1602.04020 [hep-ph]} \BibitemShut
  {NoStop}%
\bibitem [{\citenamefont {Jubb}\ \emph {et~al.}(2017)\citenamefont {Jubb},
  \citenamefont {Kirk}, \citenamefont {Lenz},\ and\ \citenamefont
  {Tetlalmatzi-Xolocotzi}}]{Jubb:2016mvq}%
  \BibitemOpen
  \bibfield  {author} {\bibinfo {author} {\bibfnamefont {T.}~\bibnamefont
  {Jubb}}, \bibinfo {author} {\bibfnamefont {M.}~\bibnamefont {Kirk}}, \bibinfo
  {author} {\bibfnamefont {A.}~\bibnamefont {Lenz}}, \ and\ \bibinfo {author}
  {\bibfnamefont {G.}~\bibnamefont {Tetlalmatzi-Xolocotzi}},\ }\href {\doibase
  10.1016/j.nuclphysb.2016.12.020} {\bibfield  {journal} {\bibinfo  {journal}
  {Nucl. Phys.}\ }\textbf {\bibinfo {volume} {B915}},\ \bibinfo {pages} {431}
  (\bibinfo {year} {2017})},\ \Eprint {http://arxiv.org/abs/1603.07770}
  {arXiv:1603.07770 [hep-ph]} \BibitemShut {NoStop}%
\bibitem [{\citenamefont {Buras}\ and\ \citenamefont
  {De~Fazio}(2016)}]{Buras:2016dxz}%
  \BibitemOpen
  \bibfield  {author} {\bibinfo {author} {\bibfnamefont {A.~J.}\ \bibnamefont
  {Buras}}\ and\ \bibinfo {author} {\bibfnamefont {F.}~\bibnamefont
  {De~Fazio}},\ }\href {\doibase 10.1007/JHEP08(2016)115} {\bibfield  {journal}
  {\bibinfo  {journal} {JHEP}\ }\textbf {\bibinfo {volume} {08}},\ \bibinfo
  {pages} {115} (\bibinfo {year} {2016})},\ \Eprint
  {http://arxiv.org/abs/1604.02344} {arXiv:1604.02344 [hep-ph]} \BibitemShut
  {NoStop}%
\bibitem [{\citenamefont {Altmannshofer}\ \emph
  {et~al.}(2017{\natexlab{c}})\citenamefont {Altmannshofer}, \citenamefont
  {Gori}, \citenamefont {Robinson},\ and\ \citenamefont
  {Tuckler}}]{Altmannshofer:2017uvs}%
  \BibitemOpen
  \bibfield  {author} {\bibinfo {author} {\bibfnamefont {W.}~\bibnamefont
  {Altmannshofer}}, \bibinfo {author} {\bibfnamefont {S.}~\bibnamefont {Gori}},
  \bibinfo {author} {\bibfnamefont {D.~J.}\ \bibnamefont {Robinson}}, \ and\
  \bibinfo {author} {\bibfnamefont {D.}~\bibnamefont {Tuckler}},\ }\href@noop
  {} {\  (\bibinfo {year} {2017}{\natexlab{c}})},\ \Eprint
  {http://arxiv.org/abs/1712.01847} {arXiv:1712.01847 [hep-ph]} \BibitemShut
  {NoStop}%
\bibitem [{\citenamefont {Kirk}\ \emph {et~al.}(2017)\citenamefont {Kirk},
  \citenamefont {Lenz},\ and\ \citenamefont {Rauh}}]{Kirk:2017juj}%
  \BibitemOpen
  \bibfield  {author} {\bibinfo {author} {\bibfnamefont {M.}~\bibnamefont
  {Kirk}}, \bibinfo {author} {\bibfnamefont {A.}~\bibnamefont {Lenz}}, \ and\
  \bibinfo {author} {\bibfnamefont {T.}~\bibnamefont {Rauh}},\ }\href@noop {}
  {\  (\bibinfo {year} {2017})},\ \Eprint {http://arxiv.org/abs/1711.02100}
  {arXiv:1711.02100 [hep-ph]} \BibitemShut {NoStop}%
\bibitem [{\citenamefont {Bobrowski}\ \emph {et~al.}(2009)\citenamefont
  {Bobrowski}, \citenamefont {Lenz}, \citenamefont {Riedl},\ and\ \citenamefont
  {Rohrwild}}]{Bobrowski:2009ng}%
  \BibitemOpen
  \bibfield  {author} {\bibinfo {author} {\bibfnamefont {M.}~\bibnamefont
  {Bobrowski}}, \bibinfo {author} {\bibfnamefont {A.}~\bibnamefont {Lenz}},
  \bibinfo {author} {\bibfnamefont {J.}~\bibnamefont {Riedl}}, \ and\ \bibinfo
  {author} {\bibfnamefont {J.}~\bibnamefont {Rohrwild}},\ }\href {\doibase
  10.1103/PhysRevD.79.113006} {\bibfield  {journal} {\bibinfo  {journal} {Phys.
  Rev.}\ }\textbf {\bibinfo {volume} {D79}},\ \bibinfo {pages} {113006}
  (\bibinfo {year} {2009})},\ \Eprint {http://arxiv.org/abs/0902.4883}
  {arXiv:0902.4883 [hep-ph]} \BibitemShut {NoStop}%
\bibitem [{\citenamefont {Golowich}\ \emph {et~al.}(2011)\citenamefont
  {Golowich}, \citenamefont {Hewett}, \citenamefont {Pakvasa}, \citenamefont
  {Petrov},\ and\ \citenamefont {Yeghiyan}}]{Golowich:2011cx}%
  \BibitemOpen
  \bibfield  {author} {\bibinfo {author} {\bibfnamefont {E.}~\bibnamefont
  {Golowich}}, \bibinfo {author} {\bibfnamefont {J.}~\bibnamefont {Hewett}},
  \bibinfo {author} {\bibfnamefont {S.}~\bibnamefont {Pakvasa}}, \bibinfo
  {author} {\bibfnamefont {A.~A.}\ \bibnamefont {Petrov}}, \ and\ \bibinfo
  {author} {\bibfnamefont {G.~K.}\ \bibnamefont {Yeghiyan}},\ }\href {\doibase
  10.1103/PhysRevD.83.114017} {\bibfield  {journal} {\bibinfo  {journal} {Phys.
  Rev.}\ }\textbf {\bibinfo {volume} {D83}},\ \bibinfo {pages} {114017}
  (\bibinfo {year} {2011})},\ \Eprint {http://arxiv.org/abs/1102.0009}
  {arXiv:1102.0009 [hep-ph]} \BibitemShut {NoStop}%
\bibitem [{\citenamefont {Ciuchini}\ \emph {et~al.}(1998)\citenamefont
  {Ciuchini}, \citenamefont {Franco}, \citenamefont {Lubicz}, \citenamefont
  {Martinelli}, \citenamefont {Scimemi},\ and\ \citenamefont
  {Silvestrini}}]{Ciuchini:1997bw}%
  \BibitemOpen
  \bibfield  {author} {\bibinfo {author} {\bibfnamefont {M.}~\bibnamefont
  {Ciuchini}}, \bibinfo {author} {\bibfnamefont {E.}~\bibnamefont {Franco}},
  \bibinfo {author} {\bibfnamefont {V.}~\bibnamefont {Lubicz}}, \bibinfo
  {author} {\bibfnamefont {G.}~\bibnamefont {Martinelli}}, \bibinfo {author}
  {\bibfnamefont {I.}~\bibnamefont {Scimemi}}, \ and\ \bibinfo {author}
  {\bibfnamefont {L.}~\bibnamefont {Silvestrini}},\ }\href {\doibase
  10.1016/S0550-3213(98)00161-8} {\bibfield  {journal} {\bibinfo  {journal}
  {Nucl. Phys.}\ }\textbf {\bibinfo {volume} {B523}},\ \bibinfo {pages} {501}
  (\bibinfo {year} {1998})},\ \Eprint {http://arxiv.org/abs/hep-ph/9711402}
  {arXiv:hep-ph/9711402 [hep-ph]} \BibitemShut {NoStop}%
\bibitem [{\citenamefont {Buras}\ \emph {et~al.}(2000)\citenamefont {Buras},
  \citenamefont {Misiak},\ and\ \citenamefont {Urban}}]{Buras:2000if}%
  \BibitemOpen
  \bibfield  {author} {\bibinfo {author} {\bibfnamefont {A.~J.}\ \bibnamefont
  {Buras}}, \bibinfo {author} {\bibfnamefont {M.}~\bibnamefont {Misiak}}, \
  and\ \bibinfo {author} {\bibfnamefont {J.}~\bibnamefont {Urban}},\ }\href
  {\doibase 10.1016/S0550-3213(00)00437-5} {\bibfield  {journal} {\bibinfo
  {journal} {Nucl. Phys.}\ }\textbf {\bibinfo {volume} {B586}},\ \bibinfo
  {pages} {397} (\bibinfo {year} {2000})},\ \Eprint
  {http://arxiv.org/abs/hep-ph/0005183} {arXiv:hep-ph/0005183 [hep-ph]}
  \BibitemShut {NoStop}%
\bibitem [{\citenamefont {Alok}\ \emph
  {et~al.}(2017{\natexlab{b}})\citenamefont {Alok}, \citenamefont
  {Bhattacharya}, \citenamefont {Kumar}, \citenamefont {Kumar}, \citenamefont
  {London},\ and\ \citenamefont {Sankar}}]{Alok:2017jgr}%
  \BibitemOpen
  \bibfield  {author} {\bibinfo {author} {\bibfnamefont {A.~K.}\ \bibnamefont
  {Alok}}, \bibinfo {author} {\bibfnamefont {B.}~\bibnamefont {Bhattacharya}},
  \bibinfo {author} {\bibfnamefont {D.}~\bibnamefont {Kumar}}, \bibinfo
  {author} {\bibfnamefont {J.}~\bibnamefont {Kumar}}, \bibinfo {author}
  {\bibfnamefont {D.}~\bibnamefont {London}}, \ and\ \bibinfo {author}
  {\bibfnamefont {S.~U.}\ \bibnamefont {Sankar}},\ }\href {\doibase
  10.1103/PhysRevD.96.015034} {\bibfield  {journal} {\bibinfo  {journal} {Phys.
  Rev.}\ }\textbf {\bibinfo {volume} {D96}},\ \bibinfo {pages} {015034}
  (\bibinfo {year} {2017}{\natexlab{b}})},\ \Eprint
  {http://arxiv.org/abs/1703.09247} {arXiv:1703.09247 [hep-ph]} \BibitemShut
  {NoStop}%
\bibitem [{\citenamefont {Altmannshofer}\ \emph
  {et~al.}(2014{\natexlab{b}})\citenamefont {Altmannshofer}, \citenamefont
  {Gori}, \citenamefont {Pospelov},\ and\ \citenamefont
  {Yavin}}]{Altmannshofer:2014pba}%
  \BibitemOpen
  \bibfield  {author} {\bibinfo {author} {\bibfnamefont {W.}~\bibnamefont
  {Altmannshofer}}, \bibinfo {author} {\bibfnamefont {S.}~\bibnamefont {Gori}},
  \bibinfo {author} {\bibfnamefont {M.}~\bibnamefont {Pospelov}}, \ and\
  \bibinfo {author} {\bibfnamefont {I.}~\bibnamefont {Yavin}},\ }\href
  {\doibase 10.1103/PhysRevLett.113.091801} {\bibfield  {journal} {\bibinfo
  {journal} {Phys. Rev. Lett.}\ }\textbf {\bibinfo {volume} {113}},\ \bibinfo
  {pages} {091801} (\bibinfo {year} {2014}{\natexlab{b}})},\ \Eprint
  {http://arxiv.org/abs/1406.2332} {arXiv:1406.2332 [hep-ph]} \BibitemShut
  {NoStop}%
\bibitem [{\citenamefont {Falkowski}\ \emph {et~al.}(2018)\citenamefont
  {Falkowski}, \citenamefont {King}, \citenamefont {Perdomo},\ and\
  \citenamefont {Pierre}}]{Falkowski:2018dsl}%
  \BibitemOpen
  \bibfield  {author} {\bibinfo {author} {\bibfnamefont {A.}~\bibnamefont
  {Falkowski}}, \bibinfo {author} {\bibfnamefont {S.~F.}\ \bibnamefont {King}},
  \bibinfo {author} {\bibfnamefont {E.}~\bibnamefont {Perdomo}}, \ and\
  \bibinfo {author} {\bibfnamefont {M.}~\bibnamefont {Pierre}},\ }\href@noop {}
  {\  (\bibinfo {year} {2018})},\ \Eprint {http://arxiv.org/abs/1803.04430}
  {arXiv:1803.04430 [hep-ph]} \BibitemShut {NoStop}%
\bibitem [{\citenamefont {Di~Luzio}\ and\ \citenamefont
  {Nardecchia}(2017)}]{DiLuzio:2017chi}%
  \BibitemOpen
  \bibfield  {author} {\bibinfo {author} {\bibfnamefont {L.}~\bibnamefont
  {Di~Luzio}}\ and\ \bibinfo {author} {\bibfnamefont {M.}~\bibnamefont
  {Nardecchia}},\ }\href {\doibase 10.1140/epjc/s10052-017-5118-9} {\bibfield
  {journal} {\bibinfo  {journal} {Eur. Phys. J.}\ }\textbf {\bibinfo {volume}
  {C77}},\ \bibinfo {pages} {536} (\bibinfo {year} {2017})},\ \Eprint
  {http://arxiv.org/abs/1706.01868} {arXiv:1706.01868 [hep-ph]} \BibitemShut
  {NoStop}%
\bibitem [{\citenamefont {Di~Luzio}\ \emph
  {et~al.}(2017{\natexlab{a}})\citenamefont {Di~Luzio}, \citenamefont
  {Kamenik},\ and\ \citenamefont {Nardecchia}}]{DiLuzio:2016sur}%
  \BibitemOpen
  \bibfield  {author} {\bibinfo {author} {\bibfnamefont {L.}~\bibnamefont
  {Di~Luzio}}, \bibinfo {author} {\bibfnamefont {J.~F.}\ \bibnamefont
  {Kamenik}}, \ and\ \bibinfo {author} {\bibfnamefont {M.}~\bibnamefont
  {Nardecchia}},\ }\href {\doibase 10.1140/epjc/s10052-017-4594-2} {\bibfield
  {journal} {\bibinfo  {journal} {Eur. Phys. J.}\ }\textbf {\bibinfo {volume}
  {C77}},\ \bibinfo {pages} {30} (\bibinfo {year} {2017}{\natexlab{a}})},\
  \Eprint {http://arxiv.org/abs/1604.05746} {arXiv:1604.05746 [hep-ph]}
  \BibitemShut {NoStop}%
\bibitem [{\citenamefont {Aaboud}\ \emph {et~al.}(2017)\citenamefont {Aaboud}
  \emph {et~al.}}]{Aaboud:2017buh}%
  \BibitemOpen
  \bibfield  {author} {\bibinfo {author} {\bibfnamefont {M.}~\bibnamefont
  {Aaboud}} \emph {et~al.} (\bibinfo {collaboration} {ATLAS}),\ }\href
  {\doibase 10.1007/JHEP10(2017)182} {\bibfield  {journal} {\bibinfo  {journal}
  {JHEP}\ }\textbf {\bibinfo {volume} {10}},\ \bibinfo {pages} {182} (\bibinfo
  {year} {2017})},\ \Eprint {http://arxiv.org/abs/1707.02424} {arXiv:1707.02424
  [hep-ex]} \BibitemShut {NoStop}%
\bibitem [{\citenamefont {Allanach}\ \emph {et~al.}(2017)\citenamefont
  {Allanach}, \citenamefont {Gripaios},\ and\ \citenamefont
  {You}}]{Allanach:2017bta}%
  \BibitemOpen
  \bibfield  {author} {\bibinfo {author} {\bibfnamefont {B.~C.}\ \bibnamefont
  {Allanach}}, \bibinfo {author} {\bibfnamefont {B.}~\bibnamefont {Gripaios}},
  \ and\ \bibinfo {author} {\bibfnamefont {T.}~\bibnamefont {You}},\
  }\href@noop {} {\  (\bibinfo {year} {2017})},\ \Eprint
  {http://arxiv.org/abs/1710.06363} {arXiv:1710.06363 [hep-ph]} \BibitemShut
  {NoStop}%
\bibitem [{\citenamefont {Hiller}\ and\ \citenamefont
  {Schmaltz}(2014)}]{Hiller:2014yaa}%
  \BibitemOpen
  \bibfield  {author} {\bibinfo {author} {\bibfnamefont {G.}~\bibnamefont
  {Hiller}}\ and\ \bibinfo {author} {\bibfnamefont {M.}~\bibnamefont
  {Schmaltz}},\ }\href {\doibase 10.1103/PhysRevD.90.054014} {\bibfield
  {journal} {\bibinfo  {journal} {Phys. Rev.}\ }\textbf {\bibinfo {volume}
  {D90}},\ \bibinfo {pages} {054014} (\bibinfo {year} {2014})},\ \Eprint
  {http://arxiv.org/abs/1408.1627} {arXiv:1408.1627 [hep-ph]} \BibitemShut
  {NoStop}%
\bibitem [{\citenamefont {Gripaios}\ \emph {et~al.}(2015)\citenamefont
  {Gripaios}, \citenamefont {Nardecchia},\ and\ \citenamefont
  {Renner}}]{Gripaios:2014tna}%
  \BibitemOpen
  \bibfield  {author} {\bibinfo {author} {\bibfnamefont {B.}~\bibnamefont
  {Gripaios}}, \bibinfo {author} {\bibfnamefont {M.}~\bibnamefont
  {Nardecchia}}, \ and\ \bibinfo {author} {\bibfnamefont {S.~A.}\ \bibnamefont
  {Renner}},\ }\href {\doibase 10.1007/JHEP05(2015)006} {\bibfield  {journal}
  {\bibinfo  {journal} {JHEP}\ }\textbf {\bibinfo {volume} {05}},\ \bibinfo
  {pages} {006} (\bibinfo {year} {2015})},\ \Eprint
  {http://arxiv.org/abs/1412.1791} {arXiv:1412.1791 [hep-ph]} \BibitemShut
  {NoStop}%
\bibitem [{\citenamefont {de~Medeiros~Varzielas}\ and\ \citenamefont
  {Hiller}(2015)}]{Varzielas:2015iva}%
  \BibitemOpen
  \bibfield  {author} {\bibinfo {author} {\bibfnamefont {I.}~\bibnamefont
  {de~Medeiros~Varzielas}}\ and\ \bibinfo {author} {\bibfnamefont
  {G.}~\bibnamefont {Hiller}},\ }\href {\doibase 10.1007/JHEP06(2015)072}
  {\bibfield  {journal} {\bibinfo  {journal} {JHEP}\ }\textbf {\bibinfo
  {volume} {06}},\ \bibinfo {pages} {072} (\bibinfo {year} {2015})},\ \Eprint
  {http://arxiv.org/abs/1503.01084} {arXiv:1503.01084 [hep-ph]} \BibitemShut
  {NoStop}%
\bibitem [{\citenamefont {Becirevic}\ \emph {et~al.}(2015)\citenamefont
  {Becirevic}, \citenamefont {Fajfer},\ and\ \citenamefont
  {Kosnik}}]{Becirevic:2015asa}%
  \BibitemOpen
  \bibfield  {author} {\bibinfo {author} {\bibfnamefont {D.}~\bibnamefont
  {Becirevic}}, \bibinfo {author} {\bibfnamefont {S.}~\bibnamefont {Fajfer}}, \
  and\ \bibinfo {author} {\bibfnamefont {N.}~\bibnamefont {Kosnik}},\ }\href
  {\doibase 10.1103/PhysRevD.92.014016} {\bibfield  {journal} {\bibinfo
  {journal} {Phys. Rev.}\ }\textbf {\bibinfo {volume} {D92}},\ \bibinfo {pages}
  {014016} (\bibinfo {year} {2015})},\ \Eprint
  {http://arxiv.org/abs/1503.09024} {arXiv:1503.09024 [hep-ph]} \BibitemShut
  {NoStop}%
\bibitem [{\citenamefont {Alonso}\ \emph {et~al.}(2015)\citenamefont {Alonso},
  \citenamefont {Grinstein},\ and\ \citenamefont
  {Martin~Camalich}}]{Alonso:2015sja}%
  \BibitemOpen
  \bibfield  {author} {\bibinfo {author} {\bibfnamefont {R.}~\bibnamefont
  {Alonso}}, \bibinfo {author} {\bibfnamefont {B.}~\bibnamefont {Grinstein}}, \
  and\ \bibinfo {author} {\bibfnamefont {J.}~\bibnamefont {Martin~Camalich}},\
  }\href {\doibase 10.1007/JHEP10(2015)184} {\bibfield  {journal} {\bibinfo
  {journal} {JHEP}\ }\textbf {\bibinfo {volume} {10}},\ \bibinfo {pages} {184}
  (\bibinfo {year} {2015})},\ \Eprint {http://arxiv.org/abs/1505.05164}
  {arXiv:1505.05164 [hep-ph]} \BibitemShut {NoStop}%
\bibitem [{\citenamefont {Bauer}\ and\ \citenamefont
  {Neubert}(2016)}]{Bauer:2015knc}%
  \BibitemOpen
  \bibfield  {author} {\bibinfo {author} {\bibfnamefont {M.}~\bibnamefont
  {Bauer}}\ and\ \bibinfo {author} {\bibfnamefont {M.}~\bibnamefont
  {Neubert}},\ }\href {\doibase 10.1103/PhysRevLett.116.141802} {\bibfield
  {journal} {\bibinfo  {journal} {Phys. Rev. Lett.}\ }\textbf {\bibinfo
  {volume} {116}},\ \bibinfo {pages} {141802} (\bibinfo {year} {2016})},\
  \Eprint {http://arxiv.org/abs/1511.01900} {arXiv:1511.01900 [hep-ph]}
  \BibitemShut {NoStop}%
\bibitem [{\citenamefont {Fajfer}\ and\ \citenamefont
  {Kosnik}(2016)}]{Fajfer:2015ycq}%
  \BibitemOpen
  \bibfield  {author} {\bibinfo {author} {\bibfnamefont {S.}~\bibnamefont
  {Fajfer}}\ and\ \bibinfo {author} {\bibfnamefont {N.}~\bibnamefont
  {Kosnik}},\ }\href {\doibase 10.1016/j.physletb.2016.02.018} {\bibfield
  {journal} {\bibinfo  {journal} {Phys. Lett.}\ }\textbf {\bibinfo {volume}
  {B755}},\ \bibinfo {pages} {270} (\bibinfo {year} {2016})},\ \Eprint
  {http://arxiv.org/abs/1511.06024} {arXiv:1511.06024 [hep-ph]} \BibitemShut
  {NoStop}%
\bibitem [{\citenamefont {Barbieri}\ \emph {et~al.}(2016)\citenamefont
  {Barbieri}, \citenamefont {Isidori}, \citenamefont {Pattori},\ and\
  \citenamefont {Senia}}]{Barbieri:2015yvd}%
  \BibitemOpen
  \bibfield  {author} {\bibinfo {author} {\bibfnamefont {R.}~\bibnamefont
  {Barbieri}}, \bibinfo {author} {\bibfnamefont {G.}~\bibnamefont {Isidori}},
  \bibinfo {author} {\bibfnamefont {A.}~\bibnamefont {Pattori}}, \ and\
  \bibinfo {author} {\bibfnamefont {F.}~\bibnamefont {Senia}},\ }\href
  {\doibase 10.1140/epjc/s10052-016-3905-3} {\bibfield  {journal} {\bibinfo
  {journal} {Eur. Phys. J.}\ }\textbf {\bibinfo {volume} {C76}},\ \bibinfo
  {pages} {67} (\bibinfo {year} {2016})},\ \Eprint
  {http://arxiv.org/abs/1512.01560} {arXiv:1512.01560 [hep-ph]} \BibitemShut
  {NoStop}%
\bibitem [{\citenamefont {Becirevic}\ \emph
  {et~al.}(2016{\natexlab{a}})\citenamefont {Becirevic}, \citenamefont
  {Kosnik}, \citenamefont {Sumensari},\ and\ \citenamefont
  {Zukanovich~Funchal}}]{Becirevic:2016oho}%
  \BibitemOpen
  \bibfield  {author} {\bibinfo {author} {\bibfnamefont {D.}~\bibnamefont
  {Becirevic}}, \bibinfo {author} {\bibfnamefont {N.}~\bibnamefont {Kosnik}},
  \bibinfo {author} {\bibfnamefont {O.}~\bibnamefont {Sumensari}}, \ and\
  \bibinfo {author} {\bibfnamefont {R.}~\bibnamefont {Zukanovich~Funchal}},\
  }\href {\doibase 10.1007/JHEP11(2016)035} {\bibfield  {journal} {\bibinfo
  {journal} {JHEP}\ }\textbf {\bibinfo {volume} {11}},\ \bibinfo {pages} {035}
  (\bibinfo {year} {2016}{\natexlab{a}})},\ \Eprint
  {http://arxiv.org/abs/1608.07583} {arXiv:1608.07583 [hep-ph]} \BibitemShut
  {NoStop}%
\bibitem [{\citenamefont {Becirevic}\ \emph
  {et~al.}(2016{\natexlab{b}})\citenamefont {Becirevic}, \citenamefont
  {Fajfer}, \citenamefont {Kosnik},\ and\ \citenamefont
  {Sumensari}}]{Becirevic:2016yqi}%
  \BibitemOpen
  \bibfield  {author} {\bibinfo {author} {\bibfnamefont {D.}~\bibnamefont
  {Becirevic}}, \bibinfo {author} {\bibfnamefont {S.}~\bibnamefont {Fajfer}},
  \bibinfo {author} {\bibfnamefont {N.}~\bibnamefont {Kosnik}}, \ and\ \bibinfo
  {author} {\bibfnamefont {O.}~\bibnamefont {Sumensari}},\ }\href {\doibase
  10.1103/PhysRevD.94.115021} {\bibfield  {journal} {\bibinfo  {journal} {Phys.
  Rev.}\ }\textbf {\bibinfo {volume} {D94}},\ \bibinfo {pages} {115021}
  (\bibinfo {year} {2016}{\natexlab{b}})},\ \Eprint
  {http://arxiv.org/abs/1608.08501} {arXiv:1608.08501 [hep-ph]} \BibitemShut
  {NoStop}%
\bibitem [{\citenamefont {Crivellin}\ \emph
  {et~al.}(2017{\natexlab{b}})\citenamefont {Crivellin}, \citenamefont
  {Muller},\ and\ \citenamefont {Ota}}]{Crivellin:2017zlb}%
  \BibitemOpen
  \bibfield  {author} {\bibinfo {author} {\bibfnamefont {A.}~\bibnamefont
  {Crivellin}}, \bibinfo {author} {\bibfnamefont {D.}~\bibnamefont {Muller}}, \
  and\ \bibinfo {author} {\bibfnamefont {T.}~\bibnamefont {Ota}},\ }\href
  {\doibase 10.1007/JHEP09(2017)040} {\bibfield  {journal} {\bibinfo  {journal}
  {JHEP}\ }\textbf {\bibinfo {volume} {09}},\ \bibinfo {pages} {040} (\bibinfo
  {year} {2017}{\natexlab{b}})},\ \Eprint {http://arxiv.org/abs/1703.09226}
  {arXiv:1703.09226 [hep-ph]} \BibitemShut {NoStop}%
\bibitem [{\citenamefont {Hiller}\ and\ \citenamefont
  {Nisandzic}(2017)}]{Hiller:2017bzc}%
  \BibitemOpen
  \bibfield  {author} {\bibinfo {author} {\bibfnamefont {G.}~\bibnamefont
  {Hiller}}\ and\ \bibinfo {author} {\bibfnamefont {I.}~\bibnamefont
  {Nisandzic}},\ }\href {\doibase 10.1103/PhysRevD.96.035003} {\bibfield
  {journal} {\bibinfo  {journal} {Phys. Rev.}\ }\textbf {\bibinfo {volume}
  {D96}},\ \bibinfo {pages} {035003} (\bibinfo {year} {2017})},\ \Eprint
  {http://arxiv.org/abs/1704.05444} {arXiv:1704.05444 [hep-ph]} \BibitemShut
  {NoStop}%
\bibitem [{\citenamefont {Becirevic}\ and\ \citenamefont
  {Sumensari}(2017)}]{Becirevic:2017jtw}%
  \BibitemOpen
  \bibfield  {author} {\bibinfo {author} {\bibfnamefont {D.}~\bibnamefont
  {Becirevic}}\ and\ \bibinfo {author} {\bibfnamefont {O.}~\bibnamefont
  {Sumensari}},\ }\href {\doibase 10.1007/JHEP08(2017)104} {\bibfield
  {journal} {\bibinfo  {journal} {JHEP}\ }\textbf {\bibinfo {volume} {08}},\
  \bibinfo {pages} {104} (\bibinfo {year} {2017})},\ \Eprint
  {http://arxiv.org/abs/1704.05835} {arXiv:1704.05835 [hep-ph]} \BibitemShut
  {NoStop}%
\bibitem [{\citenamefont {Dorsner}\ \emph {et~al.}(2017)\citenamefont
  {Dorsner}, \citenamefont {Fajfer}, \citenamefont {Faroughy},\ and\
  \citenamefont {Kosnik}}]{Dorsner:2017ufx}%
  \BibitemOpen
  \bibfield  {author} {\bibinfo {author} {\bibfnamefont {I.}~\bibnamefont
  {Dorsner}}, \bibinfo {author} {\bibfnamefont {S.}~\bibnamefont {Fajfer}},
  \bibinfo {author} {\bibfnamefont {D.~A.}\ \bibnamefont {Faroughy}}, \ and\
  \bibinfo {author} {\bibfnamefont {N.}~\bibnamefont {Kosnik}},\ }\href
  {\doibase 10.1007/JHEP10(2017)188} {\  (\bibinfo {year} {2017}),\
  10.1007/JHEP10(2017)188},\ \bibinfo {note} {[JHEP10,188(2017)]},\ \Eprint
  {http://arxiv.org/abs/1706.07779} {arXiv:1706.07779 [hep-ph]} \BibitemShut
  {NoStop}%
\bibitem [{\citenamefont {Assad}\ \emph {et~al.}(2017)\citenamefont {Assad},
  \citenamefont {Fornal},\ and\ \citenamefont {Grinstein}}]{Assad:2017iib}%
  \BibitemOpen
  \bibfield  {author} {\bibinfo {author} {\bibfnamefont {N.}~\bibnamefont
  {Assad}}, \bibinfo {author} {\bibfnamefont {B.}~\bibnamefont {Fornal}}, \
  and\ \bibinfo {author} {\bibfnamefont {B.}~\bibnamefont {Grinstein}},\
  }\href@noop {} {\  (\bibinfo {year} {2017})},\ \Eprint
  {http://arxiv.org/abs/1708.06350} {arXiv:1708.06350 [hep-ph]} \BibitemShut
  {NoStop}%
\bibitem [{\citenamefont {Di~Luzio}\ \emph
  {et~al.}(2017{\natexlab{b}})\citenamefont {Di~Luzio}, \citenamefont
  {Greljo},\ and\ \citenamefont {Nardecchia}}]{DiLuzio:2017vat}%
  \BibitemOpen
  \bibfield  {author} {\bibinfo {author} {\bibfnamefont {L.}~\bibnamefont
  {Di~Luzio}}, \bibinfo {author} {\bibfnamefont {A.}~\bibnamefont {Greljo}}, \
  and\ \bibinfo {author} {\bibfnamefont {M.}~\bibnamefont {Nardecchia}},\
  }\href {\doibase 10.1103/PhysRevD.96.115011} {\bibfield  {journal} {\bibinfo
  {journal} {Phys. Rev.}\ }\textbf {\bibinfo {volume} {D96}},\ \bibinfo {pages}
  {115011} (\bibinfo {year} {2017}{\natexlab{b}})},\ \Eprint
  {http://arxiv.org/abs/1708.08450} {arXiv:1708.08450 [hep-ph]} \BibitemShut
  {NoStop}%
\bibitem [{\citenamefont {Calibbi}\ \emph {et~al.}(2017)\citenamefont
  {Calibbi}, \citenamefont {Crivellin},\ and\ \citenamefont
  {Li}}]{Calibbi:2017qbu}%
  \BibitemOpen
  \bibfield  {author} {\bibinfo {author} {\bibfnamefont {L.}~\bibnamefont
  {Calibbi}}, \bibinfo {author} {\bibfnamefont {A.}~\bibnamefont {Crivellin}},
  \ and\ \bibinfo {author} {\bibfnamefont {T.}~\bibnamefont {Li}},\ }\href@noop
  {} {\  (\bibinfo {year} {2017})},\ \Eprint {http://arxiv.org/abs/1709.00692}
  {arXiv:1709.00692 [hep-ph]} \BibitemShut {NoStop}%
\bibitem [{\citenamefont {Bordone}\ \emph {et~al.}(2017)\citenamefont
  {Bordone}, \citenamefont {Cornella}, \citenamefont {Fuentes-Martin},\ and\
  \citenamefont {Isidori}}]{Bordone:2017bld}%
  \BibitemOpen
  \bibfield  {author} {\bibinfo {author} {\bibfnamefont {M.}~\bibnamefont
  {Bordone}}, \bibinfo {author} {\bibfnamefont {C.}~\bibnamefont {Cornella}},
  \bibinfo {author} {\bibfnamefont {J.}~\bibnamefont {Fuentes-Martin}}, \ and\
  \bibinfo {author} {\bibfnamefont {G.}~\bibnamefont {Isidori}},\ }\href@noop
  {} {\  (\bibinfo {year} {2017})},\ \Eprint {http://arxiv.org/abs/1712.01368}
  {arXiv:1712.01368 [hep-ph]} \BibitemShut {NoStop}%
\bibitem [{\citenamefont {Davidson}\ \emph {et~al.}(1994)\citenamefont
  {Davidson}, \citenamefont {Bailey},\ and\ \citenamefont
  {Campbell}}]{Davidson:1993qk}%
  \BibitemOpen
  \bibfield  {author} {\bibinfo {author} {\bibfnamefont {S.}~\bibnamefont
  {Davidson}}, \bibinfo {author} {\bibfnamefont {D.~C.}\ \bibnamefont
  {Bailey}}, \ and\ \bibinfo {author} {\bibfnamefont {B.~A.}\ \bibnamefont
  {Campbell}},\ }\href {\doibase 10.1007/BF01552629} {\bibfield  {journal}
  {\bibinfo  {journal} {Z. Phys.}\ }\textbf {\bibinfo {volume} {C61}},\
  \bibinfo {pages} {613} (\bibinfo {year} {1994})},\ \Eprint
  {http://arxiv.org/abs/hep-ph/9309310} {arXiv:hep-ph/9309310 [hep-ph]}
  \BibitemShut {NoStop}%
\bibitem [{\citenamefont {Dor¨ner}\ \emph {et~al.}(2016)\citenamefont
  {Dor¨ner}, \citenamefont {Fajfer}, \citenamefont {Greljo}, \citenamefont
  {Kamenik},\ and\ \citenamefont {Ko¨nik}}]{Dorsner:2016wpm}%
  \BibitemOpen
  \bibfield  {author} {\bibinfo {author} {\bibfnamefont {I.}~\bibnamefont
  {Dor¨ner}}, \bibinfo {author} {\bibfnamefont {S.}~\bibnamefont {Fajfer}},
  \bibinfo {author} {\bibfnamefont {A.}~\bibnamefont {Greljo}}, \bibinfo
  {author} {\bibfnamefont {J.~F.}\ \bibnamefont {Kamenik}}, \ and\ \bibinfo
  {author} {\bibfnamefont {N.}~\bibnamefont {Ko¨nik}},\ }\href {\doibase
  10.1016/j.physrep.2016.06.001} {\bibfield  {journal} {\bibinfo  {journal}
  {Phys. Rept.}\ }\textbf {\bibinfo {volume} {641}},\ \bibinfo {pages} {1}
  (\bibinfo {year} {2016})},\ \Eprint {http://arxiv.org/abs/1603.04993}
  {arXiv:1603.04993 [hep-ph]} \BibitemShut {NoStop}%
\bibitem [{\citenamefont {Biggio}\ \emph {et~al.}(2016)\citenamefont {Biggio},
  \citenamefont {Bordone}, \citenamefont {Di~Luzio},\ and\ \citenamefont
  {Ridolfi}}]{Biggio:2016wyy}%
  \BibitemOpen
  \bibfield  {author} {\bibinfo {author} {\bibfnamefont {C.}~\bibnamefont
  {Biggio}}, \bibinfo {author} {\bibfnamefont {M.}~\bibnamefont {Bordone}},
  \bibinfo {author} {\bibfnamefont {L.}~\bibnamefont {Di~Luzio}}, \ and\
  \bibinfo {author} {\bibfnamefont {G.}~\bibnamefont {Ridolfi}},\ }\href
  {\doibase 10.1007/JHEP10(2016)002} {\bibfield  {journal} {\bibinfo  {journal}
  {JHEP}\ }\textbf {\bibinfo {volume} {10}},\ \bibinfo {pages} {002} (\bibinfo
  {year} {2016})},\ \Eprint {http://arxiv.org/abs/1607.07621} {arXiv:1607.07621
  [hep-ph]} \BibitemShut {NoStop}%
\bibitem [{\citenamefont {Bobeth}\ and\ \citenamefont
  {Buras}(2017)}]{Bobeth:2017ecx}%
  \BibitemOpen
  \bibfield  {author} {\bibinfo {author} {\bibfnamefont {C.}~\bibnamefont
  {Bobeth}}\ and\ \bibinfo {author} {\bibfnamefont {A.~J.}\ \bibnamefont
  {Buras}},\ }\href@noop {} {\  (\bibinfo {year} {2017})},\ \Eprint
  {http://arxiv.org/abs/1712.01295} {arXiv:1712.01295 [hep-ph]} \BibitemShut
  {NoStop}%
\bibitem [{\citenamefont {Sirunyan}\ \emph
  {et~al.}(2017{\natexlab{b}})\citenamefont {Sirunyan} \emph
  {et~al.}}]{Sirunyan:2017yrk}%
  \BibitemOpen
  \bibfield  {author} {\bibinfo {author} {\bibfnamefont {A.~M.}\ \bibnamefont
  {Sirunyan}} \emph {et~al.} (\bibinfo {collaboration} {CMS}),\ }\href
  {\doibase 10.1007/JHEP07(2017)121} {\bibfield  {journal} {\bibinfo  {journal}
  {JHEP}\ }\textbf {\bibinfo {volume} {07}},\ \bibinfo {pages} {121} (\bibinfo
  {year} {2017}{\natexlab{b}})},\ \Eprint {http://arxiv.org/abs/1703.03995}
  {arXiv:1703.03995 [hep-ex]} \BibitemShut {NoStop}%
\bibitem [{\citenamefont {Lees}\ \emph {et~al.}(2013)\citenamefont {Lees} \emph
  {et~al.}}]{Lees:2013uzd}%
  \BibitemOpen
  \bibfield  {author} {\bibinfo {author} {\bibfnamefont {J.~P.}\ \bibnamefont
  {Lees}} \emph {et~al.} (\bibinfo {collaboration} {BaBar}),\ }\href {\doibase
  10.1103/PhysRevD.88.072012} {\bibfield  {journal} {\bibinfo  {journal} {Phys.
  Rev.}\ }\textbf {\bibinfo {volume} {D88}},\ \bibinfo {pages} {072012}
  (\bibinfo {year} {2013})},\ \Eprint {http://arxiv.org/abs/1303.0571}
  {arXiv:1303.0571 [hep-ex]} \BibitemShut {NoStop}%
\bibitem [{\citenamefont {Aaij}\ \emph
  {et~al.}(2015{\natexlab{b}})\citenamefont {Aaij} \emph
  {et~al.}}]{Aaij:2015yra}%
  \BibitemOpen
  \bibfield  {author} {\bibinfo {author} {\bibfnamefont {R.}~\bibnamefont
  {Aaij}} \emph {et~al.} (\bibinfo {collaboration} {LHCb}),\ }\href {\doibase
  10.1103/PhysRevLett.115.159901, 10.1103/PhysRevLett.115.111803} {\bibfield
  {journal} {\bibinfo  {journal} {Phys. Rev. Lett.}\ }\textbf {\bibinfo
  {volume} {115}},\ \bibinfo {pages} {111803} (\bibinfo {year}
  {2015}{\natexlab{b}})},\ \bibinfo {note} {[Erratum: Phys. Rev.
  Lett.115,no.15,159901(2015)]},\ \Eprint {http://arxiv.org/abs/1506.08614}
  {arXiv:1506.08614 [hep-ex]} \BibitemShut {NoStop}%
\bibitem [{\citenamefont {Hirose}\ \emph {et~al.}(2017)\citenamefont {Hirose}
  \emph {et~al.}}]{Hirose:2016wfn}%
  \BibitemOpen
  \bibfield  {author} {\bibinfo {author} {\bibfnamefont {S.}~\bibnamefont
  {Hirose}} \emph {et~al.} (\bibinfo {collaboration} {Belle}),\ }\href
  {\doibase 10.1103/PhysRevLett.118.211801} {\bibfield  {journal} {\bibinfo
  {journal} {Phys. Rev. Lett.}\ }\textbf {\bibinfo {volume} {118}},\ \bibinfo
  {pages} {211801} (\bibinfo {year} {2017})},\ \Eprint
  {http://arxiv.org/abs/1612.00529} {arXiv:1612.00529 [hep-ex]} \BibitemShut
  {NoStop}%
\bibitem [{\citenamefont {Feruglio}\ \emph
  {et~al.}(2017{\natexlab{a}})\citenamefont {Feruglio}, \citenamefont
  {Paradisi},\ and\ \citenamefont {Pattori}}]{Feruglio:2016gvd}%
  \BibitemOpen
  \bibfield  {author} {\bibinfo {author} {\bibfnamefont {F.}~\bibnamefont
  {Feruglio}}, \bibinfo {author} {\bibfnamefont {P.}~\bibnamefont {Paradisi}},
  \ and\ \bibinfo {author} {\bibfnamefont {A.}~\bibnamefont {Pattori}},\ }\href
  {\doibase 10.1103/PhysRevLett.118.011801} {\bibfield  {journal} {\bibinfo
  {journal} {Phys. Rev. Lett.}\ }\textbf {\bibinfo {volume} {118}},\ \bibinfo
  {pages} {011801} (\bibinfo {year} {2017}{\natexlab{a}})},\ \Eprint
  {http://arxiv.org/abs/1606.00524} {arXiv:1606.00524 [hep-ph]} \BibitemShut
  {NoStop}%
\bibitem [{\citenamefont {Feruglio}\ \emph
  {et~al.}(2017{\natexlab{b}})\citenamefont {Feruglio}, \citenamefont
  {Paradisi},\ and\ \citenamefont {Pattori}}]{Feruglio:2017rjo}%
  \BibitemOpen
  \bibfield  {author} {\bibinfo {author} {\bibfnamefont {F.}~\bibnamefont
  {Feruglio}}, \bibinfo {author} {\bibfnamefont {P.}~\bibnamefont {Paradisi}},
  \ and\ \bibinfo {author} {\bibfnamefont {A.}~\bibnamefont {Pattori}},\ }\href
  {\doibase 10.1007/JHEP09(2017)061} {\bibfield  {journal} {\bibinfo  {journal}
  {JHEP}\ }\textbf {\bibinfo {volume} {09}},\ \bibinfo {pages} {061} (\bibinfo
  {year} {2017}{\natexlab{b}})},\ \Eprint {http://arxiv.org/abs/1705.00929}
  {arXiv:1705.00929 [hep-ph]} \BibitemShut {NoStop}%
\bibitem [{\citenamefont {Buttazzo}\ \emph {et~al.}(2017)\citenamefont
  {Buttazzo}, \citenamefont {Greljo}, \citenamefont {Isidori},\ and\
  \citenamefont {Marzocca}}]{Buttazzo:2017ixm}%
  \BibitemOpen
  \bibfield  {author} {\bibinfo {author} {\bibfnamefont {D.}~\bibnamefont
  {Buttazzo}}, \bibinfo {author} {\bibfnamefont {A.}~\bibnamefont {Greljo}},
  \bibinfo {author} {\bibfnamefont {G.}~\bibnamefont {Isidori}}, \ and\
  \bibinfo {author} {\bibfnamefont {D.}~\bibnamefont {Marzocca}},\ }\href
  {\doibase 10.1007/JHEP11(2017)044} {\bibfield  {journal} {\bibinfo  {journal}
  {JHEP}\ }\textbf {\bibinfo {volume} {11}},\ \bibinfo {pages} {044} (\bibinfo
  {year} {2017})},\ \Eprint {http://arxiv.org/abs/1706.07808} {arXiv:1706.07808
  [hep-ph]} \BibitemShut {NoStop}%
\bibitem [{\citenamefont {Greljo}\ \emph {et~al.}(2015)\citenamefont {Greljo},
  \citenamefont {Isidori},\ and\ \citenamefont {Marzocca}}]{Greljo:2015mma}%
  \BibitemOpen
  \bibfield  {author} {\bibinfo {author} {\bibfnamefont {A.}~\bibnamefont
  {Greljo}}, \bibinfo {author} {\bibfnamefont {G.}~\bibnamefont {Isidori}}, \
  and\ \bibinfo {author} {\bibfnamefont {D.}~\bibnamefont {Marzocca}},\ }\href
  {\doibase 10.1007/JHEP07(2015)142} {\bibfield  {journal} {\bibinfo  {journal}
  {JHEP}\ }\textbf {\bibinfo {volume} {07}},\ \bibinfo {pages} {142} (\bibinfo
  {year} {2015})},\ \Eprint {http://arxiv.org/abs/1506.01705} {arXiv:1506.01705
  [hep-ph]} \BibitemShut {NoStop}%
\bibitem [{\citenamefont {Lenz}\ and\ \citenamefont
  {Nierste}(2007)}]{Lenz:2006hd}%
  \BibitemOpen
  \bibfield  {author} {\bibinfo {author} {\bibfnamefont {A.}~\bibnamefont
  {Lenz}}\ and\ \bibinfo {author} {\bibfnamefont {U.}~\bibnamefont {Nierste}},\
  }\href {\doibase 10.1088/1126-6708/2007/06/072} {\bibfield  {journal}
  {\bibinfo  {journal} {JHEP}\ }\textbf {\bibinfo {volume} {06}},\ \bibinfo
  {pages} {072} (\bibinfo {year} {2007})},\ \Eprint
  {http://arxiv.org/abs/hep-ph/0612167} {arXiv:hep-ph/0612167 [hep-ph]}
  \BibitemShut {NoStop}%
\bibitem [{\citenamefont {Charles}\ \emph {et~al.}(2005)\citenamefont
  {Charles}, \citenamefont {Hocker}, \citenamefont {Lacker}, \citenamefont
  {Laplace}, \citenamefont {Le~Diberder}, \citenamefont {Malcles},
  \citenamefont {Ocariz}, \citenamefont {Pivk},\ and\ \citenamefont
  {Roos}}]{Charles:2004jd}%
  \BibitemOpen
  \bibfield  {author} {\bibinfo {author} {\bibfnamefont {J.}~\bibnamefont
  {Charles}}, \bibinfo {author} {\bibfnamefont {A.}~\bibnamefont {Hocker}},
  \bibinfo {author} {\bibfnamefont {H.}~\bibnamefont {Lacker}}, \bibinfo
  {author} {\bibfnamefont {S.}~\bibnamefont {Laplace}}, \bibinfo {author}
  {\bibfnamefont {F.~R.}\ \bibnamefont {Le~Diberder}}, \bibinfo {author}
  {\bibfnamefont {J.}~\bibnamefont {Malcles}}, \bibinfo {author} {\bibfnamefont
  {J.}~\bibnamefont {Ocariz}}, \bibinfo {author} {\bibfnamefont
  {M.}~\bibnamefont {Pivk}}, \ and\ \bibinfo {author} {\bibfnamefont
  {L.}~\bibnamefont {Roos}} (\bibinfo {collaboration} {CKMfitter Group}),\
  }\href {\doibase 10.1140/epjc/s2005-02169-1} {\bibfield  {journal} {\bibinfo
  {journal} {Eur. Phys. J.}\ }\textbf {\bibinfo {volume} {C41}},\ \bibinfo
  {pages} {1} (\bibinfo {year} {2005})},\ \Eprint
  {http://arxiv.org/abs/hep-ph/0406184} {arXiv:hep-ph/0406184 [hep-ph]}
  \BibitemShut {NoStop}%
\bibitem [{\citenamefont {Buras}\ \emph {et~al.}(2001)\citenamefont {Buras},
  \citenamefont {Jager},\ and\ \citenamefont {Urban}}]{Buras:2001ra}%
  \BibitemOpen
  \bibfield  {author} {\bibinfo {author} {\bibfnamefont {A.~J.}\ \bibnamefont
  {Buras}}, \bibinfo {author} {\bibfnamefont {S.}~\bibnamefont {Jager}}, \ and\
  \bibinfo {author} {\bibfnamefont {J.}~\bibnamefont {Urban}},\ }\href
  {\doibase 10.1016/S0550-3213(01)00207-3} {\bibfield  {journal} {\bibinfo
  {journal} {Nucl. Phys.}\ }\textbf {\bibinfo {volume} {B605}},\ \bibinfo
  {pages} {600} (\bibinfo {year} {2001})},\ \Eprint
  {http://arxiv.org/abs/hep-ph/0102316} {arXiv:hep-ph/0102316 [hep-ph]}
  \BibitemShut {NoStop}%
\bibitem [{\citenamefont {Patrignani}\ \emph {et~al.}(2016)\citenamefont
  {Patrignani} \emph {et~al.}}]{Patrignani:2016xqp}%
  \BibitemOpen
  \bibfield  {author} {\bibinfo {author} {\bibfnamefont {C.}~\bibnamefont
  {Patrignani}} \emph {et~al.} (\bibinfo {collaboration} {Particle Data
  Group}),\ }\href {\doibase 10.1088/1674-1137/40/10/100001} {\bibfield
  {journal} {\bibinfo  {journal} {Chin. Phys.}\ }\textbf {\bibinfo {volume}
  {C40}},\ \bibinfo {pages} {100001} (\bibinfo {year} {2016})}\BibitemShut
  {NoStop}%
\bibitem [{\citenamefont {Beneke}\ \emph {et~al.}(2015)\citenamefont {Beneke},
  \citenamefont {Maier}, \citenamefont {Piclum},\ and\ \citenamefont
  {Rauh}}]{Beneke:2014pta}%
  \BibitemOpen
  \bibfield  {author} {\bibinfo {author} {\bibfnamefont {M.}~\bibnamefont
  {Beneke}}, \bibinfo {author} {\bibfnamefont {A.}~\bibnamefont {Maier}},
  \bibinfo {author} {\bibfnamefont {J.}~\bibnamefont {Piclum}}, \ and\ \bibinfo
  {author} {\bibfnamefont {T.}~\bibnamefont {Rauh}},\ }\href {\doibase
  10.1016/j.nuclphysb.2014.12.001} {\bibfield  {journal} {\bibinfo  {journal}
  {Nucl. Phys.}\ }\textbf {\bibinfo {volume} {B891}},\ \bibinfo {pages} {42}
  (\bibinfo {year} {2015})},\ \Eprint {http://arxiv.org/abs/1411.3132}
  {arXiv:1411.3132 [hep-ph]} \BibitemShut {NoStop}%
\bibitem [{\citenamefont {Beneke}\ \emph {et~al.}(2016)\citenamefont {Beneke},
  \citenamefont {Maier}, \citenamefont {Piclum},\ and\ \citenamefont
  {Rauh}}]{Beneke:2016oox}%
  \BibitemOpen
  \bibfield  {author} {\bibinfo {author} {\bibfnamefont {M.}~\bibnamefont
  {Beneke}}, \bibinfo {author} {\bibfnamefont {A.}~\bibnamefont {Maier}},
  \bibinfo {author} {\bibfnamefont {J.}~\bibnamefont {Piclum}}, \ and\ \bibinfo
  {author} {\bibfnamefont {T.}~\bibnamefont {Rauh}},\ }\bibfield  {booktitle}
  {\emph {\bibinfo {booktitle} {{Proceedings, 12th International Symposium on
  Radiative Corrections (Radcor 2015) and LoopFest XIV (Radiative Corrections
  for the LHC and Future Colliders): Los Angeles, CA, USA, June 15-19,
  2015}}},\ }\href@noop {} {\bibfield  {journal} {\bibinfo  {journal} {PoS}\
  }\textbf {\bibinfo {volume} {RADCOR2015}},\ \bibinfo {pages} {035} (\bibinfo
  {year} {2016})},\ \Eprint {http://arxiv.org/abs/1601.02949} {arXiv:1601.02949
  [hep-ph]} \BibitemShut {NoStop}%
\bibitem [{\citenamefont {Bona}\ \emph {et~al.}(2006)\citenamefont {Bona} \emph
  {et~al.}}]{Bona:2006ah}%
  \BibitemOpen
  \bibfield  {author} {\bibinfo {author} {\bibfnamefont {M.}~\bibnamefont
  {Bona}} \emph {et~al.} (\bibinfo {collaboration} {UTfit}),\ }\href {\doibase
  10.1088/1126-6708/2006/10/081} {\bibfield  {journal} {\bibinfo  {journal}
  {JHEP}\ }\textbf {\bibinfo {volume} {10}},\ \bibinfo {pages} {081} (\bibinfo
  {year} {2006})},\ \Eprint {http://arxiv.org/abs/hep-ph/0606167}
  {arXiv:hep-ph/0606167 [hep-ph]} \BibitemShut {NoStop}%
\bibitem [{\citenamefont {Herren}\ and\ \citenamefont
  {Steinhauser}(2017)}]{Herren:2017osy}%
  \BibitemOpen
  \bibfield  {author} {\bibinfo {author} {\bibfnamefont {F.}~\bibnamefont
  {Herren}}\ and\ \bibinfo {author} {\bibfnamefont {M.}~\bibnamefont
  {Steinhauser}},\ }\href@noop {} {\  (\bibinfo {year} {2017})},\ \Eprint
  {http://arxiv.org/abs/1703.03751} {arXiv:1703.03751 [hep-ph]} \BibitemShut
  {NoStop}%
\bibitem [{\citenamefont {Baikov}\ \emph {et~al.}(2017)\citenamefont {Baikov},
  \citenamefont {Chetyrkin},\ and\ \citenamefont {Kühn}}]{Baikov:2016tgj}%
  \BibitemOpen
  \bibfield  {author} {\bibinfo {author} {\bibfnamefont {P.~A.}\ \bibnamefont
  {Baikov}}, \bibinfo {author} {\bibfnamefont {K.~G.}\ \bibnamefont
  {Chetyrkin}}, \ and\ \bibinfo {author} {\bibfnamefont {J.~H.}\ \bibnamefont
  {Kühn}},\ }\href {\doibase 10.1103/PhysRevLett.118.082002} {\bibfield
  {journal} {\bibinfo  {journal} {Phys. Rev. Lett.}\ }\textbf {\bibinfo
  {volume} {118}},\ \bibinfo {pages} {082002} (\bibinfo {year} {2017})},\
  \Eprint {http://arxiv.org/abs/1606.08659} {arXiv:1606.08659 [hep-ph]}
  \BibitemShut {NoStop}%
\bibitem [{\citenamefont {Herzog}\ \emph {et~al.}(2017)\citenamefont {Herzog},
  \citenamefont {Ruijl}, \citenamefont {Ueda}, \citenamefont {Vermaseren},\
  and\ \citenamefont {Vogt}}]{Herzog:2017ohr}%
  \BibitemOpen
  \bibfield  {author} {\bibinfo {author} {\bibfnamefont {F.}~\bibnamefont
  {Herzog}}, \bibinfo {author} {\bibfnamefont {B.}~\bibnamefont {Ruijl}},
  \bibinfo {author} {\bibfnamefont {T.}~\bibnamefont {Ueda}}, \bibinfo {author}
  {\bibfnamefont {J.~A.~M.}\ \bibnamefont {Vermaseren}}, \ and\ \bibinfo
  {author} {\bibfnamefont {A.}~\bibnamefont {Vogt}},\ }\href {\doibase
  10.1007/JHEP02(2017)090} {\bibfield  {journal} {\bibinfo  {journal} {JHEP}\
  }\textbf {\bibinfo {volume} {02}},\ \bibinfo {pages} {090} (\bibinfo {year}
  {2017})},\ \Eprint {http://arxiv.org/abs/1701.01404} {arXiv:1701.01404
  [hep-ph]} \BibitemShut {NoStop}%
\bibitem [{\citenamefont {Luthe}\ \emph
  {et~al.}(2017{\natexlab{a}})\citenamefont {Luthe}, \citenamefont {Maier},
  \citenamefont {Marquard},\ and\ \citenamefont {Schroder}}]{Luthe:2017ttc}%
  \BibitemOpen
  \bibfield  {author} {\bibinfo {author} {\bibfnamefont {T.}~\bibnamefont
  {Luthe}}, \bibinfo {author} {\bibfnamefont {A.}~\bibnamefont {Maier}},
  \bibinfo {author} {\bibfnamefont {P.}~\bibnamefont {Marquard}}, \ and\
  \bibinfo {author} {\bibfnamefont {Y.}~\bibnamefont {Schroder}},\ }\href
  {\doibase 10.1007/JHEP03(2017)020} {\bibfield  {journal} {\bibinfo  {journal}
  {JHEP}\ }\textbf {\bibinfo {volume} {03}},\ \bibinfo {pages} {020} (\bibinfo
  {year} {2017}{\natexlab{a}})},\ \Eprint {http://arxiv.org/abs/1701.07068}
  {arXiv:1701.07068 [hep-ph]} \BibitemShut {NoStop}%
\bibitem [{\citenamefont {Luthe}\ \emph
  {et~al.}(2017{\natexlab{b}})\citenamefont {Luthe}, \citenamefont {Maier},
  \citenamefont {Marquard},\ and\ \citenamefont {Schroder}}]{Luthe:2017ttg}%
  \BibitemOpen
  \bibfield  {author} {\bibinfo {author} {\bibfnamefont {T.}~\bibnamefont
  {Luthe}}, \bibinfo {author} {\bibfnamefont {A.}~\bibnamefont {Maier}},
  \bibinfo {author} {\bibfnamefont {P.}~\bibnamefont {Marquard}}, \ and\
  \bibinfo {author} {\bibfnamefont {Y.}~\bibnamefont {Schroder}},\ }\href
  {\doibase 10.1007/JHEP10(2017)166} {\bibfield  {journal} {\bibinfo  {journal}
  {JHEP}\ }\textbf {\bibinfo {volume} {10}},\ \bibinfo {pages} {166} (\bibinfo
  {year} {2017}{\natexlab{b}})},\ \Eprint {http://arxiv.org/abs/1709.07718}
  {arXiv:1709.07718 [hep-ph]} \BibitemShut {NoStop}%
\bibitem [{\citenamefont {Chetyrkin}\ \emph {et~al.}(2017)\citenamefont
  {Chetyrkin}, \citenamefont {Falcioni}, \citenamefont {Herzog},\ and\
  \citenamefont {Vermaseren}}]{Chetyrkin:2017bjc}%
  \BibitemOpen
  \bibfield  {author} {\bibinfo {author} {\bibfnamefont {K.~G.}\ \bibnamefont
  {Chetyrkin}}, \bibinfo {author} {\bibfnamefont {G.}~\bibnamefont {Falcioni}},
  \bibinfo {author} {\bibfnamefont {F.}~\bibnamefont {Herzog}}, \ and\ \bibinfo
  {author} {\bibfnamefont {J.~A.~M.}\ \bibnamefont {Vermaseren}},\ }\href
  {\doibase 10.1007/JHEP12(2017)006, 10.1007/JHEP10(2017)179} {\bibfield
  {journal} {\bibinfo  {journal} {JHEP}\ }\textbf {\bibinfo {volume} {10}},\
  \bibinfo {pages} {179} (\bibinfo {year} {2017})},\ \bibinfo {note}
  {[Addendum: JHEP12,006(2017)]},\ \Eprint {http://arxiv.org/abs/1709.08541}
  {arXiv:1709.08541 [hep-ph]} \BibitemShut {NoStop}%
\bibitem [{\citenamefont {Dowdall}\ \emph {et~al.}(2014)\citenamefont
  {Dowdall}, \citenamefont {Davies}, \citenamefont {Horgan}, \citenamefont
  {Lepage}, \citenamefont {Monahan},\ and\ \citenamefont
  {Shigemitsu}}]{Dowdall:2014qka}%
  \BibitemOpen
  \bibfield  {author} {\bibinfo {author} {\bibfnamefont {R.~J.}\ \bibnamefont
  {Dowdall}}, \bibinfo {author} {\bibfnamefont {C.~T.~H.}\ \bibnamefont
  {Davies}}, \bibinfo {author} {\bibfnamefont {R.~R.}\ \bibnamefont {Horgan}},
  \bibinfo {author} {\bibfnamefont {G.~P.}\ \bibnamefont {Lepage}}, \bibinfo
  {author} {\bibfnamefont {C.~J.}\ \bibnamefont {Monahan}}, \ and\ \bibinfo
  {author} {\bibfnamefont {J.}~\bibnamefont {Shigemitsu}},\ }\href@noop {} {\
  (\bibinfo {year} {2014})},\ \Eprint {http://arxiv.org/abs/1411.6989}
  {arXiv:1411.6989 [hep-lat]} \BibitemShut {NoStop}%
\bibitem [{\citenamefont {Carrasco}\ \emph {et~al.}(2014)\citenamefont
  {Carrasco} \emph {et~al.}}]{Carrasco:2013zta}%
  \BibitemOpen
  \bibfield  {author} {\bibinfo {author} {\bibfnamefont {N.}~\bibnamefont
  {Carrasco}} \emph {et~al.} (\bibinfo {collaboration} {ETM}),\ }\href
  {\doibase 10.1007/JHEP03(2014)016} {\bibfield  {journal} {\bibinfo  {journal}
  {JHEP}\ }\textbf {\bibinfo {volume} {03}},\ \bibinfo {pages} {016} (\bibinfo
  {year} {2014})},\ \Eprint {http://arxiv.org/abs/1308.1851} {arXiv:1308.1851
  [hep-lat]} \BibitemShut {NoStop}%
\bibitem [{\citenamefont {Gamiz}\ \emph {et~al.}(2009)\citenamefont {Gamiz},
  \citenamefont {Davies}, \citenamefont {Lepage}, \citenamefont {Shigemitsu},\
  and\ \citenamefont {Wingate}}]{Gamiz:2009ku}%
  \BibitemOpen
  \bibfield  {author} {\bibinfo {author} {\bibfnamefont {E.}~\bibnamefont
  {Gamiz}}, \bibinfo {author} {\bibfnamefont {C.~T.~H.}\ \bibnamefont
  {Davies}}, \bibinfo {author} {\bibfnamefont {G.~P.}\ \bibnamefont {Lepage}},
  \bibinfo {author} {\bibfnamefont {J.}~\bibnamefont {Shigemitsu}}, \ and\
  \bibinfo {author} {\bibfnamefont {M.}~\bibnamefont {Wingate}} (\bibinfo
  {collaboration} {HPQCD}),\ }\href {\doibase 10.1103/PhysRevD.80.014503}
  {\bibfield  {journal} {\bibinfo  {journal} {Phys. Rev.}\ }\textbf {\bibinfo
  {volume} {D80}},\ \bibinfo {pages} {014503} (\bibinfo {year} {2009})},\
  \Eprint {http://arxiv.org/abs/0902.1815} {arXiv:0902.1815 [hep-lat]}
  \BibitemShut {NoStop}%
\bibitem [{\citenamefont {Aoki}\ \emph {et~al.}(2014)\citenamefont {Aoki} \emph
  {et~al.}}]{Aoki:2013ldr}%
  \BibitemOpen
  \bibfield  {author} {\bibinfo {author} {\bibfnamefont {S.}~\bibnamefont
  {Aoki}} \emph {et~al.},\ }\href {\doibase 10.1140/epjc/s10052-014-2890-7}
  {\bibfield  {journal} {\bibinfo  {journal} {Eur. Phys. J.}\ }\textbf
  {\bibinfo {volume} {C74}},\ \bibinfo {pages} {2890} (\bibinfo {year}
  {2014})},\ \Eprint {http://arxiv.org/abs/1310.8555} {arXiv:1310.8555
  [hep-lat]} \BibitemShut {NoStop}%
\bibitem [{\citenamefont {Gelhausen}\ \emph {et~al.}(2013)\citenamefont
  {Gelhausen}, \citenamefont {Khodjamirian}, \citenamefont {Pivovarov},\ and\
  \citenamefont {Rosenthal}}]{Gelhausen:2013wia}%
  \BibitemOpen
  \bibfield  {author} {\bibinfo {author} {\bibfnamefont {P.}~\bibnamefont
  {Gelhausen}}, \bibinfo {author} {\bibfnamefont {A.}~\bibnamefont
  {Khodjamirian}}, \bibinfo {author} {\bibfnamefont {A.~A.}\ \bibnamefont
  {Pivovarov}}, \ and\ \bibinfo {author} {\bibfnamefont {D.}~\bibnamefont
  {Rosenthal}},\ }\href {\doibase 10.1103/PhysRevD.88.014015,
  10.1103/PhysRevD.91.099901, 10.1103/PhysRevD.89.099901} {\bibfield  {journal}
  {\bibinfo  {journal} {Phys. Rev.}\ }\textbf {\bibinfo {volume} {D88}},\
  \bibinfo {pages} {014015} (\bibinfo {year} {2013})},\ \bibinfo {note}
  {[Erratum: Phys. Rev.D91,099901(2015)]},\ \Eprint
  {http://arxiv.org/abs/1305.5432} {arXiv:1305.5432 [hep-ph]} \BibitemShut
  {NoStop}%
\bibitem [{\citenamefont {Dalgic}\ \emph {et~al.}(2007)\citenamefont {Dalgic},
  \citenamefont {Gray}, \citenamefont {Gamiz}, \citenamefont {Davies},
  \citenamefont {Lepage}, \citenamefont {Shigemitsu}, \citenamefont
  {Trottier},\ and\ \citenamefont {Wingate}}]{Dalgic:2006gp}%
  \BibitemOpen
  \bibfield  {author} {\bibinfo {author} {\bibfnamefont {E.}~\bibnamefont
  {Dalgic}}, \bibinfo {author} {\bibfnamefont {A.}~\bibnamefont {Gray}},
  \bibinfo {author} {\bibfnamefont {E.}~\bibnamefont {Gamiz}}, \bibinfo
  {author} {\bibfnamefont {C.~T.~H.}\ \bibnamefont {Davies}}, \bibinfo {author}
  {\bibfnamefont {G.~P.}\ \bibnamefont {Lepage}}, \bibinfo {author}
  {\bibfnamefont {J.}~\bibnamefont {Shigemitsu}}, \bibinfo {author}
  {\bibfnamefont {H.}~\bibnamefont {Trottier}}, \ and\ \bibinfo {author}
  {\bibfnamefont {M.}~\bibnamefont {Wingate}},\ }\href {\doibase
  10.1103/PhysRevD.76.011501} {\bibfield  {journal} {\bibinfo  {journal} {Phys.
  Rev.}\ }\textbf {\bibinfo {volume} {D76}},\ \bibinfo {pages} {011501}
  (\bibinfo {year} {2007})},\ \Eprint {http://arxiv.org/abs/hep-lat/0610104}
  {arXiv:hep-lat/0610104 [hep-lat]} \BibitemShut {NoStop}%
\bibitem [{\citenamefont {Aoki}\ \emph {et~al.}(2015)\citenamefont {Aoki},
  \citenamefont {Ishikawa}, \citenamefont {Izubuchi}, \citenamefont {Lehner},\
  and\ \citenamefont {Soni}}]{Aoki:2014nga}%
  \BibitemOpen
  \bibfield  {author} {\bibinfo {author} {\bibfnamefont {Y.}~\bibnamefont
  {Aoki}}, \bibinfo {author} {\bibfnamefont {T.}~\bibnamefont {Ishikawa}},
  \bibinfo {author} {\bibfnamefont {T.}~\bibnamefont {Izubuchi}}, \bibinfo
  {author} {\bibfnamefont {C.}~\bibnamefont {Lehner}}, \ and\ \bibinfo {author}
  {\bibfnamefont {A.}~\bibnamefont {Soni}},\ }\href {\doibase
  10.1103/PhysRevD.91.114505} {\bibfield  {journal} {\bibinfo  {journal} {Phys.
  Rev.}\ }\textbf {\bibinfo {volume} {D91}},\ \bibinfo {pages} {114505}
  (\bibinfo {year} {2015})},\ \Eprint {http://arxiv.org/abs/1406.6192}
  {arXiv:1406.6192 [hep-lat]} \BibitemShut {NoStop}%
\bibitem [{\citenamefont {Bouchard}\ \emph {et~al.}(2011)\citenamefont
  {Bouchard}, \citenamefont {Freeland}, \citenamefont {Bernard}, \citenamefont
  {El-Khadra}, \citenamefont {Gamiz}, \citenamefont {Kronfeld}, \citenamefont
  {Laiho},\ and\ \citenamefont {Van~de Water}}]{Bouchard:2011xj}%
  \BibitemOpen
  \bibfield  {author} {\bibinfo {author} {\bibfnamefont {C.~M.}\ \bibnamefont
  {Bouchard}}, \bibinfo {author} {\bibfnamefont {E.~D.}\ \bibnamefont
  {Freeland}}, \bibinfo {author} {\bibfnamefont {C.}~\bibnamefont {Bernard}},
  \bibinfo {author} {\bibfnamefont {A.~X.}\ \bibnamefont {El-Khadra}}, \bibinfo
  {author} {\bibfnamefont {E.}~\bibnamefont {Gamiz}}, \bibinfo {author}
  {\bibfnamefont {A.~S.}\ \bibnamefont {Kronfeld}}, \bibinfo {author}
  {\bibfnamefont {J.}~\bibnamefont {Laiho}}, \ and\ \bibinfo {author}
  {\bibfnamefont {R.~S.}\ \bibnamefont {Van~de Water}},\ }\bibfield
  {booktitle} {\emph {\bibinfo {booktitle} {{Proceedings, 29th International
  Symposium on Lattice field theory (Lattice 2011): Squaw Valley, Lake Tahoe,
  USA, July 10-16, 2011}}},\ }\href@noop {} {\bibfield  {journal} {\bibinfo
  {journal} {PoS}\ }\textbf {\bibinfo {volume} {LATTICE2011}},\ \bibinfo
  {pages} {274} (\bibinfo {year} {2011})},\ \Eprint
  {http://arxiv.org/abs/1112.5642} {arXiv:1112.5642 [hep-lat]} \BibitemShut
  {NoStop}%
\bibitem [{\citenamefont {Caprini}\ \emph {et~al.}(1998)\citenamefont
  {Caprini}, \citenamefont {Lellouch},\ and\ \citenamefont
  {Neubert}}]{Caprini:1997mu}%
  \BibitemOpen
  \bibfield  {author} {\bibinfo {author} {\bibfnamefont {I.}~\bibnamefont
  {Caprini}}, \bibinfo {author} {\bibfnamefont {L.}~\bibnamefont {Lellouch}}, \
  and\ \bibinfo {author} {\bibfnamefont {M.}~\bibnamefont {Neubert}},\ }\href
  {\doibase 10.1016/S0550-3213(98)00350-2} {\bibfield  {journal} {\bibinfo
  {journal} {Nucl. Phys.}\ }\textbf {\bibinfo {volume} {B530}},\ \bibinfo
  {pages} {153} (\bibinfo {year} {1998})},\ \Eprint
  {http://arxiv.org/abs/hep-ph/9712417} {arXiv:hep-ph/9712417 [hep-ph]}
  \BibitemShut {NoStop}%
\bibitem [{\citenamefont {Boyd}\ \emph {et~al.}(1995)\citenamefont {Boyd},
  \citenamefont {Grinstein},\ and\ \citenamefont {Lebed}}]{Boyd:1994tt}%
  \BibitemOpen
  \bibfield  {author} {\bibinfo {author} {\bibfnamefont {C.~G.}\ \bibnamefont
  {Boyd}}, \bibinfo {author} {\bibfnamefont {B.}~\bibnamefont {Grinstein}}, \
  and\ \bibinfo {author} {\bibfnamefont {R.~F.}\ \bibnamefont {Lebed}},\ }\href
  {\doibase 10.1103/PhysRevLett.74.4603} {\bibfield  {journal} {\bibinfo
  {journal} {Phys. Rev. Lett.}\ }\textbf {\bibinfo {volume} {74}},\ \bibinfo
  {pages} {4603} (\bibinfo {year} {1995})},\ \Eprint
  {http://arxiv.org/abs/hep-ph/9412324} {arXiv:hep-ph/9412324 [hep-ph]}
  \BibitemShut {NoStop}%
\bibitem [{\citenamefont {Grinstein}\ and\ \citenamefont
  {Kobach}(2017)}]{Grinstein:2017nlq}%
  \BibitemOpen
  \bibfield  {author} {\bibinfo {author} {\bibfnamefont {B.}~\bibnamefont
  {Grinstein}}\ and\ \bibinfo {author} {\bibfnamefont {A.}~\bibnamefont
  {Kobach}},\ }\href {\doibase 10.1016/j.physletb.2017.05.078} {\bibfield
  {journal} {\bibinfo  {journal} {Phys. Lett.}\ }\textbf {\bibinfo {volume}
  {B771}},\ \bibinfo {pages} {359} (\bibinfo {year} {2017})},\ \Eprint
  {http://arxiv.org/abs/1703.08170} {arXiv:1703.08170 [hep-ph]} \BibitemShut
  {NoStop}%
\bibitem [{\citenamefont {Bigi}\ \emph {et~al.}(2017)\citenamefont {Bigi},
  \citenamefont {Gambino},\ and\ \citenamefont {Schacht}}]{Bigi:2017jbd}%
  \BibitemOpen
  \bibfield  {author} {\bibinfo {author} {\bibfnamefont {D.}~\bibnamefont
  {Bigi}}, \bibinfo {author} {\bibfnamefont {P.}~\bibnamefont {Gambino}}, \
  and\ \bibinfo {author} {\bibfnamefont {S.}~\bibnamefont {Schacht}},\ }\href
  {\doibase 10.1007/JHEP11(2017)061} {\bibfield  {journal} {\bibinfo  {journal}
  {JHEP}\ }\textbf {\bibinfo {volume} {11}},\ \bibinfo {pages} {061} (\bibinfo
  {year} {2017})},\ \Eprint {http://arxiv.org/abs/1707.09509} {arXiv:1707.09509
  [hep-ph]} \BibitemShut {NoStop}%
\bibitem [{\citenamefont {Bernlochner}\ \emph {et~al.}(2017)\citenamefont
  {Bernlochner}, \citenamefont {Ligeti}, \citenamefont {Papucci},\ and\
  \citenamefont {Robinson}}]{Bernlochner:2017xyx}%
  \BibitemOpen
  \bibfield  {author} {\bibinfo {author} {\bibfnamefont {F.~U.}\ \bibnamefont
  {Bernlochner}}, \bibinfo {author} {\bibfnamefont {Z.}~\bibnamefont {Ligeti}},
  \bibinfo {author} {\bibfnamefont {M.}~\bibnamefont {Papucci}}, \ and\
  \bibinfo {author} {\bibfnamefont {D.~J.}\ \bibnamefont {Robinson}},\ }\href
  {\doibase 10.1103/PhysRevD.96.091503} {\bibfield  {journal} {\bibinfo
  {journal} {Phys. Rev.}\ }\textbf {\bibinfo {volume} {D96}},\ \bibinfo {pages}
  {091503} (\bibinfo {year} {2017})},\ \Eprint
  {http://arxiv.org/abs/1708.07134} {arXiv:1708.07134 [hep-ph]} \BibitemShut
  {NoStop}%
\bibitem [{\citenamefont {Jaiswal}\ \emph {et~al.}(2017)\citenamefont
  {Jaiswal}, \citenamefont {Nandi},\ and\ \citenamefont
  {Patra}}]{Jaiswal:2017rve}%
  \BibitemOpen
  \bibfield  {author} {\bibinfo {author} {\bibfnamefont {S.}~\bibnamefont
  {Jaiswal}}, \bibinfo {author} {\bibfnamefont {S.}~\bibnamefont {Nandi}}, \
  and\ \bibinfo {author} {\bibfnamefont {S.~K.}\ \bibnamefont {Patra}},\ }\href
  {\doibase 10.1007/JHEP12(2017)060} {\bibfield  {journal} {\bibinfo  {journal}
  {JHEP}\ }\textbf {\bibinfo {volume} {12}},\ \bibinfo {pages} {060} (\bibinfo
  {year} {2017})},\ \Eprint {http://arxiv.org/abs/1707.09977} {arXiv:1707.09977
  [hep-ph]} \BibitemShut {NoStop}%
\bibitem [{\citenamefont {Descotes-Genon}\ and\ \citenamefont
  {Koppenburg}(2017)}]{Koppenburg:2017mad}%
  \BibitemOpen
  \bibfield  {author} {\bibinfo {author} {\bibfnamefont {S.}~\bibnamefont
  {Descotes-Genon}}\ and\ \bibinfo {author} {\bibfnamefont {P.}~\bibnamefont
  {Koppenburg}},\ }\href {\doibase 10.1146/annurev-nucl-101916-123109}
  {\bibfield  {journal} {\bibinfo  {journal} {Ann. Rev. Nucl. Part. Sci.}\
  }\textbf {\bibinfo {volume} {67}},\ \bibinfo {pages} {97} (\bibinfo {year}
  {2017})},\ \Eprint {http://arxiv.org/abs/1702.08834} {arXiv:1702.08834
  [hep-ex]} \BibitemShut {NoStop}%
\end{thebibliography}%

\end{document}